\documentclass[
		openright,
		twoside,
		a4paper,
		12pt, 
		]{memoir}
\usepackage{graphicx} %
\usepackage[utf8]{inputenc} %
\usepackage{color} %
\usepackage{listings} %
\usepackage{hyperref}
\usepackage{amsfonts} %
\usepackage{amsmath} %
\usepackage{amssymb} %
\usepackage{mathrsfs} %
\usepackage{latexsym} %
\usepackage{euscript} %
\usepackage{stmaryrd} %
\usepackage{cite} %
\usepackage{mciteplus} %
\usepackage{tocbasic} %
\usepackage[full]{textcomp} %
\usepackage[final, babel=true]{microtype} %
\usepackage{pifont} %
\usepackage{textcomp} %

\hypersetup{
	colorlinks,
	citecolor=black,
	filecolor=black,
	linkcolor=black,
	urlcolor=black,
	bookmarks=true,
	bookmarksopen=true,
	pdftitle=Alexandre Payez\hspace{1pt}---\hspace{1pt}Ph.D. thesis,
	pdfauthor={Alexandre Jean-Gerald Payez},
	pdfdisplaydoctitle
}

\author{A.~Payez\\\footnotesize{\emph{IFPA group, AGO Dept., U.~of Li\`ege, B4000 Li\`ege, Belgium}}
}

	\copypagestyle{myRuled}{ruled}
	\makeoddhead{myRuled}{}{}{\scshape\MakeLowercase\rightmark}
	\makeevenhead{myRuled}{\scshape\MakeLowercase\leftmark}{}{}

	\copypagestyle{myBiblioHeaders}{myRuled}
	\makeoddhead{myBiblioHeaders}{}{}{}

\usepackage{titlesec}
\titleformat{\section}
	{
		\Large\bfseries
		\fillast
	}
	{\thesection}
	{1em}
	{}
	\renewcommand{\thesubsubsection}{%
		\alph{subsubsection})
	}
\titleformat{\subsubsection}
	{
		\normalsize
		\bfseries
	}
	{\thesubsubsection}
	{0.em}
	{}
\titleformat{\paragraph}[runin]
	{
		\normalsize
		\itshape
	}
	{\theparagraph}
	{0em}
	{}
	[---]
\titlespacing{\paragraph}
{0pt}
{1.5ex plus .1ex minus .2ex}
{1pt}

\usepackage{titletoc}
\titlecontents{subsubsection}
              [9.85em]
              {}
              {\contentslabel{2.25em}}
              {\hspace*{-2.25em}}
              {\titlerule*[1pc]{.}\contentspage}
\titlecontents{paragraph}
              [14em]
              {}
              {\contentslabel{2.25em}}
              {\hspace*{-2.25em}}
              {\titlerule*[1pc]{.}\contentspage}

\maxtocdepth{all}

	\newcommand{\afterchapskipcustomvalue}{140pt}
	\newcommand{\afterchapskipcustomvalueother}{140pt}

\makechapterstyle{myChapStyle}{

	\setlength{\midchapskip}{7pt}
	\setlength{\afterchapskip}{\afterchapskipcustomvalue}

}

\makechapterstyle{myOtherStyle}{

	\setlength{\midchapskip}{0pt}

	\setlength{\afterchapskip}{\afterchapskipcustomvalueother}

}

\makechapterstyle{myFrontmatterStyle}{

	\setlength{\midchapskip}{0pt}

	\setlength{\afterchapskip}{0pt}

}

\settrimmedsize{297mm}{210mm}{*}
\settypeblocksize{634pt}{428.13pt}{*}
\setulmargins{4cm}{*}{*}
\setlrmargins{2.8cm}{*}{1.618}
\setmarginnotes{17pt}{51pt}{\onelineskip}
\setheadfoot{\onelineskip}{2\onelineskip}
\setheaderspaces{*}{2\onelineskip}{*}

\checkandfixthelayout

\newenvironment{note}{\noindent
		\rule{\linewidth}{0.1pt}
		\\*\begin{small}\begin{itshape}\textsc{Note:}}%
    {\end{itshape}\end{small}\\*\rule[2.5pt]{\linewidth}{0.1pt}}
\newenvironment{proof}{\noindent
		\rule{\linewidth}{0.1pt}\\*
		\begin{small}
		\begin{itshape}
		\textsc{Proof:}
		}%
		{
		\end{itshape}
		\end{small}
		\\*
		\rule[2.5pt]{\linewidth}{0.1pt}
}

\usepackage{enumitem}
\setitemize{noitemsep}
\setenumerate{noitemsep}
\setlist{leftmargin=*,labelindent=\parindent}

\newenvironment{introccl}
{
	\chapterstyle{myOtherStyle}
	\setcounter{secnumdepth}{-1}
	
	\setcounter{footnote}{0}
}
{
	\setcounter{secnumdepth}{3}
}

	\newcommand \be{\begin{equation}} %
	\newcommand \ee{\end{equation}} %
	\newcommand \bea{\begin{eqnarray}} %
	\newcommand \eea{\end{eqnarray}} %
	\newcommand \ba{\begin{array}} %
	\newcommand \ea{\end{array}} %
	\newcommand \degr{\textrm{\textdegree{}}}
	\newcommand \atan{\textrm{atan}}

\renewcommand{\parallel}{{\mkern3mu\vphantom{\perp}\vrule depth 0pt\mkern2mu\vrule depth 0pt\mkern3mu}}

\let\originalleft\left
\let\originalright\right
\renewcommand{\left}{\mathopen{}\mathclose\bgroup\originalleft}
\renewcommand{\right}{\aftergroup\egroup\originalright}

\newcommand \titrethese{Large-scale alignments of quasar polarisations: a detailed study of the spinless-particle scenario}

	\newcommand \mE{\mathcal{E}^{\vphantom{*}}}
	\newcommand \mEs{\mathcal{E}^{*}}

	\newcommand \vmE{\vec{\mathcal{E}}^{\vphantom{*}}}

	\newcommand \vmB{\vec{\mathcal{B}}^{\vphantom{*}}}
	\newcommand \mB{\mathcal{B}^{\vphantom{*}}}

\usepackage{xcolor}
\definecolor{draftcolorun}{HTML}{1A72FF}
\newcommand \draftun[1]{#1}
\definecolor{draftcolor}{HTML}{FF3333}
\newcommand \draft[1]{#1}

\begin{document}

\chapterstyle{myFrontmatterStyle}
\pagestyle{myRuled}

\frontmatter

~
    \vspace{8.6cm}
\begin{center}
\thispagestyle{empty}
    {\textbf{\Large{Large-scale alignments of quasar polarisations}\\[.2cm]{\large{A detailed study of the spinless-particle scenario}}}}
\end{center}

\cleardoublepage

\begin{center}
\thispagestyle{empty}
    {\small{\textsc {Université de Liège \\Faculté des Sciences}}}
    \vskip 0.1cm
    \includegraphics[height=3cm]{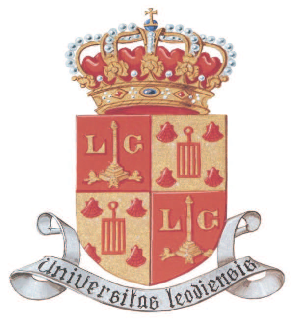}
    \vskip 0.1cm
    \vskip 5.2cm
    {\textbf{\Large{Large-scale alignments of quasar polarisations}\\[.2cm]{\large{A detailed study of the spinless-particle scenario}}}}
    \vskip 0.65cm
    \textbf{{Alexandre Payez}}
\vfill
    \textsc{A dissertation submitted\\ in partial fulfillment of the requirements\\ for the degree of Docteur en Sciences (Ph.D.)}
\end{center}

\thispagestyle{empty}
\vphantom{~}
\vfill

\begin{center}
	Copyright 2013 {{\textcopyright}} Alexandre Payez
\end{center}

{\scriptsize{

	\noindent{}This work is licensed under the Creative Commons Attribution-NonCommercial-ShareAlike 3.0 Unported License.
	To view a copy of this license, visit \texttt{\tiny{\href{http://creativecommons.org/licenses/by-nc-sa/3.0/}{http://creativecommons.org/licenses/by-nc-sa/3.0/}}} or send a letter to Creative Commons, 444 Castro Street, Suite 900, Mountain View, California, 94041, USA.
}}

	\begin{figure}[h!!]
		\centering
		\includegraphics[trim=0mm 0mm 0mm 0mm, clip]{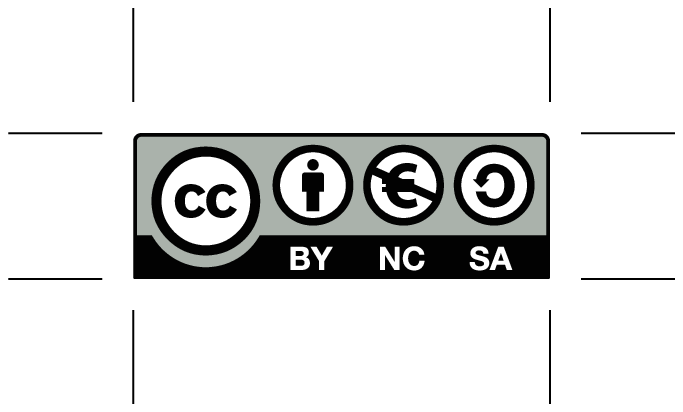}
	\end{figure}

{\scriptsize{
	\noindent{}\textbf{Attribution}\hspace{1em}Alexandre Payez, ``\titrethese'' (Ph.D. thesis).
\\[.2cm]
{\setlength{\baselineskip}{1.25\baselineskip}
	\noindent{}Whenever this manuscript\hspace{1pt}---\hspace{1pt}or part of this manuscript\hspace{1pt}---\hspace{1pt}is modified, it must be in accordance with the conditions detailed in this licence, and a valid link to retrieve the original work must also clearly appear.
Under these conditions, any part of this thesis may be reprinted or reproduced or utilised in any form or by any electronic, mechanical, or other means, now known or hereafter invented, including photocopying and recording, or in any information storage or retrieval system, with the permission of the writer.
\par}
}}

~
\vspace{-5\baselineskip}
\vspace{0.38196601\textheight}
\begin{flushright}
\thispagestyle{empty}
\emph{Like arrows in the sky\\ I can't believe my eyes\\\vspace{\baselineskip} But it's true}
\\[.2cm]\textsc{Gojira}\hspace{1pt}---\hspace{1pt}Flying Whales
\end{flushright}

\cleardoublepage

	\chapter*{Abstract}

	The main motivation for our work has been a puzzling observation concerning quasars. No one expected the existence of correlations in the polarisation of visible light coming from objects separated by gigaparsecs, until they were first reported in the form of a redshift-dependent effect that has become more and more significant with the growth of the data sample.

	In close connection with the observational group, we have studied in detail the most widely considered scenario, involving axion-photon mixing in extragalactic magnetic fields.

	After a systematic investigation, we conclude that it is very unlikely that these observations can be accounted for by axion-like particles, given current data and constraints. We also derive new limits on the parameters describing these particles.

	This thesis gives in particular a detailed account of the consequences of axion-photon mixing on polarisation, studies the influence of averages over the frequency (including a wave-packet treatment of the mixing), and discusses the consequences of different magnetic-field morphologies.

	\chapter*{Résumé}

	La motivation principale de notre travail a été une observation déroutante concernant les quasars. Personne ne s'attendait à l'existence de corrélations dans la polarisation de la lumière visible provenant d'objets séparés par des gigaparsecs, jusqu'à ce qu'elles soient observées sous la forme d'un effet dépendant du redshift et devenu de plus en plus significatif avec l'accroissement de l'échantillon de données.

	En relation étroite avec le groupe observationnel, nous avons étudié en détail le scénario le plus largement considéré, impliquant le mélange axion-photon dans des champs magnétiques extragalactiques.

	Après une étude systématique, nous concluons qu'il est très peu plausible que ces observations puissent être dues à des particules similaires aux axions, étant donné les données et contraintes actuelles. Nous dérivons également de nouvelles limites sur les paramètres décrivant ces particules.

	Cette thèse donne notamment un compte rendu détaillé des conséquences du mélange axion-photon sur la polarisation, étudie l'influence de moyennes sur la fréquence (incluant un traitement en termes de paquets d'ondes du mélange), et discute les conséquences de différentes morphologies de champs magnétiques.

	\chapter*{Acknowledgments}

	I would like to take a moment and express my heartfelt gratitude to all of you who provided me with the generous support and help I needed to make this thesis possible.

	First and foremost, I would like to thank my supervisor Jean-René Cudell for his support, as this thesis would simply not exist without him.
	I really hope that I have inherited, at least partially, your ever-questioning attitude towards the origin of physical results.
	Thank you for the useful advise you have given me during these years, and especially for sharing your insight and experience of the research world.
	Thank you finally for your dedication during the proof-reading stages, and for simplifying most of my never-ending\hspace{1pt}---\hspace{1pt}yet grammatically correct\hspace{1pt}---\hspace{1pt}sentences. 

	It is also a pleasure to thank Damien Hutsemékers. During these years, you have been like an additional supervisor to me, and I thank you for your time, for you kindness, and for always being open to physics discussions. Thank you also for your patience when exposing me the subtleties among various quasar classes and their polarisation properties.

	Special thanks go to Davide Mancusi. I am clearly indebted to you for the countless discussions we had about basically everything, including physics, programming, or linguistics (wtf?). I also thank you a lot for your time, and for your moral support when I was writing my first paper. 
	The discussion I give on unpolarised light at the amplitude level (Eqs.~\eqref{eq:quasimono} and~\eqref{eq:quasimonounpollight}) is based on one of the exchanges we had.

	In direct connection, and because physics is not all that matters, I warmly thank the members of ``la vie au quatrième'' for all the great times together and the good memories.
	This was clearly the best period of this thesis as, during a few years, we have had the perfect mix of pure fun and serious physics discussions during lunches, coffee breaks, and more. Thank you so much to all of you.

	My gratitude goes also to Joseph Cugnon, the Head of the IFPA group at the time I arrived, and thanks to whom I received an I.I.S.N. funding; I am equally grateful to the University of Liège for its prompt financial support following the contract breach by the FNRS that occurred suddenly at the end of the thesis.

	I would also take this opportunity to thank the other members of the IFPA group, including the former postdocs, for all the time we have spent together.
	Likewise, I thank all my collaborators, Rémi Cabanac and Hervé Lamy in particular, my colleagues from the AGO and Physics departments (especially the third floor), as well as anyone with whom I have had the chance to discuss physics, be it on a blackboard, on a tablecloth, or in a pub.

	Finally, I would like to thank Damien Hutsemékers, Cédric Lorcé, Davide Mancusi, Javier Redondo, and Jean Surdej for accepting to be part of my thesis committee.
	\bigskip
	
	As I do not want to forget anyone, I will not single out friends and family. I will thus simply say a massive ``thank you'' to all of you for cheering me up or simply expressing your sympathy during all these years. If you want, we will settle this around a beer later.

	Last but certainly not least, I would like to thank Audeline. You have been my daily dose of support throughout this equally exciting and frustrating experience. This thesis is dedicated to you.

\chapterstyle{myOtherStyle}

\cleardoublepage
\tableofcontents* 

\mainmatter

\chapterstyle{myChapStyle}
\pagestyle{myRuled}

\begin{introccl}

\chapter{Introduction}

\section{Looking beyond the Standard Model}

For decades, physicists have been discussing possible extensions of the Standard Model of particle physics: a theory describing the electromagnetic, weak, and strong interactions with unprecedented accuracy, as relentlessly tested at the supercolliders used to probe the fundamental behaviour of Nature at ever smaller distances. Given such an impressive agreement with data for any center-of-mass energy we produced up to now, one might wonder why it is widely believed that there is more to be discovered.

	\subsection{Motivations for new physics}

	From the theoretical point of view, the motivations are mostly driven by aesthetics: \textit{e.g.}, symmetries, different types of unification, and issues such as naturalness and hierarchy.
The Standard Model also has many free parameters, together with a complicated structure, which leads to many open questions; not to mention that one might think that gravity should be included somehow.
In an empirical science, theoretical arguments alone are not sufficient however; unambiguous experimental deviations from the predictions of the current theory are needed before a definite claim can be made.

Currently, one of the best experimental hints for new physics might be related to neutrino oscillations, which imply that these particles are massive~\cite{Fukuda:1998sk}. As neutrinos are strictly massless in the framework of the Standard Model, some modification must be made. The question is then how much new physics this change would really imply.
Most mechanisms able to generate their masses involve completely new energy scales and new particles, but
a minimal (\draftun{rather unexciting}) extension involving only right-handed neutrinos is also possible; for a review see \textit{e.g.} Ref.~\cite{Mohapatra:1991ng}.

While challenges to the theory from experimental particle physics are few, it did not go unnoticed that there are many open problems in cosmology. For instance, we do not understand most of what seems to fill the Universe: the so-called dark matter and dark energy. Nor do we know how to explain the asymmetry between matter and antimatter, without which we would not be here at all, able to think about such things.
Now, if we assume that these problems find their solutions in particle physics, then physics beyond the Standard Model is clearly needed. 
For more information, Ref.~\cite{Langacker:2010bk} gives a thorough discussion which goes along these lines.

	\subsection{Extended particle content}

	Any scientific theory must provide falsifiable predictions; in the case of an extension of the Standard Model, this typically\hspace{1pt}---\hspace{1pt}but not exclusively\hspace{1pt}---\hspace{1pt}takes the form of new particles that could be detected and studied.
	There is an obvious constraint however: for any new particle, there must be a reason why it has evaded detection so far.

	The reason could of course be that some of those new states are very heavy, with masses so large that the amount of energy required to produce them was simply too high for previous experiments. Their masses would then define energy scales where there is new physics.
	Whereas a lot of attention is given to that kind of particles at supercolliders, where they could be directly produced, this is not the only possibility.
	It could simply be that some of the new particles have particularly small couplings. Even without kinematical constraints for their production, these states would then not be expected at colliders, just because of their tiny production or interaction cross sections.\footnote{An electroweak counterpart of these are neutrinos, which require dedicated experiments often making use of huge quantities of matter to be studied.}

\section{The Universe as a laboratory}

	In the context of astrophysics however, even such especially weakly interacting particles could lead to sizeable effects on what we observe from far-away sources, as the distances involved can be sufficient for their imprint to accumulate. If this looks promising, we also must recognise that (extragalactic) astroparticle physics suffers some drawbacks with respect to terrestrial experiments, among which the impossibility to access directly the initial conditions and the absence of control on the external conditions, which are to be estimated somehow.

	Nonetheless, the benefit of astrophysical data is still twofold in this case: while constraints can be derived to exclude some theoretical models when no deviation from standard astrophysics is found, would-be hints of the existence of new particles can also be searched for in the form of unexpected patterns.

	As a matter of fact, signals of this sort have already been reported in the literature, yet not unambiguously identified as being related to particle physics. 
	One of them in particular is an extremely puzzling phenomenon that could indicate the existence of correlations over distances larger than the most extended structures presently known in the Universe.

	\subsection{Quasar polarisations coherently oriented on large scales}

		That surprising effect has been uncovered by observing the polarisation of visible light coming from high-luminosity active galactic nuclei, hereafter simply referred to as quasars. It concerns the distribution of their polarisation position angles, which are used to indicate the direction of maximum polarisation on the sky for each source with respect to an arbitrary direction (often chosen as the North--South axis in equatorial coordinates).
		
		What has been reported~\cite{Hutsemekers:1998, Hutsemekers:2001, Hutsemekers:2005} is that the distribution of these individual preferred directions is in fact not random for quasars scattered in extremely large-scale regions of the sky ($\sim$~1~Gpc).
		For the latest all-sky sample available, namely 355 high-quality measurements of linear polarisation, global statistical tests indicate that the probability for the observed distribution to be due to coincidence is between $3\times10^{-5}$ and $2\times10^{-3}$, depending on the specifics of the test applied.
		In other words, as far as linear polarisation is concerned, the polarisations of visible light from quasars projected on the sky tend to be aligned within large regions.
		Clearly, this observation is remarkable as there is no reason \textit{a priori} why one should expect such correlations of polarisation over cosmological distances.

\begin{figure}[h]
	\centering
	\includegraphics[width=.68\textwidth,trim = 0mm 6.7cm 8cm 0mm, clip]{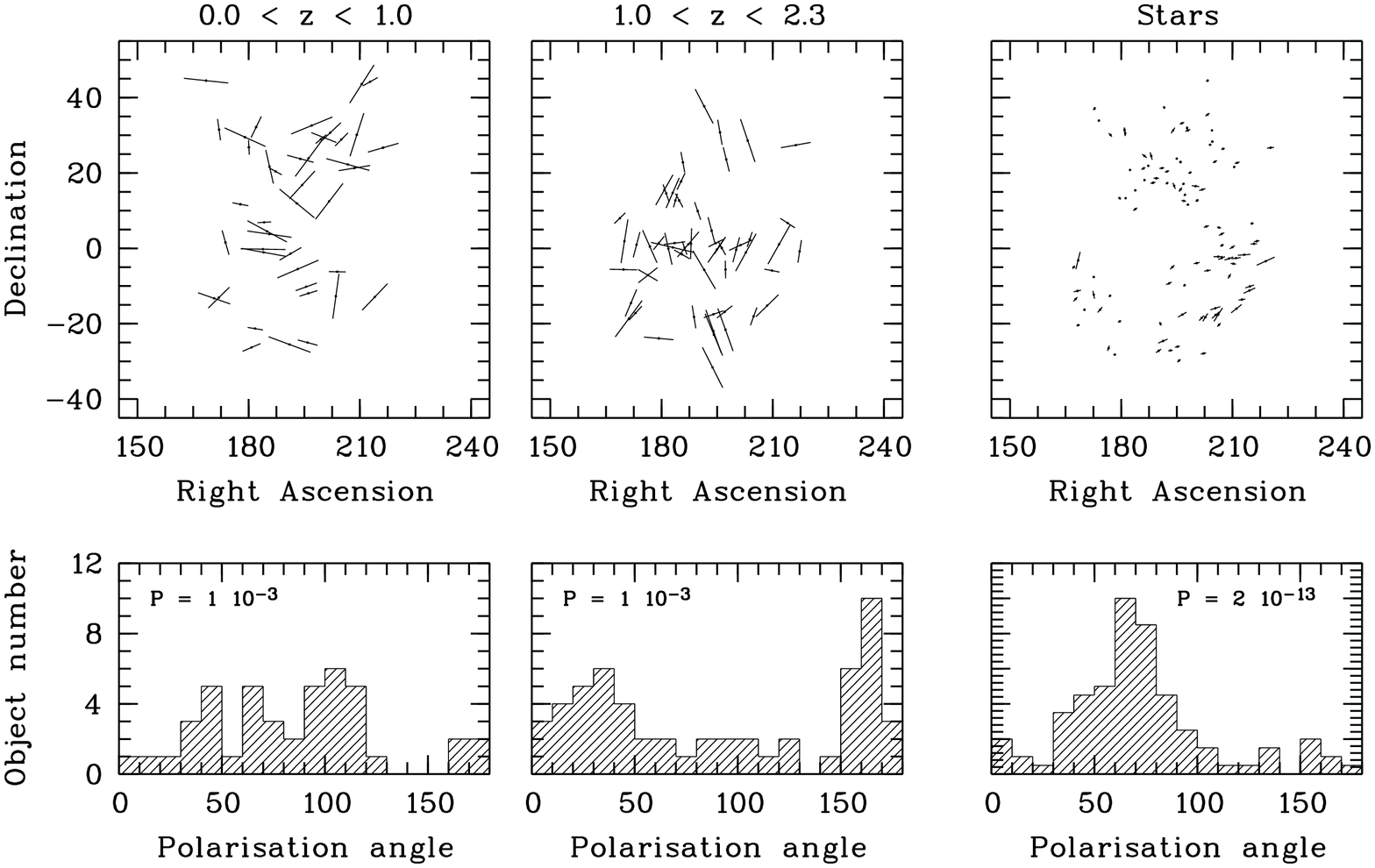}
	\caption{Maps of the polarisation vectors for sources located at low- (\emph{left}) and high- (\emph{right}) redshifts, and otherwise similar declinations and right ascensions (in degrees). Each vector is pictured with a length proportional to the linear polarisation degree of its source, up to a value of 2\%, above which it remains constant for readability purposes.}
	\label{fig:align_intro}
\end{figure}

		The alignment tendency is illustrated for two regions in Fig.~\ref{fig:align_intro}, where the polarisation position angle of each source is indicated with a polarisation vector.
		We clearly see that the orientations are not random, and also that the preferred direction for objects along similar lines of sight appears to be redshift-dependent.
		Full analyses~\cite{Hutsemekers:1998, Hutsemekers:2001, Jain:2003sg, Hutsemekers:2005} indeed indicate that the effect is not likely to be explained by local causes (influence of our galaxy, dust, etc.), \draftun{which would similarly affect both low- and high-redshift samples,} and suggest that it requires something more exotic.

	\subsection{How can we understand this effect?}

		The very fact that quasars intrinsically emit polarised light is related to their morphology: it indicates a departure from a spherical symmetry, and there are known (observational) relations between morphological axes of quasars and their preferred directions of polarisation~\cite{Rusk:1985,Joshi:2007yf,Borguet:2007kb,Borguet:2008tn}.
		In this light, the existence of coherent orientations of polarisation could therefore mean that quasars themselves, namely their morphological axes, are coherently oriented over cosmological distances; if this was the case,
it would no doubt be a striking problem for the current cosmology.

		Attempts to explain the patterns observed in quasar polarisations can therefore essentially be divided into two categories, as already stated in the original paper~\cite{Hutsemekers:1998}.
		\begin{description}[style=nextline]
			\item[Mechanisms leading to alignments of quasar axes]
			One might suppose that the observations indeed reflect the existence of an alignment of qua\-sar axes across the universe, and search for the mechanism responsible for these physical orientations.
	As there are relations between quasar morphology and polarisation, not only in optical, but also in radio waves, a similar effect could be expected at those wavelengths. 
			\item[Mechanisms affecting light during its propagation]
			One might think instead that there is no reason why quasar axes should be correlated. The observations might be explained if the properties of light are altered as it propagates towards Earth, due to a mechanism able to modify polarisation.
			As the existence of correlations is then not intrinsic to the sources, but due to the environment encountered on the way, whatever affects light from quasars would then also formally affect other sources from these regions.
		\end{description}
		Hypotheses from the first category are however severely disfavoured according to a study based on a sample of 4290 quasars (redshift information being available for 1273 of them), since \emph{no} evidence for alignments of polarisations could be found in radio waves~\cite{Joshi:2007yf}.\footnote{Note that 52 of these objects are part of the 355-quasar sample of optical measurements~\cite{Hutsemekers:2005}.} \draftun{This is still controversial~\cite{Tiwari:2012rr}; we will come back to this in the conclusion of this work.}
		\bigskip

		\draftun{Among the various ideas proposed to explain the observations, the one that has been the most widely considered in the literature is related to particle physics and involves the existence of new light spinless particles, see \textit{e.g.} Refs.~\cite{Jain:2002vx,Das:2004qka,Piotrovich:2008iy,Payez:2008pm,Hutsemekers:2008iv}.}
		Scalars and pseudoscalars can indeed have couplings to photons similar to that of neutral pions: this is known as the Primakoff effect~\cite{Primakoff:1951ef} and was first reported experimentally in 1965~\cite{Bellettini:1965pr}.\footnote{They simply behave differently under a parity transformation: pseudoscalars have odd parity.}

		In that scenario, a change in the polarisation of light would be induced in a frequency-dependent manner during its propagation.
		The basic idea is that, if such particles exist, photons could mix with them as they travel.
		Inside external electromagnetic fields, a mixing with particles of different spin is possible, as the direction of the external field can compensate for the spin mismatch; see \textit{e.g.} Ref.~\cite{Raffelt:1987im}.
		Given the seemingly ubiquitous presence of external magnetic fields in astrophysical environments, this is what is usually considered.

\draftun{
		Interestingly, several authors have reported other phenomena which might find a common explanation if one supposes the existence of such nearly massless spinless particles. 
		For instance, these include the transparency of the Universe to high-energy photons~\cite{Simet:2007sa,DeAngelis:2007dy,*Roncadelli:2010qg,Mirizzi:2009aj};\footnote{\draftun{The authors of Ref.~\cite{DeAngelis:2007dy,*Roncadelli:2010qg} also try to explain the spectrum of a given blazar (3C279) using such particles. It is remarkable that this object is also part of the sample of 355 polarised quasars with which the alignment effect has been highlighted.}} the luminosity relations for active galactic nuclei at different wavelengths~\cite{Burrage:2009mj}; neutral ultra-high-energy cosmic rays from blazars~\cite{Fairbairn:2009zi}; or the spectra of gamma-ray sources~\cite{SanchezConde:2009wu}.
		As the information we get in astrophysics comes mainly in the form of photons from distant sources, the mixing property of these particles indeed makes them an appealing ingredient in many astrophysical models. Even with very small couplings, their existence could be probed, as the distances travelled are huge in that context.
}

		In the specific case of the coherent orientations of polarisations, what makes those particles so appealing, in particular, is that the mixing of photons with spin-0 particles $\phi$ changes the polarisation of light in a background electromagnetic field; see for instance Refs.~\cite{Harari:1992ea,Raffelt:1987im} for classic works on the subject.
		This is because only one specific direction of polarisation feels the interaction.

		For pseudo\-scalars, the interaction Lagrangian contains a term proportional to $\phi (\vec{E}\cdot\vec{B})$. 
		Inside an external magnetic field $\vec{\mathcal{B}}_e$, that term reduces to $\phi (\vec{\mathcal{E}_r} \cdot \vec{\mathcal{B}}_e)$, with $\vec{\mathcal{E}_r}$ the electric field of the radiation from which the polarisation is defined.
Photons will thus mix with pseudo\-scalars through the projection of $\vec{\mathcal{B}}_e$ on their polarisation vector.

		Things are similar for scalars~\cite{Biggio:2006im}, the main difference being that it is then the perpendicular direction which will mix, as the interaction is then related to $\phi (\vec{B}^2 - \vec{E}^2)$, and the relevant term is $\phi (\vec{\mathcal{B}_r}\cdot\vec{\mathcal{B}}_e)$, with $\vec{\mathcal{B}_r}$ the magnetic field of the radiation, so that $\vec{\mathcal{E}_r}$ has to be perpendicular to $\vec{\mathcal{B}}_e$.

		We will stick to what happens in the pseudo\-scalar case for the developments throughout this thesis, but bear in mind that our results also hold for scalars.

		Let us now say a few words on the \draftun{motivations for} this kind of particles from theory.

\section{Axion-like particles}

		Among the list of hypothetical very weakly coupled particles, the existence of new spinless particles that are electrically neutral, stable, and characterised by very light (sub-eV) masses is a frequent prediction of extensions of the Standard Model.\footnote{The conventions and notations used throughout this thesis can be found in Appendix~\ref{app:notations}.}

		Despite the smallness of their masses and couplings, for the most part these scalar or pseudoscalar particles are actually linked to physics at extremely high-energy scales, largely beyond the reach of supercolliders. Any observation of such testable candidates would then provide valuable information about physics at these energies.
		How they are connected to such scales and how they do typically arise in theories beyond the Standard model is rooted in the Goldstone theorem and in the concept of spontaneous symmetry breaking.

	\subsection{From symmetries to pseudo-Nambu--Goldstone bosons}

		In the history of science, and of physics in particular, following the path laid down by symmetries has led to great advances. More often than not, they have helped get a unified picture of phenomena which were seemingly unrelated, and have been used as guides to discover new ones. Studying the symmetries of a problem\hspace{1pt}---\hspace{1pt}in other words, what transformations leave it unchanged\hspace{1pt}---\hspace{1pt}grants access to a deeper understanding of its underlying structure, and helps identify what are the properties that in fact really matter.
		In modern physics, they have come to play an essential role; see \textit{e.g.} Ref.~\cite{Wigner:1959bk}. While local symmetries are now associated to all known fundamental interactions via diffeomorphism~\cite{Hooft:2008sp,*Guida:2008sp,*Weinstein:1999} or gauge invariance~\cite{Weyl:1929en,*Yang:1954ek}, global symmetries are also extremely useful, to find~\cite{GellMann:1962xb,*Ne'eman:1961cd} and understand~\cite{GellMann:1964nj,*Zweig:1981pd} patterns in the ``particle zoo'' for instance. 

		Symmetries can be of great importance in physics even when they are broken or hidden; a good example is spontaneous symmetry breaking, which happens when the fundamental state of a theory is not invariant under a symmetry of its Lagrangian.
		Being hidden in the ground state, a spontaneously broken symmetry remains respected by the theory nonetheless and, in the case of a continuous transformation, the associated current conservation therefore holds.

		Now, whenever a continuous global\footnote{Note that symmetry breaking also plays a central role in the Brout--Englert--Higgs mechanism, where it is applied to local (gauge) symmetries.} symmetry is spontaneously broken, the Goldstone theorem~\cite{Goldstone:1961gs,*Goldstone:1962gs} states that, to every broken generator one must associate a new massless scalar degree of freedom with the same quantum numbers: a Nambu--Goldstone boson.
		While there is no identified massless scalar in nature, the three pions stand out as being much lighter than the rest of the meson spectrum and are well-known examples of pseudo-Nambu--Goldstone bosons~\cite{GellMann:1960sigma,Nambu:1961tp,*Nambu:1961fr}.

		These appear when the Goldstone theorem is applied to the spontaneous breaking of a continuous global transformation which leaves the theory invariant except for a small explicit breaking. If this term involves a parameter which is small compared to the scale of the spontaneous breaking (given by the expectation value of a Higgs-like field for instance) the spinless particles are not expected to be massless anymore, but acquire a small mass, related to the value of the explicitly breaking parameter; see \textit{e.g.} Ref.~\cite{Kim:1986ax}.\footnote{As the symmetry is only approximate, the associated current is partially conserved; its divergence is not zero and involves a decay constant. At the quantum level, anomalies spoil current conservation as well and can lead to non-vanishing masses even for exact symmetries of the Lagrangian at the classical level~\cite{Beringer:1900zz}.}

\begin{note}
		Let us illustrate the kind of mass term associated with pions, when described as pseudo-Nambu--Goldstone bosons.
		For pions, the global transformation involved is chiral isospin, which (in a nutshell) is spontaneously broken by quark condensates in the vacuum, and slightly spoiled by non-vanishing quark masses in the QCD Lagrangian; see \textit{e.g.} Ref.~\cite{Peskin:1995bk,*Leutwyler:2001}.
		As a result, these three mesons have a mass, approximately the same. According to the Gell-Mann--Oakes--Renner relation,
		\be
			{m_{\pi}}^2 = -\frac{1}{{f_{\pi}}^2}\frac{(m_u + m_d)}{2}\langle0|\bar{u}u + \bar{d}d|0\rangle,
			\label{eq:pionmass}
		\ee
		in the limit where up and down quark bare masses, $m_u$ and $m_d$, are equal; see \textit{e.g.} Ref.~\cite{Hatsuda:1994pi,*ChengLi:1984bk}.
		Inspecting the structure of Eq.~\eqref{eq:pionmass} is quite instructive: as expected, it involves the quark bare masses, which explicitly break the symmetry, but we see that a role is also played by the quark condensate $\langle0|\bar{q}q|0\rangle$, which triggers the spontaneous symmetry breaking. Finally, the mass of these pseudo-Nambu--Goldstone bosons is directly suppressed by the decay constant $f_{\pi}$, which is also identified as the scale of the breaking \draftun{in some simple models~\cite{GellMann:1960sigma}.}
\end{note}

		In extensions of the Standard Model, hypothetical particles collectively known as axion-like particles (ALPs) essentially arise from that very scheme, only repeated at energy scales $f\ggg f_{\pi},\Lambda_{\mathrm{QCD}}$.
		As any such new spinless particle truly belongs to an extremely high scale, to study its low-energy effects, all the other new degrees of freedom from the scale $f$ can be integrated out, leading to an effective Lagrangian with non-renormalisable couplings $g_{\mathrm{eff}}\sim1/f$.\footnote{Other particles with similar phenomenological properties can also be called axion-like particles.}

		Their common appellation traces back to the well-known axion~\cite{Kim:1979,*SVZ:1979,Zhitnitsky:1980,*DFS:1981}: the necessary pseudo-Nambu--Goldstone boson~\cite{Weinberg:1978,*Wilczek:1978} of the spontaneous breakdown of the Peccei--Quinn symmetry, introduced to solve the Strong CP Problem~\cite{PecceiQuinn:1977}.
		Among similar spinless particles, one also finds for instance majorons and familons, the Nambu--Goldstone bosons of the breaking of lepton number~\cite{Chikashige:1980ui,*Gelmini:1980re} and of family~\cite{Wilczek:1982rv} symmetries respectively. 
		Interestingly, similar particles can arise in a large variety of models, ranging from theories involving extra dimensions, such as Kaluza-Klein or superstrings (for reviews, see \textit{e.g.} Refs.~\cite{Lakic:2008zy,Jaeckel:2010ni}), to theories embedding the Standard Model with extended gauge groups, as Grand Unified Theories (GUT), where accidental symmetries could appear.\footnote{Imposing a larger gauge group, while retaining only renormalisable terms, sometimes leads to accidental global symmetries, some of which might be spontaneously broken~\cite{Weinberg:1979pi,*Weinberg:1972fn}.}

\subsection{Generic light spinless particles: searches and limits}

		If, from the theoretical point of view these particles are very well motivated, from the experimental point of view, however, they are yet to be observed. A lot of effort has been and is made to try to detect them: many dedicated experiments such as~\cite{Andriamonje:2007ew,Zavattini:2007ee,*Pugnat:2007nu,*Chou:2007zzc,*Asztalos:2009yp,*Mei:2010aq,*Ohta:2012em,*Boyce:2012ym} have been designed to probe their existence, based on their electromagnetic coupling~\cite{Sikivie:1983ip,*Sikivie:1984erratum}, and, if most of them primarily focus on the search of the QCD axion, they would also be sensitive to the presence of other kinds of particles (see \textit{e.g.} Ref.~\cite{Redondo:2010dp} for a review), including any other light spinless particles with a similar two-photon coupling~\cite{Masso:1995tw}.
		Following this approach, one does not stick to a given model when discussing axion-like particles in general but, instead, studies generic spinless particles of masses and couplings considered as free parameters.

		Looking in the literature, one can then identify different promising regions in the parameter space of axion-like particles.
		In particular, the region relevant for our work is the low-mass region: it is quite striking that various astrophysical hints, including the ones mentioned earlier, seem to point towards the same kind of spinless particle, with a mass $m\lesssim 10^{-10}$~eV and a coupling to photons $g\sim10^{-12}$--$10^{-11}$~GeV$^{-1}$, defining a whole new window of interest in the $(m,g)$ parameter space.

		To date, the most stringent constraints on the parameters of axion-like particles are coming from astrophysics; for reviews, see \textit{e.g.} Ref.~\cite{Raffelt:2006cw,*Cadamuro:2012phd}. 
		In particular, over the years, many of them have come from stellar dynamics: indeed, the existence of such particles would imply the opening of new channels to evacuate energy, which should have an impact on stellar evolution.
		For example, a limit on their coupling has been given by the theoretical bound derived from the observation of the Horizontal Branch in the Hertzsprung--Russell diagram for globular clusters~\cite{Raffelt:1987yu,*Raffelt:1996bk}; if too efficient in stellar medium, ALP production could have shortened the duration of this transitional period, and contradict the agreement between observations and theory.
		This remained the strongest limit for two decades for low-mass ALPs, until it was eventually improved in 2007 by CAST (the CERN Axion Solar Telescope), designed to convert axion-like particles coming from the Sun back into (X-Ray) photons using an LHC magnet~\cite{Andriamonje:2007ew}. The constraint obtained with this instrument is \draft{$g<8.8\times10^{-11}$~GeV$^{-1}$ for $m\lesssim 0.02$~eV.} Additionally, discussions on the absence of a gamma-ray flux associated with the SN1987A supernova have lead to an even more severe upper bound on their couplings to photons, which however only applies for nearly massless ALPs \draft{with $m\lesssim 10^{-9}$~eV}: $g\lesssim10^{-11}$~GeV$^{-1}$~\cite{Brockway:1996yr} or even $g\lesssim 3\times10^{-12}~$GeV$^{-1}$~\cite{Grifols:1996id}, the most conservative one being usually used given the associated uncertainties.
		\draft{As we shall see in Chap.~\ref{chap:constraints}, these constraints are now improved.}

\section{Structure of this thesis}

		The aim of this work is to perform a detailed study to determine whether axion-like particles can help understand the existence of large-scale correlations of polarisation of visible light from quasars in a realistic model. This will be done taking into account the existing constraints on these particles, the current understanding of magnetic fields from observations, as well as all the polarisation data currently available.

		The first chapter begins with a short reminder on polarisation, which allows us to introduce the notations that shall be used throughout. We then quickly move on and give a derivation of the mixing of photons with axion-like particles, leading to the relevant set of equations for our study. The main subject of this chapter finally follows as we review and discuss the general consequences of the mixing with these particles for the polarisation of light.

		We do not apply the mixing to discuss the alignment effect until Chap.~\ref{chap:alignments}, which starts with a summary of the main observational properties of the effect itself and of the sample that was used to highlight it. Only then do we switch to the spinless-particle scenario, which requires the existence of extended magnetic fields and of nearly massless particles to reproduce correlations over cosmological scales. As already known, it seems able to reproduce alignments of linear polarisation even within a toy model, while providing clear predictions.

		In Chap.~\ref{chap:circular}, we discuss circular polarisation and introduce a more general wave-packet treatment of the mixing, while Chap.~\ref{chap:moregenmagn} is devoted to the mixing in more general magnetic field configurations, such as the ones reported in observational papers and used in large-scale structure simulations. We then finally derive constraints on axion-like particle parameters using polarimetry in Chap.~\ref{chap:constraints}, and adopting a conservative approach.
		We then conclude with some prospects.

		This thesis is based on the following publications:~\cite{Payez:2008pm,Hutsemekers:2008iv,Payez:2008jjc,Payez:2009kc,Payez:2009vi,Payez:2010xb,Payez:2011mk,Payez:2011sh,Payez:2012rc,Payez:2012vf}.

\end{introccl}

\chapter{Mixing photons and axion-like particles}\label{chap:mixing}

	\section{How do we describe the polarisation of light?}\label{sec:pol}

		To say the least, this thesis relies heavily on polarisation-related concepts; it therefore seems appropriate to begin the discussion with a short reminder of one of the standard descriptions of the polarisation of electromagnetic waves, which will also introduce the notations used throughout.

		Maxwell's classical equations of electromagnetism tell that these wave solutions are made of mutually orthogonal electric and magnetic fields, with equal maximal amplitudes,\footnote{In natural units, see Appendix~\ref{app:notations}.} oscillating at a given frequency.
		As these two fields are not independent, it is sufficient to discuss one of them: the electric field, by convention.
		In all the cases we are interested in, the electric field of any electromagne\-tic radiation \draftun{travelling} in the $z$ direction can be written as
		\be
			\vmE_{\mathrm{r}}(z,t) = \mE_{\mathrm{r}_x}(z,t)\vec{e}_x + \mE_{\mathrm{r}_y}(z,t)\vec{e}_y,\label{eq:genlight}
		\ee
		where $\vec{e}_x$ and $\vec{e}_y$ define an orthonormal basis in the plane transverse to the direction of propagation.\footnote{Throughout, a complex exponential form will be used for convenience, as usually in classical electrodynamics; by convention, the real parts of these amplitudes correspond to the physical fields.}
		The polarisation then describes the existence of patterns in the direction in which the electric field of an electromagnetic wave oscillates over time; for more generalities and naming conventions, see Ref.~\cite{jackson:polgeneralities,*hechtzajac:polgeneralities,*bornwolf:polgeneralities,*goldstein:polgeneralities}.\footnote{\draftun{As a matter of fact, polarisation can be seen to some extent with naked eye, as in the Haidinger brush phenomenon, see \textit{e.g.} Ref.~\cite{Seliger:1965bk,*Fairbairn:2001hb}.}}

		\subsection{Fully polarised light and polarisation basis}

		Formally, one can write at the amplitude level a general fully polarised light beam of mean frequency $\omega$ and width $\Delta\omega$ as
		\be
			\vmE_{\mathrm{r}}(z,t) = \cos(\varphi_0)\vec{E}^{(x)}(z,t;\omega;\Delta\omega) + \sin(\varphi_0)\vec{E}^{(y)}(z,t;\omega;\Delta\omega),
			\label{eq:fullypollight}
		\ee
		where we introduce a polarisation basis given by $\vec{E}^{(x)}$ and $\vec{E}^{(y)}$, two fully linearly polarised beams, along $x$ and $y$ respectively, and with identical intensities.\footnote{Note that one could use a circular polarisation basis instead of a linear one.} The behaviour and shape of $\vec{E}^{(x)}\cdot\vec{e}_x$ and $\vec{E}^{(y)}\cdot\vec{e}_y$ are the same; depending on the kind of light we are interested in, they can be plane waves or wave packets, for instance.
		Any ellipticity for $\vmE_{\mathrm{r}}$ is then simply included via a relative phase between $\vec{E}^{(x)}$ and $\vec{E}^{(y)}$, and the preferred direction of $\vmE_{\mathrm{r}}$ is given by the angle $\varphi_0$ when the beam is not purely circularly polarised.

		\subsection{Unpolarised light and Stokes parameters}\label{subsec:unpol}

		On the other hand, the description of an unpolarised beam is a bit more complicated, especially if one wants something close to a representation of it at the amplitude level. In that case, the polarisation is so erratic that no preferred behaviour emerges over time\hspace{1pt}---\hspace{1pt}instantaneously, of course, the polarisation is always well-defined, due to the vectorial nature of light, but it changes incoherently. This kind of light, often referred to as ``natural light'', is typically represented as an incoherent average over $\varphi_0$ and over time of fully polarised beams of the form~\eqref{eq:fullypollight}. However, this representation, which is not the minimal one, is quite heavy and is not very satisfying. As a matter of fact, this is what motivated Stokes to introduce the parameters named after him~\cite{goldstein:pollight}: the need for an appropriate mathematical description of unpolarised light.

		By characterising the polarisation with quantities that can be built out of intensities, Stokes offered an approach describing completely the polarisation state of any light beam, and being at the same time much simpler and closer to what can be observed experimentally.
		There are different notations and conventions; here, we use the following definitions:
		\be
                \left\{
                    \begin{split}
                        I(z) &= \langle\mathcal{I}(z,t)\rangle \ \!= \langle \mE_{\mathrm{r}_y}\mEs_{\mathrm{r}_y} + \mE_{\mathrm{r}_x}\mEs_{\mathrm{r}_x}\rangle\\
                        Q(z) &= \langle\mathcal{Q}(z,t)\rangle = \langle \mE_{\mathrm{r}_y}\mEs_{\mathrm{r}_y} - \mE_{\mathrm{r}_x}\mEs_{\mathrm{r}_x}\rangle\\
                        U(z) &= \langle\mathcal{U}(z,t)\rangle = \langle \mE_{\mathrm{r}_x}\mEs_{\mathrm{r}_y} + \mEs_{\mathrm{r}_x}\mE_{\mathrm{r}_y}\rangle\\
                        V(z) &= \langle\mathcal{V}(z,t)\rangle = \langle i(\mE_{\mathrm{r}_x}\mEs_{\mathrm{r}_y} - \mEs_{\mathrm{r}_x}\mE_{\mathrm{r}_y}) \rangle;
                    \end{split} \right.
                    \label{eq:StokesE}
                \ee
	these quantities are averages (denoted by $\langle \cdot \rangle$) over the exposure time at a given distance $z$ from the source.
	In Eq.~\eqref{eq:StokesE}, $Q$ and $U$ represent the linear polarisation, and $V$ the circular one.
	One often normalises these parameters by the intensity $I$ to enable comparisons between different sources; they are then written in lowercase, \textit{e.g.} $v=\frac{V}{I}$.
	Moreover, as $Q$ and $U$ depend on the choice of axes, one can discuss the intensity of linear polarisation $P_{\mathrm{lin}}=\sqrt{Q^2+U^2}$ which is independent of such a choice, as well as the (total) intensity of polarisation $P_{\mathrm{tot}}=\sqrt{Q^2+U^2+V^2}$. One also introduces the linear polarisation degree and the (total) polarisation degree, respectively defined as
			\be
				p_{\mathrm{lin}} = \frac{\sqrt{Q^2 + U^2}}{I}\qquad \textrm{and} \qquad p_{\mathrm{tot}} = \frac{\sqrt{Q^2 + U^2 + V^2}}{I},
			\label{eq:poldegree}
			\ee
		and sometimes drops the sign of circular polarisation (which does have a physical meaning) to defines $p_{\mathrm{circ}}=|v|$ as the the circular polarisation degree.
	A light beam is then said to be unpolarised if $p_{\mathrm{tot}}=0$ or, equivalently, if $I\neq0$ while $Q=U=V=0$.
	Finally, for linearly polarised light, the polarisation angle reads
	\be
		\varphi = \frac{1}{2}\atan\left(\frac{U}{Q}\right).\label{eq:varphi}
	\ee
	For more information about the properties of the Stokes parameters and their connection to observables, see Appendix~\ref{app:stokes}.

	Interestingly, the Stokes parameters~\eqref{eq:StokesE} have a decomposition property: a radiation described by a set ($I,Q,U,V$) can always be discussed as the sum of two other incoherent ones, described by ($I',Q',U',V'$) and ($I'',Q'',U'',V''$), as long as $I'+I''=I$, etc. In particular, this means that any partially polarised beam can be described by the weighted sum of a fully polarised one and of an unpolarised one.

	Furthermore, this decomposition property implies that it is in fact sufficient to describe unpolarised light as an incoherent sum of two orthogonal linearly polarised beams with equal intensities, such as $\vec{E}^{(x)}$ and $\vec{E}^{(y)}$.
	Now, while an incoherent sum can have a clear meaning in the case of wave packets (which can be thought of as being independent), monochromatic light beams are infinite waves with definite directions, and are coherent with each other when they share the same frequency. Formally speaking, a strictly monochromatic light beam cannot therefore represent an unpolarised light beam, nor can it act as partially polarised light; see also Ref.~\cite{hechtzajac:unpol,*bornwolf:unpol}. In fact, this indicates that, to describe such kinds of light mathematically at the amplitude level, one has to consider at least a small deviation from monochromaticity. Let us illustrate this for plane waves with a simple quasi-monochromatic example:
	\be
	\begin{split}
		\vmE_{\mathrm{r}}(z,t)& = \frac{E}{\sqrt{2}}\ e^{i(kz-\omega t)}\vec{e}_x + \frac{E}{\sqrt{2}}\ e^{i\left(k'z - \omega't\right)}\vec{e}_y,
	\end{split}
	\label{eq:quasimono}
	\ee
	namely a sum of two plane waves with orthogonal polarisations of equal intensities, propagating in phase in the same direction, and of frequencies which only differ by a very small amount $|\delta\omega|=|(\omega'-\omega)|\ll\omega, \omega'$.
	Using the definition of Stokes parameters, see Eq.~\eqref{eq:StokesE}, one obtains that, for $t\gg(\delta\omega)^{-1}$, this indeed corresponds to an unpolarised light beam:
	\be
        \left\{
            \begin{split}
                I(z) &= \frac{1}{2} \langle {|E|}^2 + {|E|}^2 \rangle\\
                Q(z) &= \frac{1}{2} \langle {|E|}^2 - {|E|}^2 \rangle\\
                U(z) &= \langle {|E|}^2\left[ \cos\left(kz-\omega t\right)\cos\left(k'z - \omega't\right) + \sin\left(kz-\omega t\right)\sin\left(k'z - \omega't\right) \right] \rangle\\
                V(z) &= \langle {|E|}^2\left[ \sin\left(kz-\omega t\right)\cos\left(k'z - \omega't\right) - \cos\left(kz-\omega t\right)\sin\left(k'z - \omega't\right) \right] \rangle,\label{eq:quasimonounpollight}
            \end{split} \right.
        \ee
	where only the intensity is non-zero, after averaging over the exposure time.
	Now, in practice, what we do in general is to enforce the incoherent sum by summing the two orthogonal contributions at the Stokes-parameters level: more precisely, if we represent by $S$ any of the four Stokes parameters of a given light beam, if $\vmE_{\mathrm{r}}$ is unpolarised, we have
	\be
            \begin{split}
		S[\vmE_{\mathrm{r}}]	&= S\Big[\frac{\vec{E}^{(x)}}{\sqrt{2}}\Big] + S\Big[\frac{\vec{E}^{(y)}}{\sqrt{2}}\Big]\\
					&= \frac{1}{2}\left(S[\vec{E}^{(x)}] + S[\vec{E}^{(y)}]\right),\label{eq:unpollight}
            \end{split}
	\ee
	if we normalise the three waves to have the same intensity.

	In summary, one can simply consider two orthogonal linearly polarised beams with equal intensities, such as the ones defining our linear polarisation basis, $\vec{E}^{(x)}$ and $\vec{E}^{(y)}$, to reconstruct the polarisation state of any light beam $\vmE_{\mathrm{r}}$. Indeed, both polarised and unpolarised cases can be expressed using those\hspace{1pt}---\hspace{1pt}as a linear combination and as an incoherent sum respectively\hspace{1pt}---\hspace{1pt}and they can be combined to give partially polarised light.

	\section{Axion--photon interaction in magnetic fields}

		After this reminder, we are ready to study the mixing of photons with axion-like particles and its consequences on the polarisation of light.
		Both to be complete and to introduce our notations, let us first go through the developments leading to the relevant set of equations; this derivation goes along the lines of Refs.~\cite{Raffelt:1987im,Das:2004qka}.

		The mixing can be obtained starting from a suitable Lagrangian density taking into account the interaction. For pseudo\-scalars $\phi$, we use
			\begin{equation}
				 \mathcal{L} =  \frac{1}{2}\ (\partial_{\mu}\phi) (\partial^{\mu}\phi) - \frac{1}{2}\ m^2\phi^2 - \frac{1}{4}\ F_{\mu\nu} F^{\mu\nu} + \frac{1}{4}\ g \phi F_{\mu\nu}\widetilde{F}^{\mu\nu}, \label{eq:lagrangian}
			\end{equation}
			where $m$ is the pseudo\-scalar mass and $g$ is the dimension-minus-one coupling constant of the interaction between pseudo\-scalars and photons. The first two terms form the Klein-Gordon Lagrangian describing a free spinless field; the third one is the Maxwell Lagrangian for a free electromagnetic field; the last one gives the three-field interaction term between pseudoscalars and photons \draftun{as in the axion case}.

		The equations of motion are then derived using Euler-Lagrange equations:
		\be
			\partial^{\beta}\Big(\frac{\partial\mathcal{L}}{\partial(\partial^{\beta}\chi)}\Big) - \frac{\partial\mathcal{L}}{\partial\chi} = 0,\qquad \textrm{for a given field } \chi;
		\ee
		for the pseudoscalar and for the electromagnetic fields, one obtains
		\be
    \square\phi + m^2\phi = \frac{1}{4}gF_{\mu\nu}\widetilde{F}^{\mu\nu}\label{eq:eomaxion}
		\ee
		and
		\be
    \partial_{\beta}F^{\beta\alpha} = g(\partial_{\beta}\phi) \widetilde{F}^{\beta\alpha}.\label{eq:eomelm}
		\ee
		In both cases, the well-known equations for the free fields are modified: we see that the electromagnetic fields enter the equation in a new source term for the $\phi$ field, and \textit{vice versa}, leading to a system of coupled partial differential equations.

		Moreover, a direct consequence of the definition of the electromagnetic field strength tensor is that \draftun{its components obey} the Bianchi identity:
		\be
			\partial^{\alpha}F^{\beta\gamma} + \partial^{\beta}F^{\gamma\alpha} + \partial^{\gamma}F^{\alpha\beta} = 0,
		\ee
		that can also be written
		\be
			\partial^{\beta}\widetilde{F}_{\beta\alpha}=0.\label{eq:bianchi}
		\ee
		This, in fact, is the relativistic formulation of the homogeneous Maxwell equations. As they arise from a property of the electromagnetic tensor, these two equations remain unchanged.

		Written in terms of electric and magnetic fields, the modified Klein-Gordon equation~\eqref{eq:eomaxion} then becomes
		\be
			\frac{\partial^2\phi}{\partial t^2} - \nabla^2\phi + m^2\phi = -g\big(\vec{E}\cdot\vec{B}\big);\label{eq:kg_noap}
		\ee
		while the modified Maxwell equations, Eqs.~\eqref{eq:eomelm} and~\eqref{eq:bianchi}, read
		\be
        	\left\{
            		\begin{split}
    		\vec{\nabla}\cdot\vec{E}&=g\Big(\vec{B}\cdot\vec{\nabla}\phi\Big)\\
    		-\frac{\partial\vec{E}}{\partial t}+\vec{\nabla}\times\vec{B}&=g\Big(\vec{E}\times\vec{\nabla}\phi-\vec{B} \frac{\partial\phi}{\partial t}\Big)\\
    		\vec{\nabla} \times \vec{E} + \frac{\partial \vec{B}}{\partial t} &= 0\\
    		\vec{\nabla} \cdot \vec{B} &= 0.
			\end{split}
		\right.
		\ee
		In the following, we will be looking for \draftun{fixed-energy} solutions going in the positive $z$-direction for the electromagnetic and for the spinless fields.

		Simplifications can be made if we specify the domain of application we are interested in.
		Our goal is to discuss how the properties of light coming from distant sources are affected during its propagation in large-scale magnetic-field zones.
		Following Ref.~\cite{Raffelt:1987im}, we assume the external magnetic field to be stationary and to change slowly with spatial coordinates, compared to the radiation field. The total magnetic field is written as $\vec{B} = \vmB_{\mathrm{r}} + \vmB_{\mathrm{e}}$, where $\vmB_{\mathrm{r}}$ is the magnetic field of the radiation and $\vmB_{\mathrm{e}}$ is the external magnetic field, and, similarly, $\vec{E}=\vmE_{\mathrm{r}} + \vmE_{\mathrm{e}}$. Now, as we restrict ourselves to considering the effects induced by external magnetic fields, we set $|\vmE_{\mathrm{e}}| = 0$ and we can take $|\vmB_{\mathrm{e}}|\gg|\vmB_{\mathrm{r}}|,|\vmE_{\mathrm{r}}|$. Doing so, one obtains
		\be
        	\left\{
            		\begin{split}
    		\vec{\nabla} \cdot \vmE_{\mathrm{r}} & = g  \Big(\vmB_{\mathrm{e}} \cdot \vec{\nabla}\phi\Big)\\
    		-\frac{\partial\vmE_{\mathrm{r}}}{\partial t}+\vec{\nabla}\times\vmB_{\mathrm{r}} &= -g\vmB_{\mathrm{e}}\frac{\partial\phi}{\partial t}\\
    		\vec{\nabla} \times \vmE_{\mathrm{r}} + \frac{\partial \vmB_{\mathrm{r}}}{\partial t} &= 0\\
    		\vec{\nabla} \cdot \vmB_{\mathrm{r}} &= 0;
			\end{split}
		\right.
		\ee
		which eventually leads to
		\be
			\frac{\partial^2 \vmE_{\mathrm{r}}}{\partial t^2} - \nabla^2 \vmE_{\mathrm{r}}
			= g\vmB_{\mathrm{e}}\frac{\partial^2 \phi}{\partial t^2} - g\vec{\nabla}\big(\vmB_{\mathrm{e}} \cdot \vec{\nabla}\phi\big).
			\label{eq:eeomfull}
		\ee

		Before we move on, notice that Eq.~\eqref{eq:eeomfull} tells us that the electric field of the radiation is not strictly transverse in general, so that we formally have $\vmE_{\mathrm{r}} = \mE_{\mathrm{r}_x} \vec{e}_x + \mE_{\mathrm{r}_y} \vec{e}_y + \mE_{\mathrm{r}_z} \vec{e}_z $.
		In the case of terrestrial experiments, one can design the apparatus in such a way that the external magnetic field is purely transverse to the direction of propagation of light; when there is no longitudinal external magnetic field, the electric field of the radiation is then transverse. However, if we are interested in the mixing happening in astronomical environments, this, of course, cannot be assumed. If we write $\vmB_{\mathrm{e}} = \mB_{\mathrm{e}_x} \vec{e}_x + \mB_{\mathrm{e}_y} \vec{e}_y + \mB_{\mathrm{e}_z} \vec{e}_z$, we will have $\mB_{\mathrm{e}_x}\approx\mB_{\mathrm{e}_y}\approx\mB_{\mathrm{e}_z}$ in general. Nevertheless, we can show that this longitudinal contribution plays no role in our study; see also, \textit{e.g.} Refs.~\cite{Raffelt:1987im,Harari:1992ea,Das:2004qka}.
On the one hand, from Eq.~\eqref{eq:eeomfull}, the longitudinal part $\mE_{\mathrm{r}_z}$ obeys
		\be
		\begin{split}
			\partial^2_t \mE_{\mathrm{r}_z}(z,t) - \partial_z^2 \mE_{\mathrm{r}_z}(z,t)
			&= g\mB_{\mathrm{e}_z}\partial^2_t \phi(z,t) - g\partial_z\big(\mB_{\mathrm{e}_z}\partial_z\phi(z,t)\big)\\
			&= -g\mB_{\mathrm{e}_z}\big(m^2\phi(z,t) + \mathcal{O}(g)\big),
		\end{split}
		\ee
		where we have made use of the equation of motion for the spinless field~\eqref{eq:kg_noap}; on the other hand, for $\mE_{\mathrm{r}_x}$ and  $\mE_{\mathrm{r}_y}$ we have
		\be
		\begin{split}
			\partial^2_t \mE_{\mathrm{r}_{x,y}}(z,t) - \partial_z^2 \mE_{\mathrm{r}_{x,y}}(z,t)
			&= g\mB_{\mathrm{e}_{x,y}}\partial^2_t \phi(z,t)\\
			&= -g\mB_{\mathrm{e}_{x,y}}\omega^2\phi(z,t).
		\end{split}
		\ee
		Keeping only terms to lowest order in the coupling constant $g$, one obtains that we can indeed always neglect the longitudinal contribution, as we consider nearly massless spinless particles with $m^2\lll\omega^2$.

		These equations of motion for the fields were obtained considering a propagation in free space; however, true vacuum does not exist, even in outer space.
		We can take into account possible plasma effects via the inclusion of a term involving the plasma frequency $\omega_{\mathrm{p}}$, see \textit{e.g.} Refs.~\cite{Raffelt:1987im,Deffayet:2001pc}:
		\be
			\omega_{\mathrm{p}} \equiv \sqrt{\frac{4\pi\alpha n_{\mathrm{e}}}{m_{\mathrm{e}}}} = \left({3.7\times10^{-14}~\textrm{eV}}\right)\times\sqrt{\frac{n_{\mathrm{e}}}{10^{-6}\ \textrm{cm}^{-3}}},\label{eq:plasmafhz}
		\ee
which acts as an effective mass for the propagating electromagnetic field ($n_{\mathrm{e}}$ is the electron number density, and $m_{\mathrm{e}}$, the electron mass).\footnote{Strictly speaking, as photons travel inside an electron plasma, the electric field has three components; the longitudinal mode can be physically understood as due to the motion of electrons induced by the propagation of light inside the plasma~\cite{Anderson:1963pc,Raffelt:1996bk_plasma}.}
Note that Faraday rotation is not included in the discussion as its effect is irrelevant in the range of frequencies we are interested in.

		The appropriate set of equations for our problem finally reads:
		\be
		\begin{split}
			\frac{\partial^2\phi}{\partial t^2} - \nabla^2\phi\ + m^2\phi
			&= -g\Big(\vmE_{\mathrm{r}}\cdot\vec{\mathcal{B}}\Big),\\
			\frac{\partial^2 \vmE_{\mathrm{r}}}{\partial t^2} - \nabla^2 \vmE_{\mathrm{r}} + \omega_{\mathrm{p}}^2\vmE_{\mathrm{r}}
			&= g\vec{\mathcal{B}}\frac{\partial^2 \phi}{\partial t^2},
			\label{eq:eqstobesolved}
		\end{split}
		\ee
		where we introduce the notation $\vec{\mathcal{B}}$ for the projection of the external magnetic field onto the plane transverse to the direction of propagation, as it is the only part relevant for the mixing. Interestingly, notice that it always appears together with the coupling constant $g$ as a product in the equations. Note also that, from Eqs.~\eqref{eq:eqstobesolved}, we can anticipate that photons only couple to pseudoscalars through their polarisation along the external transverse magnetic field direction.\footnote{As briefly touched upon in the introduction, it is instead the polarisation perpendicular to $\vec{\mathcal{B}}$ in the case of scalar particles.}
\bigskip

	Due to the interaction with axion-like particles in magnetic environments, light will be affected and its polarisation will change as it propagates.
	In the following, we are interested in any initial polarisation state. Because the equations~\eqref{eq:eqstobesolved} are linear, it is sufficient to solve the mixing separately for the two initial states $\vec{E}^{(x)}$ and $\vec{E}^{(y)}$ introduced in Sec.~\ref{sec:pol}, from which any other initial polarisation state can be reconstructed.
As the polarisation will now change, the indices only refer to the initial polarisation direction of these two light beams however. As they propagate, we will have in general
	\be
		\vec{E}^{(x)}(z) = E^{(x)}_x(z) \vec{e}_x + E^{(x)}_y(z) \vec{e}_y\quad{\textrm{and}}\quad\vec{E}^{(y)}(z) = E^{(y)}_x(z) \vec{e}_x + E^{(y)}_y(z) \vec{e}_y.
	\label{eq:egen}
	\ee

	\section{Illustration with a single magnetic field region}\label{sec:mixing}
	In order to introduce the consequences of the mixing on polarisation, we first consider a region where the external magnetic field is constant.
	Inside such a region, $\vec{\mathcal{B}}$ provides a preferred direction throughout, and it is advantageous to let it define one of the axes in the transverse plane. 

	\subsection{Solutions of the mixing for the polarisation basis}\label{sec:solution_onezone_polarisationbasis}

	We choose our orthonormal basis in such a way that $\vec{e}_x$ and $\vec{e}_y$ are respectively perpendicular and parallel to this particular direction, and write them $\vec{e}_{\perp}$ and $\vec{e}_{\parallel}$. It is interesting to do so because polarisations along these two directions are completely decoupled: perpendicular ones do not feel the interaction, while the change for the parallel ones only occurs along the magnetic field direction. In particular, this implies that, for light beams with initial polarisation perfectly perpendicular or parallel to the external transverse magnetic field, the direction of polarisation will remain unchanged.
	In this case, the general form for the evolution of $E^{(x)}(z)$ and $E^{(y)}(z)$ given in Eq.~\eqref{eq:egen}, reduces to
	\be
		\vec{E}^{(x)}(z) \equiv E_{\perp}(z)\vec{e}_{\perp}
			\quad\textrm{and}\quad
		\vec{E}^{(y)}(z) \equiv E_{\parallel}(z)\vec{e}_{\parallel},
		\label{eq:polbasisoneregion}
	\ee
	and, as there is no possible ambiguity since polarisations parallel and perpendicular to $\vec{\mathcal{B}}$ are decoupled, we can solve for $E_{\perp}(z)$ and $E_{\parallel}(z)$ at once.

	Writing $\mathcal{B}\equiv|\vec{\mathcal{B}}|$, the equations of the mixing for the two fully linearly polarised beams then read:
	    \be
		    \Bigg[\Big(\omega^2 + \frac{\partial^2}{\partial z^2}\Big) -
	                                \left(
	                                \begin{array}{ccc}
	                                 {\omega_{\mathrm{p}}}^2 & 0            & 0\\
	                                0            &  {\omega_{\mathrm{p}}}^2 & - g \mathcal{B} \omega\\
	                                0            & - g \mathcal{B} \omega & m^2
	                                \end{array} \right)\Bigg] \left(\!\! \begin{array}{c}A_{\perp}(z) \\ A_{\parallel}(z) \\\phi(z)\end{array}  \!\!\right) = 0.\label{eom_planewaves}
	    \ee
This is for eigenstates of energy $\omega$, in the timelike axial gauge $A^0=0$ (so that $E_{\perp,\parallel} = i\omega A_{\perp,\parallel}$), and after a rephasing of $\phi(z)$.

	The fact that the mass matrix in Eq.~\eqref{eom_planewaves} is not diagonal means that  $A_{\parallel}$ and $\phi$ are not the eigenmodes of propagation inside $\vec{\mathcal{B}}$. These are found by diagonalisation and correspond to two new mass eigenvalues, $\mu_+$ and $\mu_-$, that depend on $\omega$:
	\be
			{\mu_{\pm}}^2 = \frac{1}{2}({\omega_{\mathrm{p}}}^2 + m^2) \pm \frac{1}{2}\Delta\mu^2,
	\label{eq:mupm}
	\ee
where we write
	\be
		\Delta\mu^2 \equiv \sqrt{{{(2g\mathcal{B}\omega)}^2 + (m^2 - {\omega_{\mathrm{p}}}^2)}^2},
	\ee
which corresponds to the difference of the masses squared of the eigenstates of the mixing.\footnote{Even though the product $g\mathcal{B}$ can be arbitrarily large, so that the right hand side of Eq.~\eqref{eq:mupm} can become negative, the group velocity for these eigenstates
	\be
		v_g=\frac{d\omega}{dk} = {\left(\frac{dk}{d\omega}\right)}^{-1}
	\ee
	is always smaller than unity.} On the other hand, the mixing angle is given by
	\be
		\theta = \frac{1}{2} \textrm{atan}\left(\frac{2g\mathcal{B}\omega}{m^2 - {\omega_{\mathrm{p}}}^2}\right).\label{eq:thetamix}
	\ee
	Let us first write $k_E = \sqrt{\omega^2 - {\omega_{\mathrm{p}}}^2}$ and $k_{\phi} =\sqrt{\omega^2 - m^2}$, the dispersion relations for photons and pseudoscalars \draftun{outside the} magnetic field regions.
	The solutions for the propagating fields then take the form:
	\begin{align}
		A_{\perp}(z) & = A_{\perp}(0)\ e^{ik_Ez},\label{eq:Aperpgen}\\
		A_{\parallel}(z)  & = A_{\parallel}(0)\ \big(\cos^2\theta\ e^{ik_C z} + \sin^2\theta\ e^{ik_D z}\big) + \phi(0)\ \frac{\sin2\theta}{2}\big(e^{ik_C z} - e^{ik_D z}\big),\label{eq:Apargen}\\
		\phi(z) & = A_{\parallel}(0)\ \frac{\sin2\theta}{2}\big(e^{ik_C z} - e^{ik_D z}\big) + \phi(0)\ \big(\sin^2\theta\ e^{ik_C z} + \cos^2\theta\ e^{ik_D z}\big),\label{eq:phigen}
	\end{align}
	where $k_C$ and $k_D$ are respectively $k_+=\sqrt{\omega^2-{\mu_+}^2}$ and $k_-=\sqrt{\omega^2-{\mu_-}^2}$ when $\omega_{\mathrm{p}}>m$, and the other way around when $m>\omega_{\mathrm{p}}$. If we are mostly interested in the small mixing case, we can remember that the heaviest eigenmode of propagation is mostly made of the heaviest state among photons and pseudo\-scalars, and conversely for the lightest one. If $\mathcal{B}$ is vanishing, $k_C=k_E$ and $k_D=k_{\phi}$, and we recover the propagation of the free fields.

	\subsection{Evolution of the polarisation due to the mixing}

		As only photons are detected, we focus on observables for the electromagnetic field, which now evolve because of the interaction with pseudoscalars.
		Using the solutions given by Eqs.~\eqref{eq:Aperpgen} and \eqref{eq:Apargen}, the Stokes parameters for any light beam can be derived given the properties discussed in Sec.~\ref{sec:pol},
		and we can learn how the intensity, as well as the linear and circular polarisations of light, behave due to the mixing. For classic references discussing this without the Stokes parameters, see~\textit{e.g.} Refs.~\cite{Maiani:1986md,Raffelt:1987im}.

		Regarding linear polarisation in particular, as the magnetic field defines a physical direction of prime importance for the mixing, it is in fact particularly useful to consider $Q$ and $U$ separately in the $(\vec{e}_{\perp},\vec{e}_{\parallel})$ basis. Indeed, doing so greatly helps get a \draftun{deeper} understanding of the physical implications of the mixing; it is not as straightforward if we only rely on the linear polarisation degree, for instance. Note also that, as the intensity $I$ will now evolve, it is more instructive to consider the unnormalised Stokes parameters $Q$, $U$ and $V$ rather than $q$, $u$ and $v$.

		In a first approach, we restrict the discussion to what happens to an initial photon beam of any polarisation; that is, if we take $\phi(0)=0$. Indeed, we can then express the Stokes parameters simply as functions of the initial ones, without terms involving products of initial photon and axion field amplitudes. 
		Doing so, we obtain that the evolution of the Stokes parameters inside a magnetic field region for a plane-wave beam $\vmE_{\mathrm{r}}$ described initially by $I_0, Q_0, U_0$ and $V_0$ reads
                \be
                \left\{
                    \begin{split}
                        I(z) &= I_0 - \frac{1}{2}\left(I_0 + Q_0\right) \sin^2 2\theta \sin^2\left(\frac{1}{2}(k_C-k_D)z\right)\\
                        Q(z) &= I(I_0 \rightleftarrows Q_0)\\
                        U(z) &= U_0\left[ \cos^2\theta\cos((k_E - k_C)z) + \sin^2\theta\cos((k_E - k_D)z) \right]\\
		& \hspace{.09cm}+ V_0\left[ \cos^2\theta\hspace{.045cm}\sin((k_E - k_C)z) + \sin^2\theta\hspace{.04cm}\sin((k_E - k_D)z) \right]\\
                        V(z) &= U(U_0\rightarrow V_0, V_0\rightarrow-U_0).
                    \end{split} \right.
                    \label{eq:Stokes}
                \ee

\begin{proof}
		These relations can for instance be derived using a fully polarised plane-wave beam
		\be
			\vmE_{\mathrm{r}}(z,t) = \cos(\varphi_0)E_{\perp}(z,t)\vec{e}_{\perp} + \sin(\varphi_0)E_{\parallel}(z,t)\vec{e}_{\parallel},
		\ee
	in which case the Stokes parameters \eqref{eq:StokesE} can be expressed, in terms of $A_{\perp,\parallel}(z)$, as
        \be
                \left\{
		\begin{split}
                I(z) &= \omega^2\sin^2(\varphi_0) {|A_{\parallel}(z)|}^2 + \omega^2\cos^2(\varphi_0) {|A_{\perp}(z)|}^2\\
                Q(z) &= \omega^2\sin^2(\varphi_0) {|A_{\parallel}(z)|}^2 - \omega^2\cos^2(\varphi_0) {|A_{\perp}(z)|}^2\\
                U(z) &= \omega^2\sin(2\varphi_0)\operatorname{Re}\{A_{\parallel}(z)A^*_{\perp}(z)\}\\
                V(z) &= \omega^2\sin(2\varphi_0)\operatorname{Im}\{A_{\parallel}(z)A^*_{\perp}(z)\},
	        \label{eq:StokesA}
		\end{split}
                \right.
        \ee
	and we then have to use both the fact that, when $\phi(0)=0$, Eqs.~\eqref{eq:Aperpgen} and \eqref{eq:Apargen} give
	\begin{align}
		A_{\parallel}(z)A_{\perp}^*(z) &= A_{\parallel}(0)A_{\perp}^*(0)\left[ \cos^2\theta\ e^{-i(k_E - k_C)z} + \sin^2\theta\ e^{-i(k_E - k_D)z} \right],\\
		|A_{\perp}(z)|^2 &= |A_{\perp}(0)|^2,\\
		|A_{\parallel}(z)|^2
		&= |A_{\parallel}(0)|^2 \left[\frac{1+\cos^2 2\theta}{2}  + \frac{\sin^2 2\theta}{2}\cos((k_C-k_D)z)\right],
	\end{align}
	and that, initially,
                \be
                \left\{
                    \begin{split}
                        I(0) \equiv \ \! I_0 &= \underbrace{\omega^2\sin^2(\varphi_0) {|A_{\parallel}(0)|}^2}_{\frac{1}{2}\left(I_0 + Q_0\right)} + \underbrace{\omega^2\cos^2(\varphi_0) {|A_{\perp\vphantom{\parallel}}(0)|}^2}_{\frac{1}{2}\left(I_0 - Q_0\right)}\\ 
                        Q(0) \equiv\! Q_0 &= \overbrace{\omega^2\sin^2(\varphi_0) {|A_{\parallel}(0)|}^2} - \overbrace{\omega^2\cos^2(\varphi_0) {|A_{\perp}(0)|}^2}\\
                        U(0) \equiv U_0 &= \omega^2\sin(2\varphi_0)\operatorname{Re}\{A_{\parallel}(0)A^*_{\perp}(0)\}\\
                        V(0) \equiv V_0 &= \omega^2\sin(2\varphi_0)\operatorname{Im}\{A_{\parallel}(0)A^*_{\perp}(0)\}.
                    \end{split} \right.
                    \label{eq:StokesA0}
                \ee

		Now, the unpolarised case can, for example, be obtained from this one simply via a normalised sum of the Stokes parameters of beams with $\varphi_0 = 0$ and $\varphi_0 = \frac{\pi}{2}$, for which $\vmE_{\mathrm{r}}$ is equal to $\vec{E}_{\perp}$ or $\vec{E}_{\parallel}$ respectively. We then see that the relations for the Stokes parameters keep the same form (the equations for $U(z)$ and $V(z)$ are then just complicated ways to write zero). Therefore, Eqs.~\eqref{eq:Stokes} hold not only for fully polarised light beams but also for unpolarised and partially polarised ones.
\end{proof}

\begin{figure}[h!!]
	\centering
	\includegraphics[width=0.3\textwidth,trim = 0mm 0mm 45mm 0mm, clip]{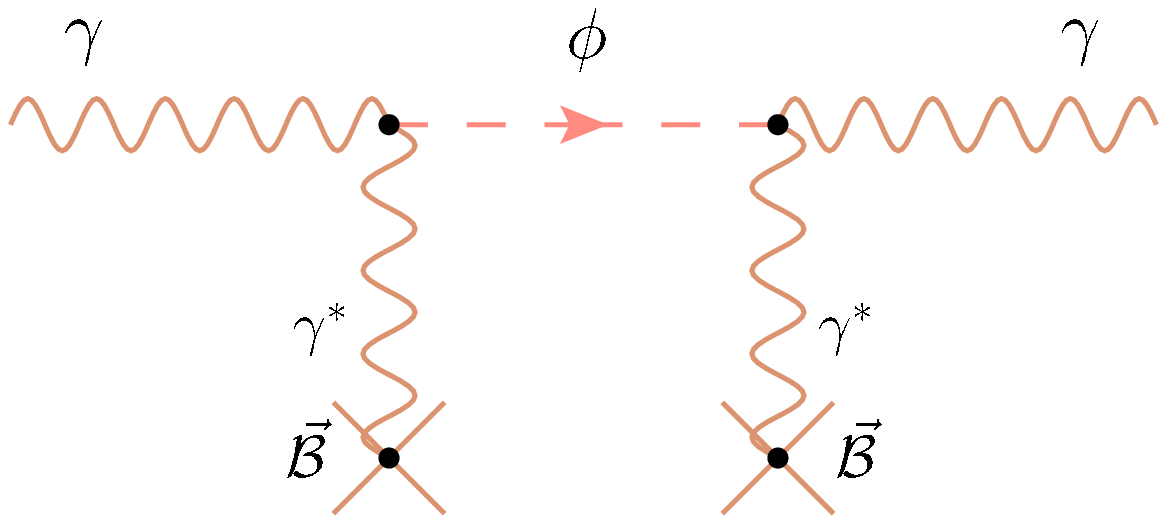}
	\caption{Process responsible for dichroism; the vertex is only non-zero for a specific direction of polarisation.}
	\label{fig:primakoffdichr}
\end{figure}

	The relations~\eqref{eq:Stokes} imply in fact dichroism and birefringence; see, \textit{e.g.} Ref.~\cite{Payez:2008pm}. Dichroism is the selective absorption of one direction of polarisation; it modifies the linear polarisation of light. This effect is clearly seen in the evolution of the Stokes parameter $Q(z)$ which compares the intensity in the two orthogonal directions.
	The total intensity $I(z)$ of course follows the same behaviour. The pair ($I$, $Q$) is directly sensitive to the modifications of the amplitude of photons due to on-shell pseudo\-scalars; see Fig.~\ref{fig:primakoffdichr}.

\begin{figure}[h!!]
	\centering
	\includegraphics[width=0.5\textwidth]{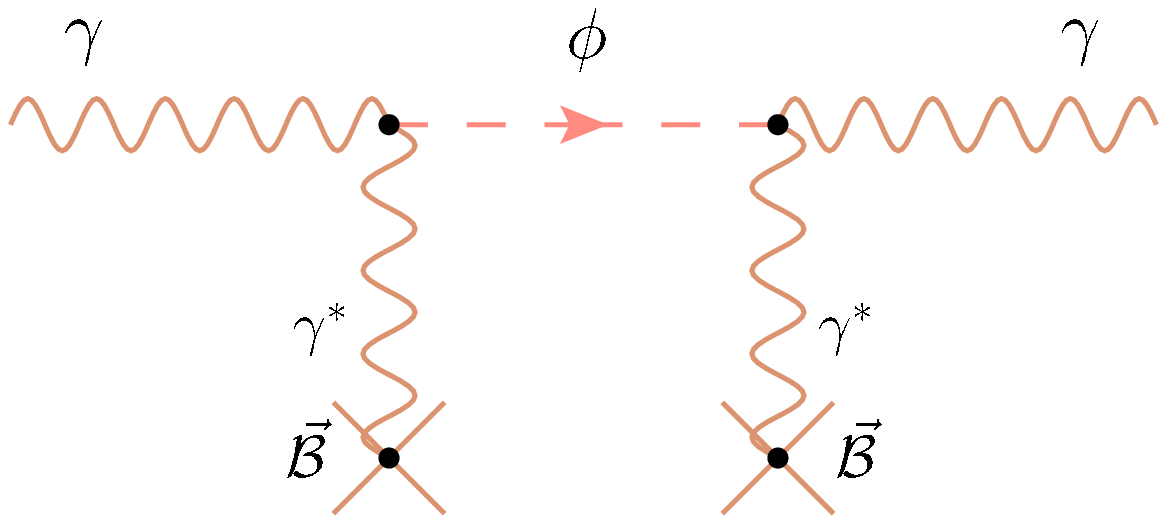}
	\caption{Process responsible for birefringence: for photons sensitive to the mixing, this process has the structure of a correction to the propagator, a modification of the mass.}
	\label{fig:dbleprimakoffbirefr}
\end{figure}

	Birefringence, on the other hand, is the existence of different refractive indices for different polarisations (along $\vec{\mathcal{B}}$ and perpendicularly to it), which then travel at different velocities; it causes linear and circular polarisations to convert into each other. This strong connection between the two is explicit in the evolution of $U(z)$ and $V(z)$. The pair ($U$, $V$) is directly sensitive to the phase shift induced by virtual pseudo\-scalars; see Fig.~\ref{fig:dbleprimakoffbirefr}.

	Note also that, for an initially unpolarised light beam, while $I(z)$ and $Q(z)$ will evolve due to pseudo\-scalar-photon mixing, $U(z)$ and $V(z)$ will remain zero. Indeed, for unpolarised light, the concept of phase shift does not make sense.
Similarly, if a linearly polarised beam points either exactly in the magnetic field direction or perpendicularly to it, \textit{i.e.} $Q_0\neq0$ and $U_0=0$, there cannot be any induced phase shift and, therefore, any induced circular polarisation.
\bigskip

	Now, even in the restricted case of a homogeneous magnetic field, and for fixed external conditions, it may seem complicated to obtain an overall picture of the phenomenological effects of the mixing.
	The evolution of the Stokes parameters is not only function of the initial ones, but also depends on many other parameters: the frequency of light, the plasma frequency, the field strength of the external transverse magnetic field, the distance travelled inside it, and, of course, the parameters of the pseudoscalar particle.
	To get a better understanding of the consequences of the mixing and get a glimpse of its rich and exciting phenomenology, we are therefore going to discuss the behaviour of the Stokes parameters in different cases; this requires assigning particular values to the external parameters, which we are going to choose as relevant to the later astrophysical applications of Sec.~\ref{sec:alpscenario}.

	\subsubsection{Behaviour of $I$ and $Q$, and connection with dichroism}

		\paragraph{Unpolarised light}

\begin{figure}
		\centering
		\includegraphics[width=\textwidth]{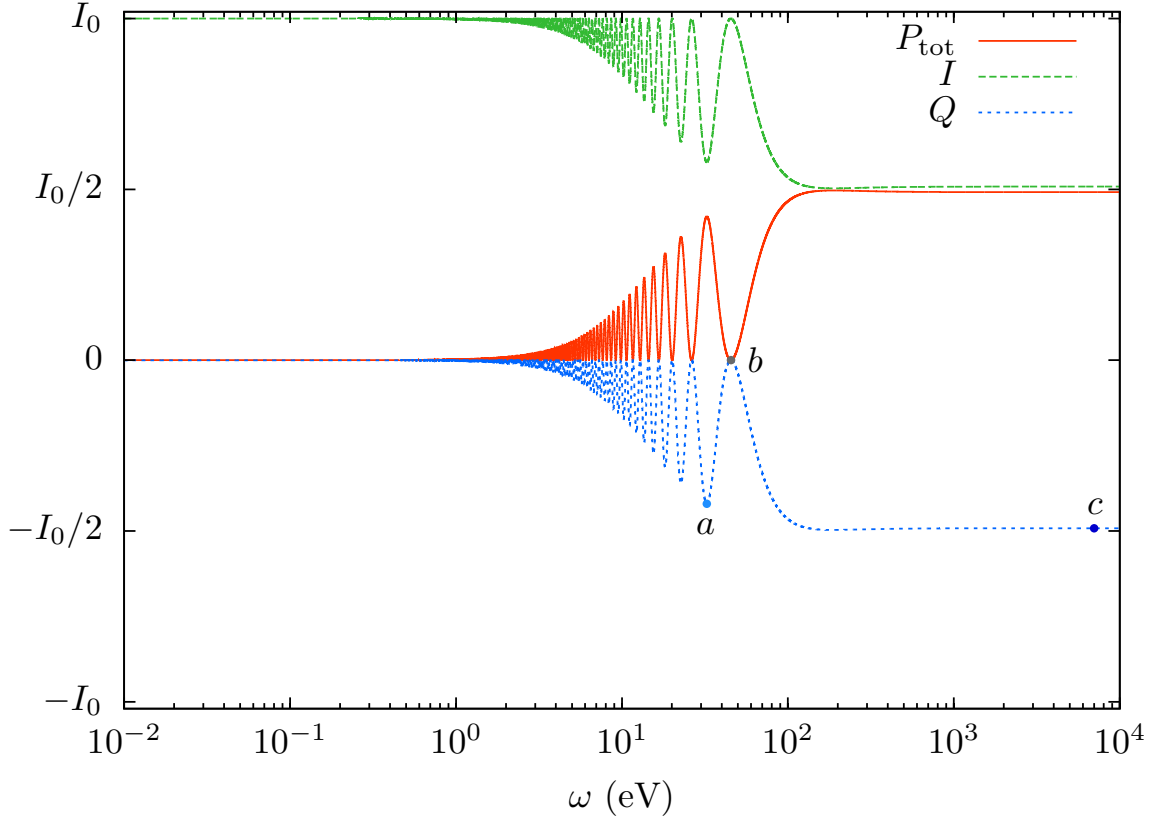}
		\caption{The total intensity of polarisation $P_{\mathrm{tot}}$, $I$ and $Q$, at the end of a $1.6\times10^{30}$~eV$^{-1}$ ($\simeq10$~Mpc) magnetic field zone in the case of initially unpolarised plane-wave light beams of different frequencies, using, as parameters: $\omega_{\mathrm{p}}=3.7\times10^{-14}$~eV, $m=10^{-14}$~eV, and $g\mathcal{B}=4.5\times10^{-29}$~eV (\textit{e.g.}, $g=7.5\times10^{-12}$~GeV$^{-1}$ and $\mathcal{B}=0.3$~$\mu$G). The Stokes parameters $U$ and $V$, not shown here, are zero at all frequencies.}
		\label{fig:stokespw_omega_npol}
	\end{figure}

		Using the solutions~\eqref{eq:Stokes}, we illustrate in Fig.~\ref{fig:stokespw_omega_npol} the values of the Stokes parameters and of the total intensity of polarisation at the end of a constant magnetic field region, for initially unpolarised beams of different frequencies.

		While these beams were all unpolarised initially, we observe a spontaneous appearance of polarisation for some values of the frequency due to the mixing. 
		We notice on this figure an oscillatory behaviour, with the general trend that the amplitude increases from small to large values of frequency, ending with a plateau.\footnote{Of course, in such graphs, the exact position of the different regimes is determined by the values of the external parameters entering the problem, but the general trend does not change.} 
		These oscillations already start at low energies but with amplitudes so suppressed that they do not appear on this graph; on the other hand, once the plateau is reached, no more oscillations remain.\footnote{Note that for other values of the parameters, this plateau will take other values.}

\begin{figure}
		\centering
		\includegraphics[width=\textwidth]{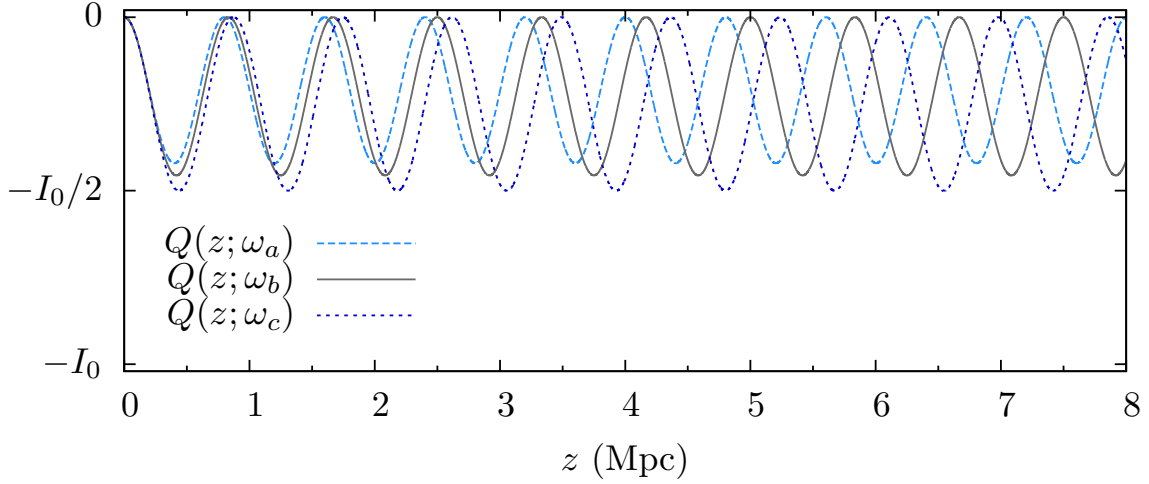}
		\caption{Illustration of the evolution of $Q(z)$ for values of the frequency corresponding to the points labelled $a$, $b$ and $c$ in Fig.~\ref{fig:stokespw_omega_npol} (respectively written $\omega_a$, $\omega_b$ and $\omega_c$); the other parameters used here are the same as in that figure.}
		\label{fig:stokespw_z_npol}
	\end{figure}

		Note that the maximum amount of polarisation achievable in this unpolarised case is actually a monotonous function of the frequency. The fact that the polarisation vanishes at some frequencies before the plateau (point $b$ in Fig~\ref{fig:stokespw_omega_npol}, for instance) simply reflects the oscillations of $I(z)$ and $Q(z)$ with the distance $z$ travelled inside the magnetic field, which have frequency-dependent wavelengths; see Fig.~\ref{fig:stokespw_z_npol}.

		Indeed, the phase controlling the propagation of these Stokes parameters is given by
		\be
			\frac{1}{2}(k_C - k_D)z = \pm\frac{1}{2}(k_--k_+)z,
		\ee
		depending on whether $m>\omega_{\mathrm{p}}$ or the other way around. As it enters an even function, one can choose
		\be
		    \frac{1}{2}(k_--k_+)z=\frac{1}{2}\frac{\sqrt{{(2g\mathcal{B}\omega)}^2 + {(m^2 - {\omega_{\mathrm{p}}}^2)}^2}}{\sqrt{\omega^2-\mu_-^2}+\sqrt{\omega^2-\mu_+^2}}z=\frac{1}{2}\frac{\Delta\mu^2 z}{\sqrt{\omega^2-\mu_-^2}+\sqrt{\omega^2-\mu_+^2}},
		\ee
		which gives
		\be
			\sin^2\left(\frac{1}{2}(k_C-k_D)z\right)  \stackrel{\omega^2\ggg {\mu_{\pm}}^2}{=}  \sin^2\left(\frac{\Delta\mu^2 z}{4 \omega}\right),\label{eq:kCkD}
		\ee
		provided that $\omega^2\ggg\mu_{\pm}^2$ (for the denominator), which is an extremely good approximation in all the cases we are interested in, where \draftun{this relation} is verified by many orders of magnitude.\footnote{For instance, using $\omega = 2.5$~eV and the parameters used for illustration in Fig.~\ref{fig:stokespw_omega_npol}, one gets:
		\be
			\frac{\omega^2}{{\mu_{\pm}}^2}\approx10^{27}.\label{eq:approxgood}
		\ee
		}

		For small enough values of $\omega$, the wavelength of oscillation along $z$ for $I(z)$ and $Q(z)$ is essentially linear in $\omega$, as $|m^2 - {\omega_{\mathrm{p}}}^2|$ dominates $\Delta\mu^2$.
		It then starts receiving contributions from $(2g\mathcal{B}\omega)$, which eventually becomes the leading term at high energies and makes the wavelength of oscillation along $z$ frequency independent, hence the plateau.

		What we also see in Fig.~\ref{fig:stokespw_omega_npol} is that the amount of total polarisation mimics the evolution of the parameter $Q$ (we indeed have $P_{\mathrm{tot}} = |Q|$) and mirrors that of $I$.
		This is actually a signature of dichroism. What happens is that we have a depletion of photons polarised parallel to $\vec{\mathcal{B}}$, leading to a loss of intensity and to a modification of the $Q$ parameter (see equations~\eqref{eq:Stokes}) by the same amount. This, in turn, leads to a net appearance of polarisation in the direction perpendicular to the magnetic field. As we know that the initial state of unpolarised light can be described as an incoherent weighted sum of beams with orthogonal polarisations, we understand that we can therefore at most lose half of the initial intensity (that along $\vec{e}_{\parallel}$), leading to light fully linearly polarised along $\vec{e}_{\perp}$ when $P_{\mathrm{tot}}=I$.

		\begin{figure}
			\centering
					\includegraphics[width=\textwidth]{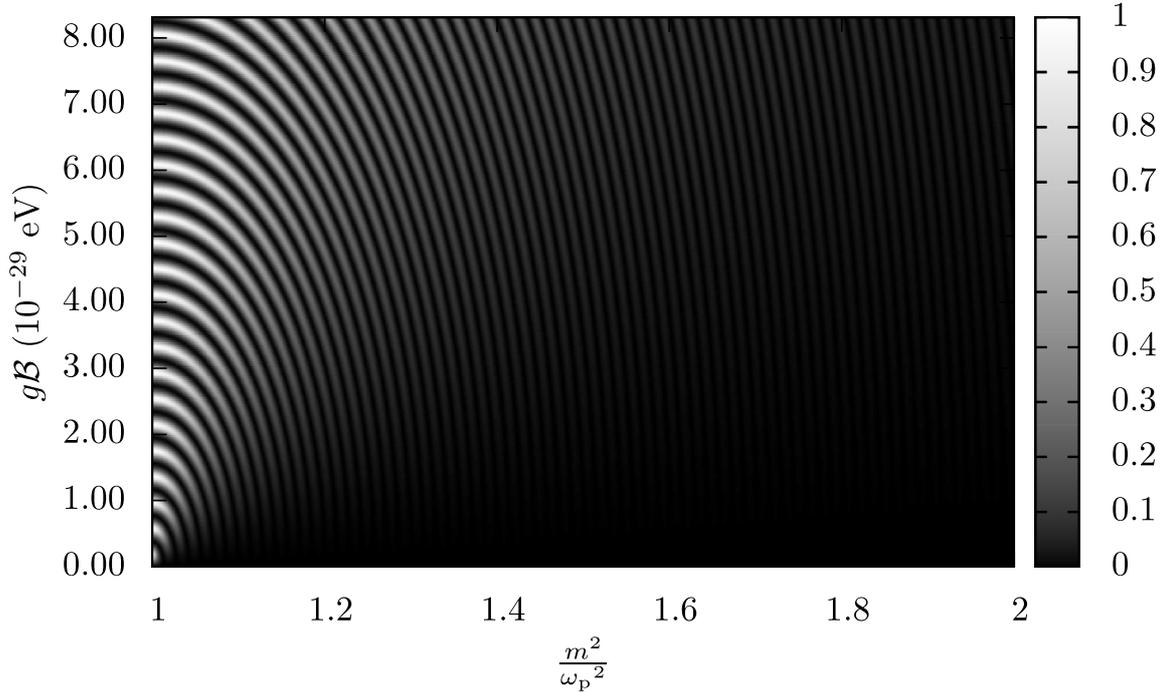}
			\caption{Linear polarisation degree (shown in the right-hand box) generated through pseudoscalar-photon mixing at the end of a 10~Mpc magnetic field region, in the case of initially unpolarised light of frequency $\omega = 2.5$~eV; in this figure, the plasma frequency has been kept fixed at $\omega_{\mathrm{p}}=3.7\times10^{-14}$~eV, corresponding to $n_{\mathrm{e}}=10^{-6}$~cm$^{-3}$.}
			\label{fig:japfan_unpol}
		\end{figure}

		This is also seen in Fig.~\ref{fig:japfan_unpol}, where we illustrate a dependence of the degree of linear polarisation $p_{\mathrm{lin}}$ on the external parameters and on the pseudoscalar properties. We do not display pseudo\-scalar masses smaller than $\omega_{\mathrm{p}}$ since the linear polarisation degree is an even function of ($m^2 - {\omega_{\mathrm{p}}}^2$) in this case, which remains true as long as there is not initial circular polarisation. Note also that the mixing depends on $m$ and $\omega_{\mathrm{p}}$ separately and that nothing special happens if we take $\omega_{\mathrm{p}} = 0$. In this figure, we see a pattern (that we will explain later) from which it is clear that, together with the frequency of light, there is a strong dependence of the mixing upon these parameters: it is larger when $m$ and $\omega_{\mathrm{p}}$ are close, as well as when $g\mathcal{B}$ increases, and it can lead to a complete polarisation of light for some values of the parameters.

		\paragraph{Linearly ($Q_0\neq0$) polarised light}
		Let us now consider initially linearly polarised beams with $Q_0\neq0$ and $U_0=V_0=0$.
		This is in fact not so different from the unpolarised situation, as only dichroism is at work.

\begin{figure}
			\centering
			\includegraphics[width=\textwidth]{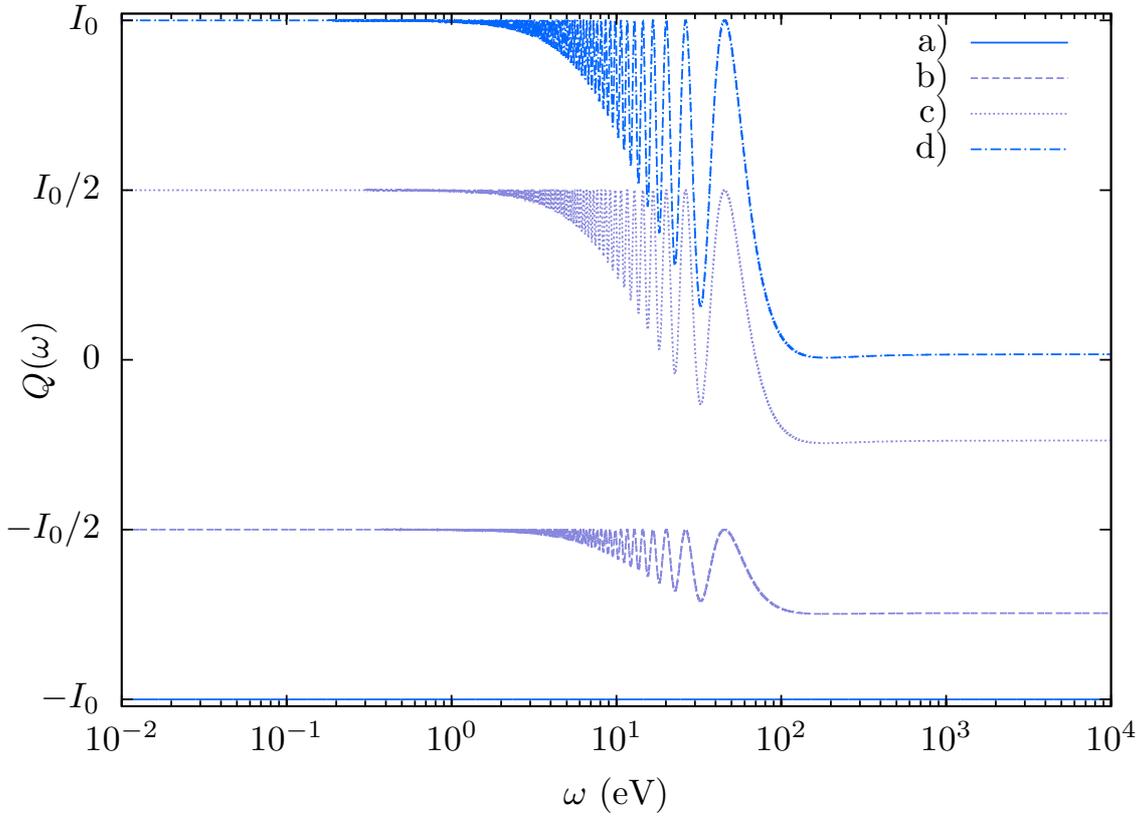}
			\caption{$Q$, at the end of a 10~Mpc magnetic field zone in the case of plane-wave light beams of different frequencies, for different initial conditions: a)~$Q_0=-I_0$; b)~$Q_0 = -I_0/2$; c)~$Q_0=I_0/2$; d)~$Q_0=I_0$. The other parameters used are the same as in Fig.~\ref{fig:stokespw_omega_npol}.}
			\label{fig:stokespw_omega_polQ}
		\end{figure}

		Given that the behaviour of $I$ and $P_{\mathrm{tot}}$ can again be easily obtained from that of $Q$ (respectively as the same figure shifted, and as the absolute value of $Q$), we show only the latter in Fig.~\ref{fig:stokespw_omega_polQ}. Note that, again, the parameters $U$ and $V$ remain zero at all frequencies.

		Once more, the intensity of light in the direction parallel to the external magnetic field is more and more attenuated with $\omega$, as axion-like particles are produced.
		When one uses Stokes parameters, it is enough to consider fully polarised beams, with $Q_0=\pm I_0$. We nonetheless also include initial partial polarisations for illustration, which indeed behave as intermediates between the fully polarised and the unpolarised cases.
		
		As expected, beams initially fully polarised perpendicularly to the direction of the magnetic field (namely with $Q_0=-I_0$) are unaffected by the mixing for any value of the frequency, while we have the opposite situation when they are fully polarised along $\vec{\mathcal{B}}$ (namely with $Q_0=I_0$), as for high enough energies the whole beam can oscillate into axion-like particles.

		Also note that, while the linear polarisation changes, there is no rotation of the plane of polarisation; this is because we are considering initial polarisations exactly parallel or perpendicular to $\vec{\mathcal{B}}$ here. A change of the polarisation angle
		\be
			\varphi(z;\omega) = \frac{1}{2}\atan\left(\frac{U(z;\omega)}{Q(z;\omega)}\right)\tag{\ref{eq:varphi}}
		\ee
		would require $U_0\neq0$ (or $V_0\neq0$ as we will now see).

	\subsubsection{Behaviour of $U$ and $V$: birefringence and dichroism}

		\paragraph{Linearly ($U_0\neq0$) and circularly polarised light}

		We now consider at the same time initially linearly polarised light beams with $U_0\neq0$ and initially circularly polarised ones ($V_0\neq0$): there indeed exist very strong connections in the evolution of the parameters $U$ and $V$.
		Note that for these initial conditions both dichroism and birefringence are now at work.
		Taking benefit of what we have already discussed, direct consequences of dichroism are that:
		\begin{itemize}
			\item[-] $I$ and $Q$ actually behave exactly as in the case of unpolarised light;
			\item[-] the polarisation angle now also evolves due to the mixing because the polarisation of light will no longer remain along $\vec{e}_{\parallel}$ or $\vec{e}_{\perp}$.
		\end{itemize}

	\begin{figure}
		\centering
		\includegraphics[width=\textwidth]{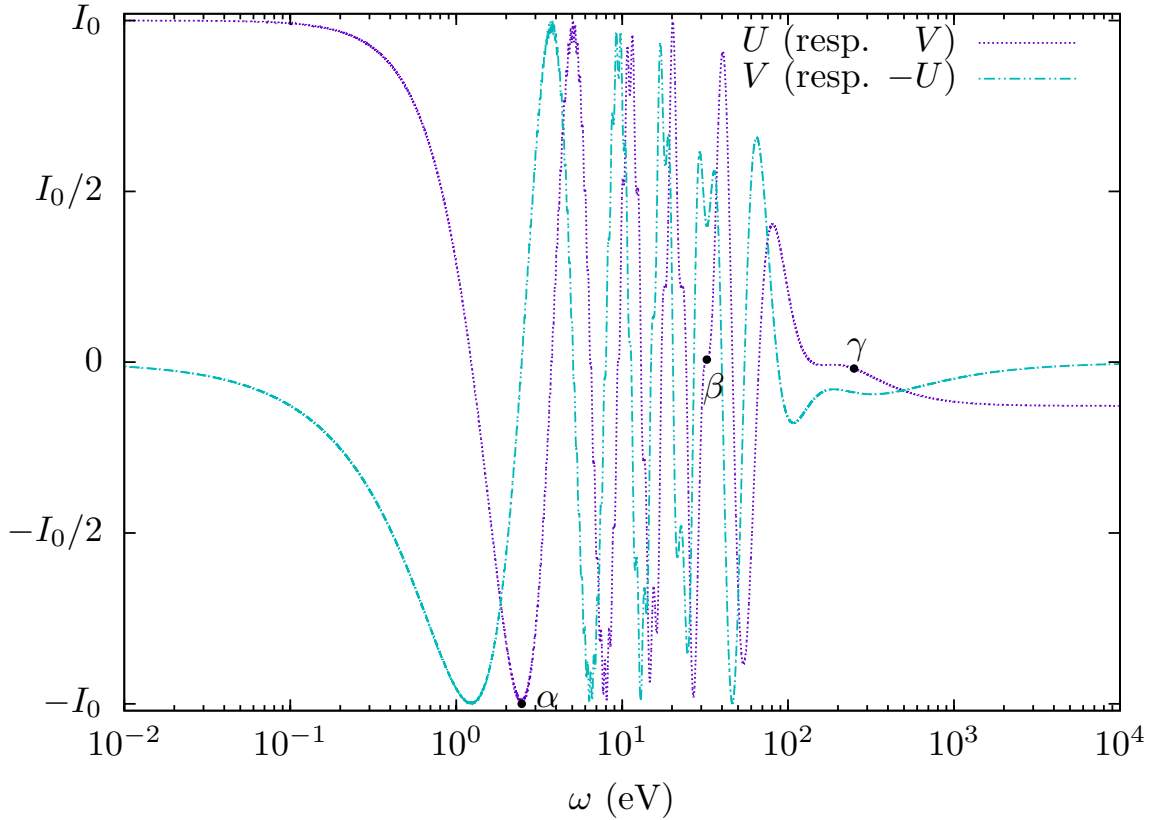}
		\caption{$U$ and $V$ at the end of a 10~Mpc magnetic field zone in the case of initially polarised plane-wave light beams of different frequencies: for these beams of initial intensity $I_0$, we have taken $U_0=I_0,$ and $Q_0=V_0=0$ (resp. $V_0=I_0,$ and $Q_0=U_0=0$). In this figure, we have used the same parameters as in Fig.~\ref{fig:stokespw_omega_npol}. The Stokes parameters $I$ and $Q$, not shown here, actually behave exactly as in the unpolarised case.}
		\label{fig:stokespw_omega_polUV}
	\end{figure}

		We can now focus on $U$ and $V$. For example, let us discuss the special case where one of the two is initially equal to zero. An illustration of this has been represented in Fig.~\ref{fig:stokespw_omega_polUV}, which can actually be interpreted as the case of light beams of different frequencies with either $U_0=I_0$ and $V_0=0$ or $U_0=0$ and $V_0=I_0$: the law of evolution of $U$ (resp. of $V$) in the first case is exactly the one for $V$ (resp. for $-U$) in the second one. This observation, which suggests some kind of symmetry in the way in which these parameters evolve due to the mixing, is a signature of birefringence, which can be understood as due to some retarder. It is indeed well-known that using retarders, one can produce a circularly polarised beam, starting with a linearly polarised one and \textit{vice versa}.

		Another interesting feature related to these two parameters is that their contribution to the total polarisation, \textit{i.e.} $U^2+V^2$, can simply be written as $U_0^2+ V_0^2$ times a function of the distance and of the other (external) parameters, \draftun{as we discuss in Ref.~\cite{Payez:2008pm}}. In other words, the only quantity that matters for the total polarisation as far as $U$ and $V$ are concerned is, in fact, the sum of their initial values squared, notwithstanding the details of their individual initial values, which confirms a close connection between them in these processes.
		Actually, in the most general situation, the behaviour of $V$ due to the mixing is that of $U$, where $U_0\rightarrow V_0$ and $V_0\rightarrow-U_0$, as one clearly sees from the expressions for $U$ and $V$ that we have written explicitly in Eqs.~\eqref{eq:Stokes}.

		Now, in Fig.~\ref{fig:stokespw_omega_polUV}, we first notice a much more complicated behaviour for $U(\omega)$ and $V(\omega)$, compared to that of $Q(\omega)$ and $I(\omega)$ in Fig.~\ref{fig:stokespw_omega_npol}. We can already isolate three distinct regimes:
		\begin{itemize}
			\item[-] first of all, the most striking observation is that for quite a wide range of frequencies, starting with pure linear polarisation, one can get a large amount of circular polarisation at the end of the magnetic field region (and conversely);
			\item[-] even though the amplitude of oscillations of $U(\omega)$ and $V(\omega)$ are not damped with decreasing values of the frequency, we also notice that, as for dichroism, the effects of the mixing get suppressed for $\omega\rightarrow0$;
			\item[-] finally, for a given travelled distance inside $\vec{\mathcal{B}}$, we see that at high enough frequencies the parameter $V$ (resp. $U$), which was initially zero, vanishes. As we will see, this is in fact generic: for fixed $z$ and $\omega\rightarrow \infty$, $V$ (resp. $U$) will always be vanishing if it was zero initially, while the amplitude of polarisation due to dichroism for unpolarised and partially polarised light tends to reach a maximum quickly and to oscillate instead.
		\end{itemize}
		What happens is that, due to the process shown in Fig.~\ref{fig:dbleprimakoffbirefr}, photons with polarisations parallel to $\vec{\mathcal{B}}$ are slowed down (or speeded up, depending on the axion mass), and are phase shifted with respect to their perpendicular counterparts. This leads to the appearance of circular polarisation, by definition.
		The reverse statement is equally true as the requirements for purely linearly or circularly polarised states are both very demanding and will be disturbed by birefringence. In the first case $\mE_{\mathrm{r},\parallel}$ and $\mE_{\mathrm{r},\perp}$ oscillate strictly in phase, while in the second they are phase shifted by exactly $\frac{\pi}{2} + k\pi\ (k\in\mathbb{Z})$.

		This overall structure as a function of the frequency is in fact the strict analogue of what has been referred to as the smoking gun of chameleon--photon mixing~\cite{Burrage:2008ii}. It is not only a feature of that kind of scalar axion-like particles: in Ref.~\cite{Bassan:2010ya}, the absence of circular polarisation is actually only due to the external conditions entering their physical problem, which are such that, at the frequencies considered, they are only scanning the third regime. The apparent difference between these results is not related to an intrinsic difference in the nature of ALPs.

	\begin{figure}
		\centering
		\includegraphics[width=\textwidth, trim = 0mm 9mm 0mm 0mm, clip]{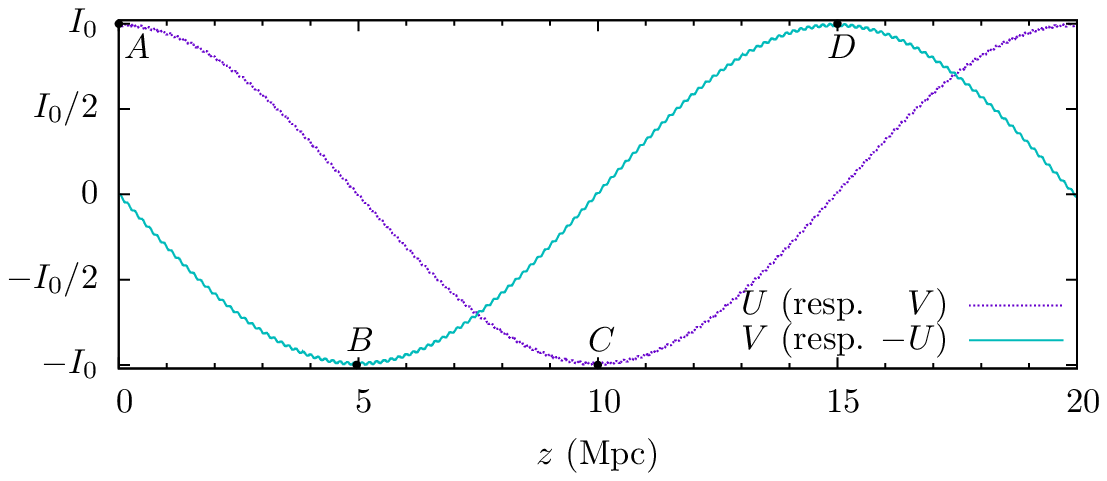}
		\includegraphics[width=\textwidth, trim = 0mm 9mm 0mm 0mm, clip]{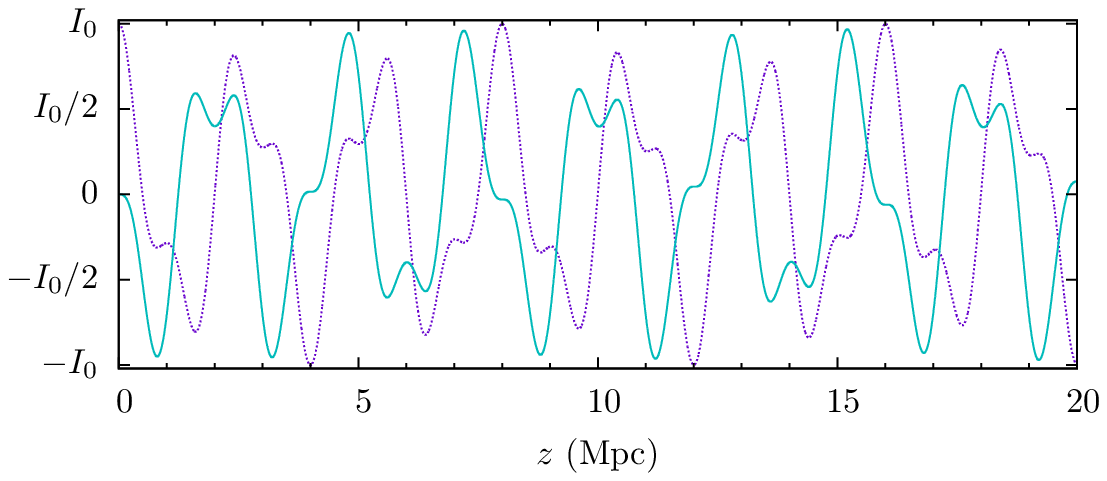}
		\includegraphics[width=\textwidth]{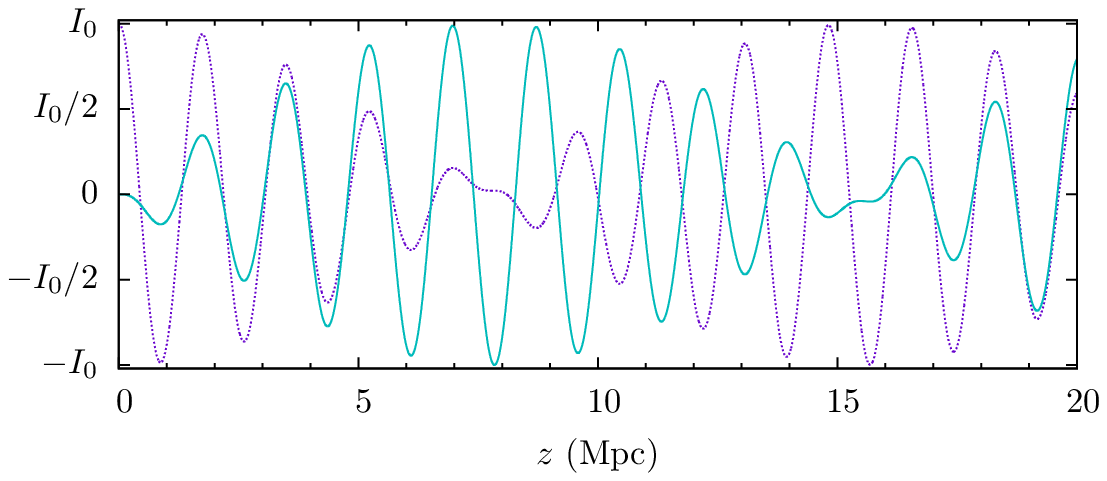}
		\caption{Illustration of the evolution of $U(z)$ and $V(z)$ for values of the frequency corresponding to the points labelled $\alpha$, $\beta$, and $\gamma$ in Fig.~\ref{fig:stokespw_omega_polUV}; the other parameters used here are the same as in that figure. With increasing values of the frequency, see respectively the \emph{top} ($\omega_{\alpha}$), \emph{middle} ($\omega_{\beta}$), and \emph{bottom} ($\omega_{\gamma}$) panels.}
		\label{fig:stokespw_z_polUV}
	\end{figure}

		We now illustrate, in Fig.~\ref{fig:stokespw_z_polUV}, for values of the frequency corresponding to the points that were labelled $\alpha$, $\beta$ and $\gamma$ in Fig.~\ref{fig:stokespw_omega_polUV}, the evolution of the Stokes parameters $U$ and $V$ of plane-wave beams initially fully linearly polarised ($U_0=I_0$, $V_0=0$) as they travel inside an external magnetic field\hspace{1pt}---\hspace{1pt}again, the same reasoning applies if we start with $U_0=0$ and $V_0=I_0$ instead.

		In particular, the small mixing case (corresponding to $\omega_{\alpha}$, see top panel) is extremely similar to what one also obtains with other birefringent media (see \textit{e.g.} Ref.~\cite{Genov:2011bf}), such as calcite crystals.\footnote{Except that, here, the Stokes parameter $Q$ also evolves at the same time due to dichroism, but we can temporarily neglect that for the sake of this discussion in this very weak mixing case.} What is shown in this panel is that, because of the mixing, the beam, as it propagates, being initially linearly polarised (point $A$, $U=I_0$, $V=0$), develops an ellipticity ($U\neq I_0$, $V\neq0$) up to the point where it is fully circularly polarised (point $B$, $U=0$, $V=-I_0$). As the propagation continues, from circularly, it becomes once again linearly polarised but with a plane of polarisation perpendicular to the initial one (point $C$, $U=-I_0$, $V=0$), then elliptical again, then, circularly polarised (point $D$, $U=0$, $V=I_0$), etc.

		While we have thus far mainly discussed birefringence, the way in which $U$ and $V$ evolve is not decoupled from dichroism; this is especially important in the strong mixing case. That can be seen in Fig.~\ref{fig:stokespw_z_polUV}, where we notice a modulation in the form of small wiggles (top panel), which becomes more and more important as the mixing increases and leads to a beating (middle and bottom panel).\footnote{Each time a photon undergoes the $\gamma_{\parallel}\rightarrow\phi\rightarrow\gamma_{\parallel}$ process, one can check that it also receives an additional $\pi$-phase shift with respect to photons with orthogonal polarisations, together with the phase related to the different masses (these are of course only relevant for $U_0\neq0$ or $V_0\neq0$), hence the modulation. This can be traced back to the unitarity of the mixing.}

		Looking at the behaviour of $U^2 + V^2$, we can actually isolate this modulation that we observe, and we find that
		\be
			U^2+V^2=\left({P_{\mathrm{tot,0}}}^2 - {Q_0}^2\right)\left[1 - \sin^2 2\theta\sin^2\left(\frac{\Delta\mu^2 z}{4 \omega}\right)\right],
		\ee
		whose amplitude and wavelength of oscillation have the same dependences as that of $I$ and $Q$; see Eqs.~\eqref{eq:Stokes} and~\eqref{eq:kCkD}.\footnote{Note that, by definition, in the case of initially polarised light, one has $P_{\mathrm{tot,0}}=I_0$.}
		One can check that it actually behaves so that $I^2\geq Q^2+U^2+V^2$ is satisfied at all times, which is of course required as $I$ represents the total intensity.\footnote{Interestingly, note also that the mixing with spinless particles at a given frequency does not depolarise fully polarised beams as long as there are photons.}
		We can relate this to the complicated behaviour seen in Fig.~\ref{fig:stokespw_omega_polUV}, which is due to the $\omega$-dependence of the wavelength of oscillation of $U$ and $V$ along $z$ and to the modulation/beat which gets bigger and bigger with $\omega$.

\begin{figure}
		\centering
\includegraphics[width=\textwidth]{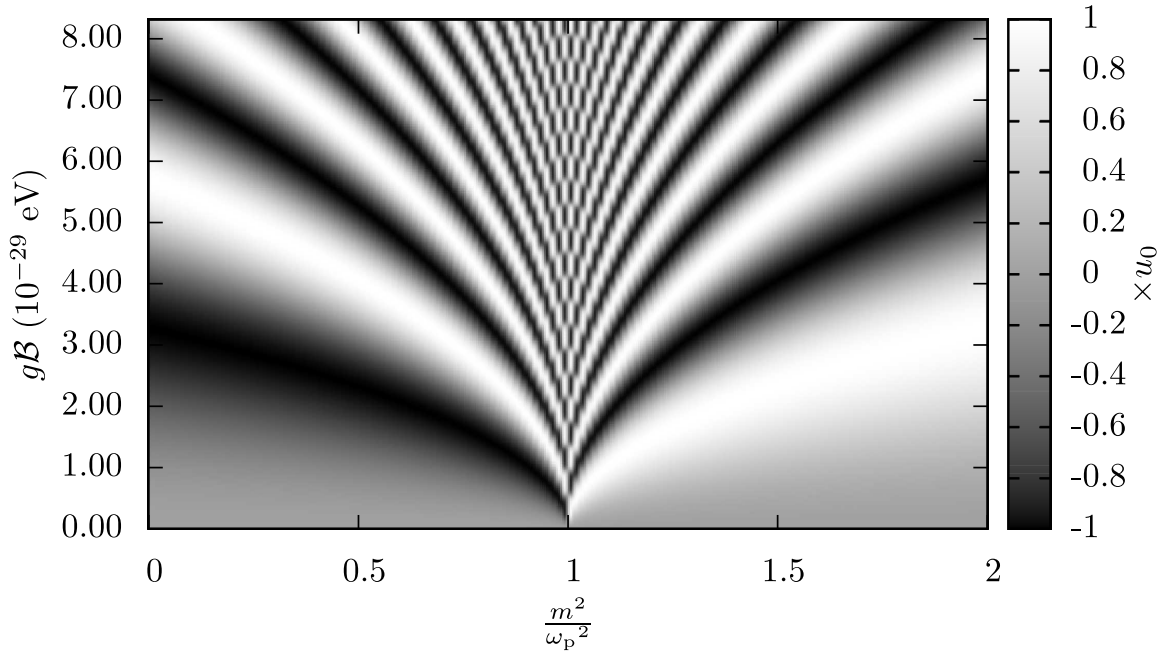}
		\caption{Circular polarisation $v=V/I$ (shown in the right-hand box) generated through pseudo\-scalar-photon mixing at the end of a 10~Mpc magnetic field region, in the case of a 2.5~eV light beam with $v_0=0$ and $q_0=0$. The plasma frequency used here is the same as in Fig.~\ref{fig:japfan_unpol}.
		}
		\label{fig:japfan_circ}
	\end{figure}

		Now, one of the most important things to remember from the mixing concerning $U$ and $V$ is that, while the \draftun{evolution with distance} and the modulation are indeed energy-dependent, the amplitude of the modulated signal is not. As seen in Fig.~\ref{fig:stokespw_z_polUV}, for very different mixing cases, starting with $U_0\neq0$ and $V_0=0$ for instance, one obtains that the maximum amount of circular polarisation attainable is always given by $U_0$.
		This is also found in Fig.~\ref{fig:japfan_circ}, which illustrates the dependence of the circular polarisation $v=V/I$ on the external parameters and on the pseudoscalar properties for initially not circularly polarised light. It is an odd function of $(m^2 - {\omega_{\mathrm{p}}}^2)$, which is not surprising as birefringence comes from the existence of different masses for $\mE_{\mathrm{r},\perp}$ and $\mE_{\mathrm{r},\parallel}$ (resp. $\omega_{\mathrm{p}}$ and either $\mu_-$ or $\mu_+$, defined in Eq.~\eqref{eq:mupm}) and the induced phase shift between them.\footnote{As for Fig.~\ref{fig:japfan_unpol}, note that nothing special happens if we take $\omega_{\mathrm{p}}=0$ as the mixing depends on the plasma frequency and on $m$ separately.}
		Again, in this figure, we clearly see that for a very large portion of the parameter space $|v|$ can be as large as $|u_0|$. What mainly differs for different sets of external and pseudoscalar parameters is actually how much time it can take\hspace{1pt}---\hspace{1pt}or how efficient it is\hspace{1pt}---\hspace{1pt}to reach this maximal amount.

		Formally there is always some induced phase shift between $\mE_{\mathrm{r},\parallel}$ and $\mE_{\mathrm{r},\perp}$. When $\omega\rightarrow0$, as the mixing is less and less efficient however, and it takes a lot of time (a long propagation inside the magnetic field) to accumulate to a sizeable amount.
		Similarly, we can understand why this effect is also negligible when $\omega\rightarrow\infty$: in this situation, $k_E, k_C, k_D\rightarrow|\omega|$, so that they all essentially travel at the same speed, approaching the speed of light in the vacuum, and it therefore also takes a lot of time to generate circular polarisation from linear polarisation (and \textit{vice versa}); we see that for instance in Fig.~\ref{fig:stokespw_z_polUV} (bottom panel).

	\subsection{Stokes parameters as functions of two pure numbers}

		Using the solutions for a single magnetic field region, we can show that the laws of evolution due to the mixing for all the Stokes parameters of a light beam can be written as a functions of two dimensionless quantities~\cite{Payez:2011sh}. The first one is the mixing angle
		\be
			\theta = \frac{1}{2} \textrm{atan}\left(\frac{2g\mathcal{B}\omega}{m^2 - {\omega_{\mathrm{p}}}^2}\right)\tag{\ref{eq:thetamix}}
		\ee
		and the other one, the quantity
		\be
			\frac{\Delta\mu^2}{\omega}z = \frac{\sqrt{{(2g\mathcal{B}\omega)}^2+{(m^2 - {\omega_{\mathrm{p}}}^2)}^2}z}{\omega};\label{eq:deltamu2z_omega}
		\ee
		which is actually very similar to what one obtains for neutrino oscillations.\footnote{In contrast, however, the difference of the squared masses of the two new eigenstates of propagation $\Delta\mu^2$ is energy-dependent for the mixing of light with spinless particles.}

		The only approximation we have made to obtain this result is to suppose that $\omega^2\ggg {\omega_{\mathrm{p}}}^2, {\mu_{\pm}}^2$, which is an excellent approximation in all the astrophysical applications we are interested in, as it holds true by many orders of magnitude; see Eq.~\eqref{eq:approxgood}. In particular, the relations we derive are valid in all the mixing cases: from weak to strong mixing, without the need for further specific assumptions or simplifications; for this reason, in situations in which our approximation holds, this generalises a discussion made in Ref.~\cite{Raffelt:1987im} and also extends it to any of the Stokes parameters.

		We are now going to give the derivation explicitly in the restricted case where $\phi(0)=0$. The dependencies are actually the same if $\phi(0)\neq0$, therefore the relevant parameters which drive the change of polarisation remain the two aforementioned dimensionless quantities even in that more general case.

		We have already discussed $\left(k_C-k_D\right)$, which enters the solutions for $I$ and $Q$, and have noted that
		\be
			\sin^2\left(\frac{1}{2}(k_C-k_D)z\right) \stackrel{\omega^2\ggg {\mu_{\pm}}^2}{=} \sin^2\left(\frac{\Delta\mu^2 z}{4 \omega}\right);\tag{\ref{eq:kCkD}}
		\ee
		we now need to focus on both $(k_E-k_C)$ and $(k_E-k_D)$, that enter the solutions for $U$ and $V$, and rewrite them in a more convenient way.
		Depending on the sign of $\theta$, or equivalently on the sign of $(m^2 - {\omega_{\mathrm{p}}}^2)$, $k_C$ and $k_D$ are equal to $k_+$ and $k_-$ or the other way around, while $k_E = \sqrt{\omega^2 - {\omega_{\mathrm{p}}}^2}$; we then have
		\be
			(k_E - k_{\pm})z = \frac{|\omega^2 - {\omega_{\mathrm{p}}}^2|-|\omega^2 - {\mu_{\pm}}^2|}{\sqrt{\omega^2 - {\omega_{\mathrm{p}}}^2} + \sqrt{\omega^2 - {\mu_{\pm}}^2}}z \stackrel{\omega^2\ggg {\omega_{\mathrm{p}}}^2, {\mu_{\pm}}^2}{=} \frac{\frac{m^2 - {\omega_{\mathrm{p}}}^2}{2} \pm \frac{{\Delta\mu^2}}{2}}{2\omega}z;
		\ee
		which enter even and odd functions.

		Therefore, we have to consider the different possibilities:
		\begin{enumerate}
		\item	if $\theta>0$, namely if $\omega_{\mathrm{p}} < m$, we have
			\be
			\begin{split}
				(k_E - k_C)z &= (k_E - k_-)z = \frac{|m^2 - {\omega_{\mathrm{p}}}^2| - {\Delta\mu^2}}{4\omega}z;\\
				(k_E - k_D)z &= (k_E - k_+)z = \frac{|m^2 - {\omega_{\mathrm{p}}}^2| + {\Delta\mu^2}}{4\omega}z;
				\label{eq:dvlp2params_num_cas1}
			\end{split}
			\ee
		\item	if $\theta<0$, namely if $\omega_{\mathrm{p}} > m$, we have
			\be
			\begin{split}
				(k_E - k_C)z &= (k_E - k_+)z = \frac{-|m^2 - {\omega_{\mathrm{p}}}^2| + {\Delta\mu^2}}{4\omega}z;\\
				(k_E - k_D)z &= (k_E - k_-)z = \frac{-|m^2 - {\omega_{\mathrm{p}}}^2| - {\Delta\mu^2}}{4\omega}z.
				\label{eq:dvlp2params_num_cas2}
			\end{split}
			\ee
		\end{enumerate}
		Note that these relations are also compatible with the trivial situation with no mixing $\theta \rightarrow 0$: indeed, that would imply
			\begin{align}
				\frac{2g\mathcal{B}\omega}{m^2 - {\omega_{\mathrm{p}}}^2} \rightarrow0;\\
				\Delta\mu^2 \rightarrow|m^2 - {\omega_{\mathrm{p}}}^2|;\label{eq:deltamucarre_theta0}
			\end{align}
		so that we recover the dispersion relation for the free fields $k_C = k_E$ and $k_D = k_{\phi}$ as expected.

		Besides, one can also prove that
		\be
			|m^2 - {\omega_{\mathrm{p}}}^2| = \Delta\mu^2\cos(2\theta),\qquad\textrm{with }\theta\in[-\pi/4, \pi/4].\label{eq:mcarre-wpcarre_rtheta}
		\ee
		\begin{proof}
		Indeed, provided that $\theta\neq0$, from the definition of the mixing angle, we have:
		\be
			m^2 - {\omega_{\mathrm{p}}}^2 = \frac{2g\mathcal{B}\omega}{\tan\left(2\theta\right)},\label{eq:mcarre-wpcarre_gB_tan2theta}
		\ee
		giving first
		\be
			{\left(\Delta\mu^2\right)}^2 = {(2g\mathcal{B}\omega)}^2[1 + \tan^{-2}(2\theta)],
		\ee
		then
		\be
			\Delta\mu^2 = \frac{(2g\mathcal{B}\omega)}{|\sin(2\theta)|},
		\ee
		and, finally, using again Eq.~\eqref{eq:mcarre-wpcarre_gB_tan2theta},
		\be
		\begin{split}
			m^2 - {\omega_{\mathrm{p}}}^2	&= +\Delta\mu^2\cos(2\theta),\qquad\textrm{for }\theta>0\\
							&= -\Delta\mu^2\cos(2\theta),\qquad\textrm{for }\theta<0.
		\end{split}
		\ee
		As for the case $\theta\rightarrow0$, the relation~\eqref{eq:mcarre-wpcarre_rtheta} is satisfied trivially; see Eq.~\eqref{eq:deltamucarre_theta0}.
		\end{proof}
		Using that last result, we can then write the numerators of the right-hand side of Eqs.~\eqref{eq:dvlp2params_num_cas1} and~\eqref{eq:dvlp2params_num_cas2} as
		\be
			\begin{split}
				|m^2 - {\omega_{\mathrm{p}}}^2| - {\Delta\mu^2} &= -2\sin^2\theta\Delta\mu^2;\\
				|m^2 - {\omega_{\mathrm{p}}}^2| + {\Delta\mu^2} &= \phantom{+}2\cos^2\theta\Delta\mu^2.
			\end{split}
		\ee

		Now, if we put everything together, we finally obtain, as announced, the laws for the evolution of the Stokes parameters in such a way that all the effects of the mixing with pseudo\-scalars in a given $\vec{\mathcal{B}}$ depend only on two dimensionless parameters: the mixing angle $\theta$ and the quantity $(\frac{\Delta\mu^2}{\omega}z)$.

		More explicitly, in the restricted case $\phi(0)=0$ also considered to get Eqs.~\eqref{eq:StokesE}, we obtain
		\be
			\begin{split}
				I(z) &= I_0 - \frac{1}{2}\left(I_0 + Q_0\right) \sin^2 2\theta \sin^2\left(\frac{1}{4}\frac{\Delta\mu^2}{\omega}z\right)\\
				Q(z) &= I(I_0 \rightleftarrows Q_0)\\
				U(z) &= U_0\left\{  {(\textrm{s}\theta)}^2 \cos\left(\frac{1}{2}{\left(\textrm{c}\theta\right)}^2\ \frac{\Delta\mu^2}{\omega}z\right) + {(\textrm{c}\theta)}^2 \cos\left(\frac{1}{2}{\left(\textrm{s}\theta\right)}^2\ \frac{\Delta\mu^2}{\omega}z\right) \right\}\\
				&
				\hspace{2.5pt}
				+V_0\left\{ {(\textrm{s}\theta)}^2 \sin\left(\frac{1}{2}{\left(\textrm{c}\theta\right)}^2\ \frac{\Delta\mu^2}{\omega}z\right) \ \!- {(\textrm{c}\theta)}^2 \sin\left(\frac{1}{2}{\left(\textrm{s}\theta\right)}^2\ \frac{\Delta\mu^2}{\omega}z\right) \right\} \mathrm{sign}(\theta)\\
				V(z) &= U(U_0\rightarrow V_0, V_0\rightarrow-U_0),\label{eq:Stokes_alternative}
			\end{split}
		\ee
		with $\textrm{c}\theta\equiv\cos(\theta)$ and $\textrm{s}\theta\equiv\sin(\theta)$.
		\bigskip

		With the realisation that these parameters can be described with only two variables, a nice consequence is that we can now visualise all the dependencies at once in three dimensions.

\begin{figure}
			\centering
					\includegraphics[width=\textwidth]{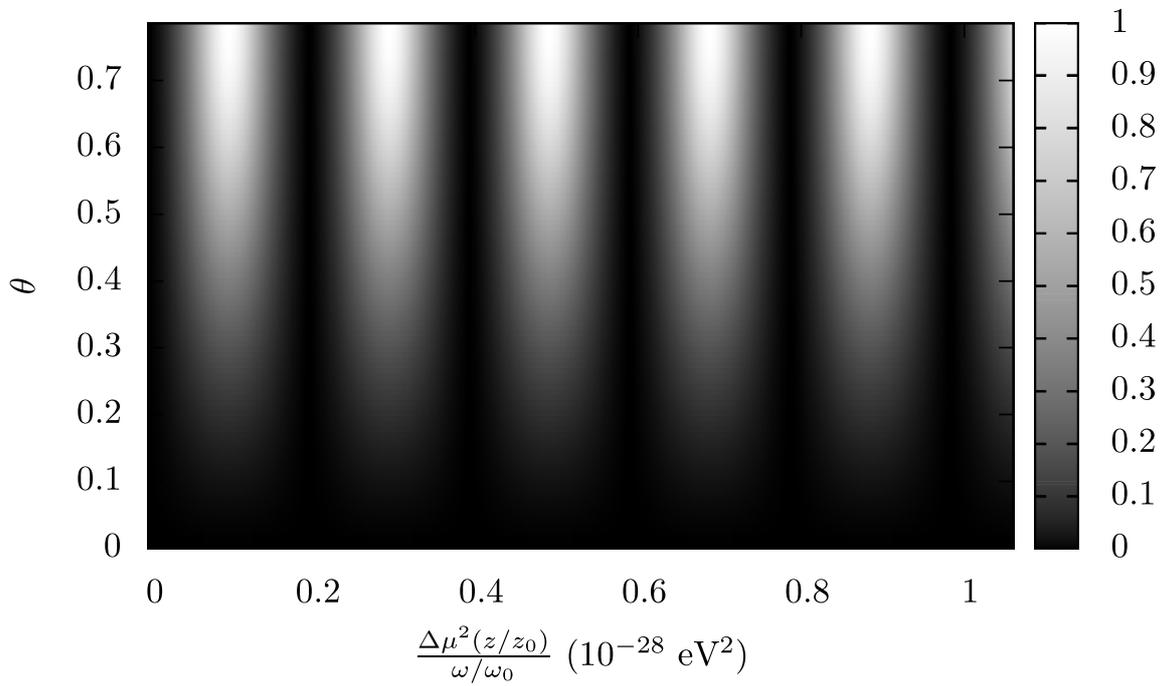}
			\caption{
Linear polarisation degree (shown in the right-hand box) generated through pseudo\-scalar-photon mixing in a transverse magnetic field region in the case of initially unpolarised light. For convenience, we have introduced $\omega_0 = 2.5$~eV (\textit{i.e.} $\lambda_0 = \frac{2\pi}{\omega_0} =500$~nm) and $z_0=1.6\times10^{30}$~eV$^{-1}$ ($\simeq10$~Mpc).
			}
			\label{fig:rtheta_plin}
		\end{figure}

		Going back to dichroism, the linear polarisation degree \eqref{eq:poldegree} in terms of $\frac{\Delta\mu^2}{\omega}z$ and $\theta$, in the case of initially unpolarised light, for instance, takes the form:
	\be
		p_{\mathrm{lin}}(z)=\frac{\frac{1}{2}\sin^2 2\theta\sin^2\left(\frac{1}{4}\frac{\Delta\mu^2}{\omega}z\right)}
				   {1 - \frac{1}{2}\sin^2 2\theta\sin^2\left(\frac{1}{4}\frac{\Delta\mu^2}{\omega}z\right)}.
		\label{eq:plinrtheta}
	\ee
		In Fig.~\ref{fig:japfan_unpol}, what exactly determined the observed pattern was not obvious; however, parametrised with $\theta$ and $\frac{\Delta\mu^2}{\omega}z$, $p_{\mathrm{lin}}$ behaves as in Fig.~\ref{fig:rtheta_plin}\hspace{1pt}---\hspace{1pt}strictly speaking $\theta\in[-\frac{\pi}{4},\frac{\pi}{4}]$ but $p_{\mathrm{lin}}$ is an even function of it. It is now clear that the maximum linear polarisation is entirely determined by $\theta$, while the details of the oscillatory behaviour with $z$ are independently controlled by $\frac{\Delta\mu^2}{\omega}z$.

		This can be physically understood as $\theta$ determines how much the particles mix, while $\frac{\Delta\mu^2}{\omega}z$ has to do with the difference of mass eigenstates and is thus related to the wavelength of the oscillation.

		As long as the Lagrangian~\eqref{eq:lagrangian} makes sense, the oscillatory pattern in $\frac{\Delta\mu^2}{\omega}z$ repeats itself unchanged to infinity: all the physics can thus be studied in a small interval.

\begin{figure}
			\centering
					\includegraphics[width=\textwidth]{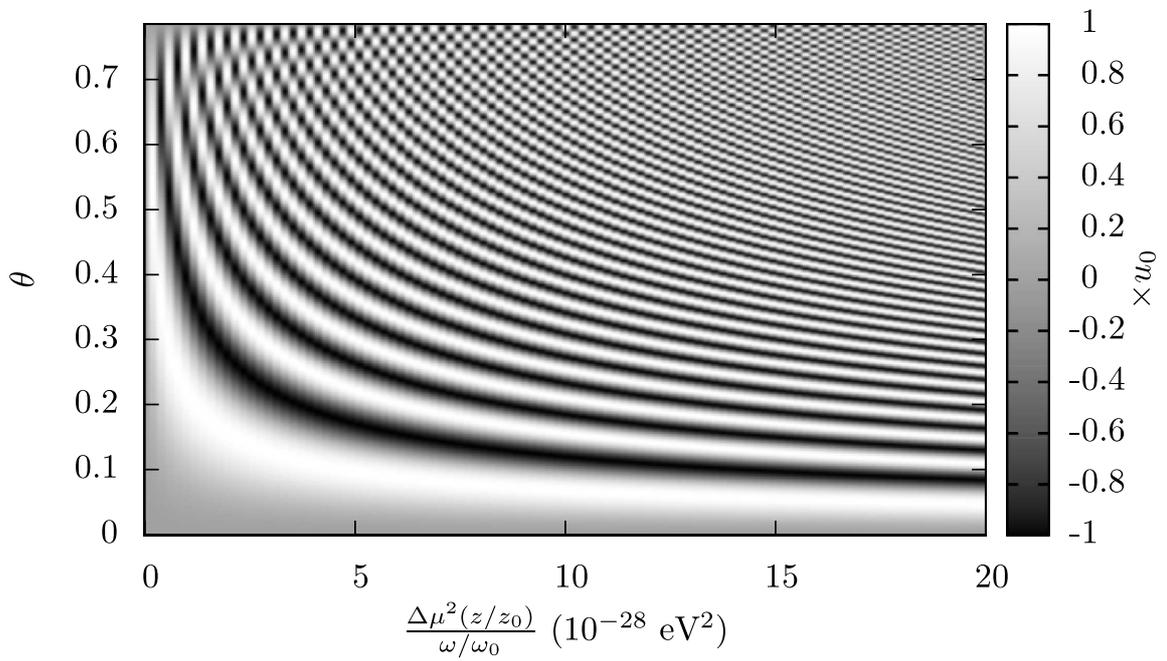}
			\caption{Circular polarisation $v=V/I$ generated through pseudo\-scalar-photon mixing in a transverse magnetic field region in the case of light beam with $v_0 = 0$ and $q_0 = 0$. The values of $\omega_0$ and $z_0$ are the same as the ones introduced in Fig.~\ref{fig:rtheta_plin}.
			}
			\label{fig:rtheta_pcirc}
		\end{figure}

		All the information on circular polarisation, which is an odd function of $\theta$, is obtained in~Fig.~\ref{fig:rtheta_pcirc}. It shows more complex dependencies with the two parameters than $p_{\mathrm{lin}}$ and is clearly a key property of the mixing, being (in absolute value) typically as large as $|u_0|$.

		This is in sharp contrast with dichroism. Indeed, in fixed external conditions, the amplitude of $Q$, for a given value of $Q_0$, is only determined by the efficiency of the mixing, \textit{i.e.} by the value of the mixing angle $\theta$. The \draftun{maximum values} of $U$ and $V$ however, for any given non-zero value of $\theta$, are only determined by $U_0$ and $V_0$; in the general case, it is equal to $\sqrt{{U_0}^2+ {V_0}^2}$.

		This difference is understood \draftun{since} the amount of polarisation from dichroism reaches a saturation (related to the conversion probability, which is the same from photons to axions and conversely; see, \textit{e.g.} Ref.~\cite{Raffelt:1987im}), while, for birefringence, it accumulates as long as there are phase shifts.

		The situations \draftun{in which} it has not yet accumulated enough are typically cases where $(2g\mathcal{B}\omega)$ and $(m^2 - {\omega_{\mathrm{p}}}^2)$ are very different, and include very special cases such as the ones that we have discussed earlier, namely:
		\begin{itemize}
			\item[-] vanishing mixing $\theta\rightarrow0$;
			\item[-] maximum mixing $\theta\rightarrow\frac{\pi}{4}$ (see Eqs.~\eqref{eq:Stokes_alternative});
			\item[-] $\frac{\Delta\mu^2z}{\omega}\rightarrow0$, which is more related to the propagation.
		\end{itemize}

	\chapter{Alignments of quasar polarisations}\label{chap:alignments}

		Best described in superlatives given how incredibly luminous they are, quasars are among the furthest astronomical sources that can be observed, with spectra often characterised by large cosmological redshifts.
		These objects are found in the centre of otherwise rather normal galaxies, and form the high-luminosity subgroup of active galactic nuclei (AGN), supplanting by up to $10^4$ times the emission of typical galaxies; see, \textit{e.g.} Ref.~\cite{Padmanabhan:2002bk}.\footnote{Let us again stress that, throughout this work, ``quasar'' and ``high-luminosity AGN'' can be used interchangeably.}
		If their luminosity is disproportionately large, quite the contrary can be said about the size of their central region: it has dimensions not so different from those of the solar system, and is not resolved angularly; see, \textit{e.g.} Ref.~\cite{Ho:2002en}. Indirect indications have therefore been used to learn about their internal structure.
		To date, quasars and their properties are not fully understood and their study is still, fifty years after their discovery~\cite{Schmidt:1963qz}, a vibrant subject of research~\cite{FiftyYearsQuasars:2012bk}.

	According to the current model, their tremendous power can be explained by the presence of a supermassive black hole, surrounded by an accretion disc; this core region being enclosed by a torus of thick dust\hspace{1pt}---\hspace{1pt}again see, \textit{e.g.} Ref.~\cite{Ho:2002en}.
	Interestingly, in direct relation with this non-spherical morphology, light from quasars is intrinsically partially polarised, involving processes such as scattering (due to dust, or to free electrons) or synchrotron emission within these sources; see, \textit{e.g.} Ref.~\cite{Borguet:2009phd}.

	For the needs of this work, it is sufficient to treat these objects as distant sources of partially polarised light; for more information about quasars and their classification, or for discussions on their internal structure, the interested reader is therefore referred to Refs.~\cite{Krolik:1999agn,*Osterbrock:1989bk,Schneider:2006bk}.
\bigskip

	It is now time to present the observations that have been the prime motivation for this thesis. In the following, we are going to summarise and emphasise some properties of the reported effect that are particularly relevant to our study.
The reader will find much more information in the original articles: for the observed effect itself and the associated sample of good-quality polarisation measurements, see Refs.~\cite{Hutsemekers:1998, Hutsemekers:2001, Hutsemekers:2005}; on the other hand, details about the acquisition and the treatment of (part of the) data\hspace{1pt}---\hspace{1pt}something that we will not discuss here\hspace{1pt}---\hspace{1pt}can be found, for instance, in Ref.~\cite{Sluse:2005}.

\section{An unexpected observation}

	Following a data analysis of linear polarisation measurements of quasars in visible light, Hutsemékers~\cite{Hutsemekers:1998} has noticed that the distribution of polarisation position angles is not random in some huge regions of the sky.
	Instead, in \draftun{each such region}, the polarisation vectors are actually preferentially oriented along some direction, and seem to indicate that the properties of objects separated by gigaparsecs are correlated.\footnote{Polarisation ``vectors'' is a misnomer since these are only defined modulo a $180\degr$ rotation, as is the polarisation angle; see Appendix~\ref{app:stokes}. As mentioned in the introduction, they simply indicate the direction of preferred polarisation on the sky\hspace{1pt}---\hspace{1pt}typically with respect to the North--South axis in equatorial coordinates. For the present data, the angles are counted clockwise if the East is to the right~\cite{Hutsemekers:1998}.}

	The first hint for the existence of such correlations came in two steps:
	\begin{enumerate}
		\item when he analysed what were new data at the time, Hutsemékers observed in particular that he could define a volume (delimited in right ascension, declination and redshift) in which all significantly polarised quasars (7 objects) had their polarisation position angles within a given interval of $80\degr$, instead of being uniformly distributed over $180\degr$;
		\item after the 3D volume was well-defined \textit{a priori}, the associated observed interval could be used to make a prediction; he therefore gathered, using major catalogs in the literature, all the optical polarimetric measurements of the other polarised quasars located in the same region (5 objects): surprisingly, all of them also had their polarisation angles within the same $80\degr$ interval, $[146\degr,46\degr]$.
	\end{enumerate}
		Already at this early stage, it was clear that an explanation involving an instrumental bias was very unlikely as these objects were observed by different groups, using different instruments; redundant measurements, on the other hand, being in excellent agreement. 
		As for the chances of this happening by sheer coincidence, they are quite low already with a small number of objects. 
		Indeed, using a simple binomial test, one obtains that the probability of having all these 5 angles exactly within this particular interval is only 1.73\% if one supposes that the polarisation position angles follow a random distribution, which is the null hypothesis~\cite{Hutsemekers:1998}.

\subsection{Defining a sample of good-quality linear polarisations}

		As this preliminary observation led to the prediction of the preferred orientation for the linear polarisation of visible light from quasars in this region, it then motivated an in-depth study of this ``alignment (tendency)'' effect, using a much larger sample to which various statistical methods were applied~\cite{Hutsemekers:1998}.

		As a starting point, the idea has been to compile all optical measurements of linear polarisation of objects classified as quasars for which the polarisation position angle and the redshift are known.
		In particular, there has been no discrimination on the filter used (or the absence thereof); this therefore implies that measurements of different objects have been taken at slightly different optical frequencies.\footnote{A parallel reasoning can be made about the redshift range of quasars within each region. Even with the same filter, measurements would correspond to different frequencies in their rest frames.} These potential differences are not crucial however as, while the polarisation degree might show a frequency dependence, the polarisation angle varies only slightly from near-infrared to near/medium UV (in the quasar rest frame), as discussed in Ref.~\cite{Hutsemekers:1998}; see also Ref.~\cite{Ogle:1999qz}.
		Alone, a frequency dependence of the polarisation angles would be unable to produce coherent orientations between different objects anyway; on the contrary, it would rather wash out the significance of such an effect.

		Hence, stringent criteria have been chosen to achieve a good balance between the sample size and the quality of the measurements, while reducing the galactic contamination as much as possible (\textit{i.e.} dichroism due to elongated dust grains aligned in the galactic magnetic field). In essence, it was then decided to keep only quasar polarisation measurements such that:
		\begin{itemize}
			\item[-]	$p_{\mathrm{lin}}\geq0.6\%$: this is the value of the linear polarisation degree above which the polarisation is mainly of intrinsic origin, and not due to the influence of the galaxy;
			\item[-]	the uncertainty on the polarisation position angle $\sigma_{\varphi}\leq14\degr$, to ensure the data are sufficiently precise;
			\item[-]	the galactic latitude $|b_{\mathrm{gal}}|\geq30\degr$, namely selecting objects away from the galactic plane, to avoid as much contamination as possible;
		\end{itemize}
		for the full discussion, see Ref.~\cite{Hutsemekers:1998} and references therein.

		Applied to all the data available at the time, obtained by different groups~\cite{Stockman:1984qz,Moore:1984qz,Berriman:1990qz,Impey:1990qz,
Impey:1991qz,Wills:1992qz,Hutsemekers:1998pp}, these criteria then led to a sample counting 170 good-quality measurements of optical linear polarisation from quasars, which gave a first indication of the existence of alignments in several large-scale regions in the northern and in the southern galactic hemispheres~\cite{Hutsemekers:1998}.

		This remains essentially the main benefit of this original sample today as, once the regions of interest delimited and their preferred range of polarisation position angles identified \textit{a priori}, it has allowed for testable predictions in subsequent studies.

		The sample has been extended twice since then~\cite{Hutsemekers:2001,Hutsemekers:2005}, and the effect has actually become more and more statistically significant. These new quasar data were taken from general surveys~\cite{Brotherton:1998qz,Visvanathan:1998qz,Schmidt:1999p,Smith:2002qz}, as well as from two campaigns dedicated to objects located in specific regions where alignments had been detected~\cite{Lamy:2000qz,Sluse:2005}. Note that, in the latter case, the objects were not taken randomly in these regions: the brightest objects were indeed preferred (the selection was based on the apparent magnitude), with an emphasis on classes of quasars more likely to be significantly polarised~\cite{Hutsemekers:2001}.

		Still subject to the quality criteria presented in Ref.~\cite{Hutsemekers:1998}, the latest all-sky sample finally counts 355 good-quality linear polarisation measurements in visible light: half of them taken from Refs.~\cite{Hutsemekers:1998pp,Lamy:2000qz,Sluse:2005} and the other half from the literature~\cite{Stockman:1984qz,Berriman:1990qz,Moore:1984qz,
Brotherton:1998qz,Schmidt:1999p,Smith:2002qz,Impey:1990qz,Impey:1991qz,Wills:1992qz,Visvanathan:1998qz}, redundant measurements always being in excellent agreement~\cite{Hutsemekers:1998,Hutsemekers:2001,Hutsemekers:2005}.\footnote{In the case of redundant measurements, it is always the one with the smallest incertitude on $p_{\mathrm{lin}}$ that was kept, as $\sigma_{\varphi}$ depends on $p_{\mathrm{lin}}$ and might be biased~\cite{Hutsemekers:1998}.}

\subsection{Some characteristics of the effect}\label{sec:descriptionalignment}

	\subsubsection{A few general properties}

		\paragraph{Various types of quasars}%
		Interestingly, in regions where preferred orientations are found, the same alignment tendency is followed by quasars of different classes, and not only by objects of a specific type~\cite{Hutsemekers:2001,Hutsemekers:2005}.
			
		\paragraph{Polarisation mostly intrinsic}%
		Moreover, known polarimetric differences between quasars of different spectroscopic types are preserved, indicating that the polarisation observed is mainly of intrinsic origin~\cite{Hutsemekers:1998,Hutsemekers:2001,Hutsemekers:2005,Hutsemekers:1998pp,Schmidt:1999p,Ogle:1999qz,Lamy:2004yz}. Note that for most quasars in the latest sample, the linear polarisation is at the 1\%-level. 

		\paragraph{Non-local effect}%
		An important question is whether there is any (potentially new) physics behind these observations, or whether it is rather due to some kind of unidentified but significant bias.
		To answer this, a crucial property of the effect is that local mechanisms simply seem unable to explain the existence of such preferred orientations.

\begin{figure}
			\centering
			\includegraphics[width=.68\textwidth,trim = 0mm 0mm 8cm 0mm, clip]{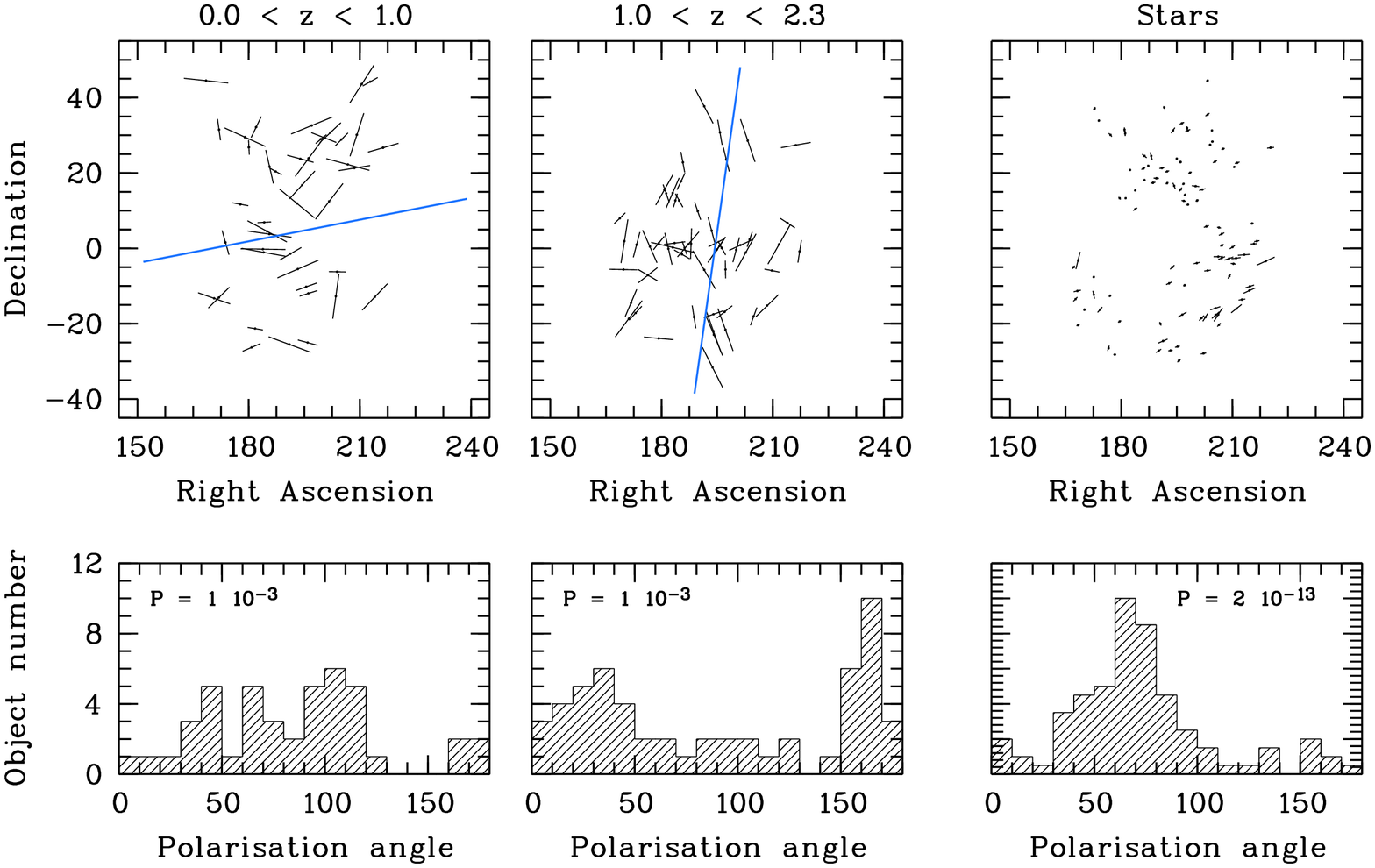}
			\caption{Same as in Fig.~\ref{fig:align_intro}, now each time with the associated distribution of polarisation position angles and its probability to be uniform~\cite{Payez:2010xb} according to a local test designed for cyclic quantities. The direction shown with a long line in each region is the preferred angle provided by this specific test: $\bar{\varphi}=79\degr$ at low-$z$ and $\bar{\varphi}=8\degr$ at high-$z$; see also Ref.~\cite{Hutsemekers:2005}.
			}
			\label{fig:align_nonlocal}
		\end{figure}

		The odds of having an instrumental bias or an extinction due to the influence of dust in the galaxy have been seriously diminished by using data from different surveys and by only considering significantly polarised objects at high galactic latitudes.
		However, the main reason is that the preferred orientations depend on the distance~\cite{Hutsemekers:1998}: the alignment tendency can indeed be very different at low and high redshifts for objects along similar lines of sight\hspace{1pt}---\hspace{1pt}namely similar right-ascensions and declinations.
		Such a situation is presented in Fig.~\ref{fig:align_nonlocal} using the latest sample. Already by mere observation, one can immediately notice that the polarisation vectors do not seem randomly oriented in these regions and show different preferred orientations.\footnote{The presence of alignments remains significant if one models and subtracts the interstellar polarisation at those latitudes from the quasar polarisations in both regions. Moreover, if one tries to suppress the effect in one region by artificially making the associated distribution almost uniform, one strongly enhances the alignment tendency in the other region. This is shown explicitly in Ref.~\cite{Payez:2010xb}; for instance, in the case where the effect is removed at low-$z$, the probability for the distribution of angles to be due to coincidence drops below $10^{-4}$ at high-$z$, where the effect is enhanced.
		}

		There are also cases in which alignments of polarisation are only found within a certain redshift range, and not for all the objects located in that given direction; see, \textit{e.g.} Ref.~\cite{Hutsemekers:2005}. This would not happen if these observations were simply due to something local, as it would then affect all measurements independently of their distance from us.
		
	\paragraph{Orientations over extremely large-scales}%
	The angular extension of these regions on the sky is such that the transverse distance between objects correlated at high redshifts is of the order of gigaparsecs, which seems to indicate a cosmological effect~\cite{Hutsemekers:1998}.

	\subsubsection{Highlighting non-random orientations of polarisation}

		Many sophisticated statistical tests have been applied by different authors to the growing sample, in particular in Refs.~\cite{Hutsemekers:1998,Hutsemekers:2001,Hutsemekers:2005,Jain:2003sg}. These include both non-parametric and parametric tests, that have been applied either in the case of specific regions (local tests), or to the sample as a whole (global tests).
		Without entering into details that are beyond the scope of this thesis, the non-uniformity of the distribution of polarisation position angles in these regions and the existence of correlations between the polarisation angles and the (3D) positions of the objects in the all-sky sample have been confirmed~\cite{Hutsemekers:2001}. Furthermore, they have become even more significant with the latest data~\cite{Hutsemekers:2005}, now typically with a probability higher than 99.9\%, as found with and without coordinate-invariant statistics~\cite{Jain:2003sg,Hutsemekers:2005}.\footnote{As expected, polarisation angles depend on the polar axis chosen to define them, and some particular choices may be unable to detect the presence of alignments, despite highly organised dependences of polarisation position angles with position; see discussion in Ref.~\cite{Hutsemekers:1998}. Note that the significance of the statistical tests is not extreme but intermediate in equatorial coordinates, and that coordinate-invariant statistics give similar results~\cite{Jain:2003sg,Hutsemekers:2005}.}

		Simpler methods have been applied as well, and also provide evidence for the existence of such an effect. As anticipated earlier, one of the simplest ones considered in Refs.~\cite{Hutsemekers:1998,Hutsemekers:2001,Hutsemekers:2005} has been to predict in advance the polarisation angles of new quasar measurements in some regions and to test (and reject) the uniformity of the distributions using a binomial test.

		\paragraph{Binomial test}%
		In particular, the high-$z$ region shown in Fig~\ref{fig:align_nonlocal} actually includes all the objects used to highlight the existence of preferred orientations.
		It was in fact realised in the original paper that these quasars are part of a more extended region in redshift, in which other objects share the same preferred range of polarisation position angles, leading to this better defined volume using 16 quasars at the time, which was then labelled region A1~\cite{Hutsemekers:1998}.

		In the latest sample~\cite{Hutsemekers:2005}, 42 quasars out of 56 from this region actually have their polarisation position angles in that given $[146\degr,46\degr]$ range.\footnote{In the corresponding low-$z$ region shown in Fig.~\ref{fig:align_nonlocal}, the authors have noted that 35 objects out of 43 have their polarisation position angles in a 90$\degr$ interval, $[30\degr,120\degr]$.}
		To apply the binomial test specifically, the authors removed the objects used to define the predicted range in this region, leaving 27 objects oriented accordingly out of 40, and found a probability of uniform distribution equal to $2.8\times10^{-3}$~\cite{Hutsemekers:2005}.

		Note that it was quickly suspected that the effect was even stronger in the center of this region~\cite{Hutsemekers:1998,Hutsemekers:2001}, which led to the definition of region A1+ prior to Ref.~\cite{Hutsemekers:2005}, with the same range of polarisation angles. With the new sample, they obtained that, out of 14 new measurements, only 1 does not fall inside the predicted $80\degr$ range, giving a probability of uniform distribution equal to $2.2\times10^{-4}$ with a similar binomial test. In total, 17 objects out of 18 have their polarisation angles within $[146\degr,46\degr]$ in this sub-region.

		\paragraph{Departure from isotropy in a $(q,u)$ space}%
		A simple method that we have proposed~\cite{Payez:2011sh} is to check how the alignment tendency in a given region looks like in a linear polarisation space $(q,u)$, and to highlight it using averages. Doing so, we drop the information about the exact position of each object, like for the binomial test, and we use the polarisation position angles and the linear polarisation degree.
		As the polarisation angle is related to the Stokes parameters $q$ and $u$ via the relation:
		\be
			\varphi= \frac{1}{2}\atan\left(\frac{u}{q}\right),
		\ee
		for a fixed value of $p_{\mathrm{lin}}$ different values of $q$ and $u$ correspond to different orientations. In particular, a random distribution of polarisation angles corresponds to an isotropic distribution in this space.
		Note also that, as $p_{\mathrm{lin}} = \sqrt{q^2+u^2}$, the distance between the origin and a given point directly gives the degree of polarisation of the associated light beam.

		\begin{figure}
				\centering
				\includegraphics[width=0.5305\textwidth]{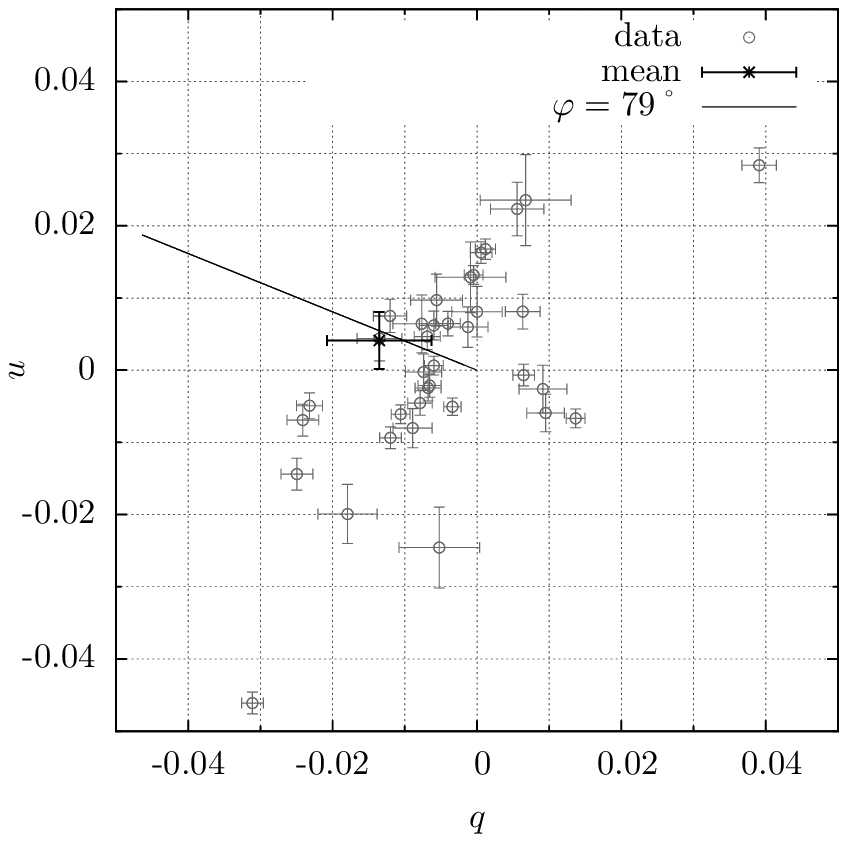}
				\includegraphics[width=0.46\textwidth]{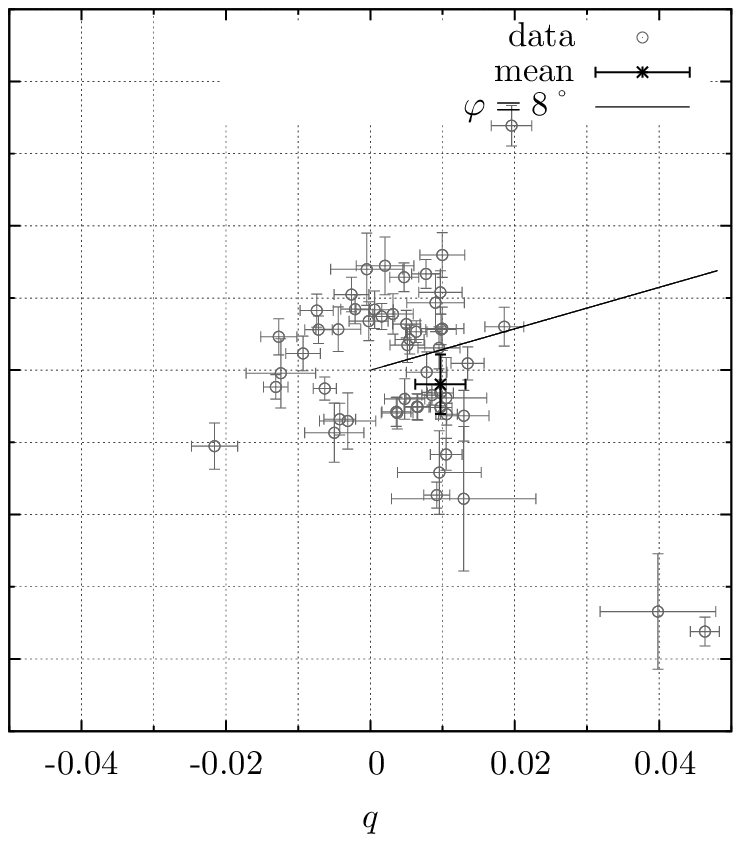}
				\caption{Experimental data: same polarisation measurements as those in Fig.~\ref{fig:align_nonlocal}. \emph{Left}: linear polarisation for the low-redshift quasars, with $0\leq z<1$ (region A0). \emph{Right}: same, but for high-redshift ones, with $1\leq z \leq 2.3$ (region A1). Some objects with higher linear polarisation degrees are not shown, but are taken into account for the mean. In total, the sample contains 43 quasars from region A0, and 56 from region A1.
				}
				\label{fig:q_u_data}
		\end{figure}

		For instance, we can present the linear polarisation of quasars located in the regions pictured in Fig.~\ref{fig:align_nonlocal}: region A1 and its (unnamed) low-$z$ counterpart, that we shall call region A0. This is illustrated in Fig.~\ref{fig:q_u_data}. As expected, we obtain that coherent orientations are translated into departures from isotropy in such graphs.
		We also clearly see that the preferred direction for the asymmetry is not the same for low and high redshifts, while these objects are along the same line of sight.
		Note that the presence of a hole in the center for each region is simply due to the criterion that quasars must satisfy $p_{\mathrm{lin}}\geq0.6\%$ to be part of the sample~\cite{Hutsemekers:1998}.

		To be more quantitative, we can calculate the mean values of $q$ and of $u$, for low- and high-redshift data, taking into account the experimental uncertainties.\footnote{For this, one takes $\sigma_q = \sigma_u = \sigma_p$, see Ref.~\cite{Serkowski:1958sa}.}
		We determine the mean values and the errors on the mean for $q$ and $u$ and plot them in Fig.~\ref{fig:q_u_data}; we obtain $(-0.0135\pm0.0072, 0.0041\pm0.0039)$ for the low-redshift region, and $(0.0097\pm0.0035, -0.0019\pm0.0041)$ for the high-redshift one. In the observational paper~\cite{Hutsemekers:2005} and in Ref.~\cite{Payez:2010xb}, another analysis was done, leading to the preferred angles one finds when considering only the angular information (those shown in Fig.~\ref{fig:align_nonlocal}), which are indicated with straight lines.

\section{Towards a spinless-particle interpretation}\label{sec:alpscenario}

		Various ideas have been proposed in the observational papers~\cite{Hutsemekers:1998,Hutsemekers:2001,Hutsemekers:2005} to address the alignment effect.
		\draftun{Discussed as an interesting possibility already in the first of these papers, the mixing of light with spinless particles has soon become the favourite scenario, and has been claimed as naturally able to account qualitatively for most of the properties of the observations of linear polarisation~\cite{Hutsemekers:2005}.}
		Actually, as for any mechanism based on a modification of polarisation during the propagation of light, the challenge was to produce different alignments for objects located along similar lines of sight.

		Finally, this model has become even more appealing as the absence of similar alignments in radio waves reported in Refs.~\cite{Joshi:2007yf,Vallee:2002} is automatic in this scenario, given the frequency-dependence of the mixing.
\bigskip

		Even if the magnetic field strengths in different subregions of the quasars themselves are big, we can be sure that this alone is not sufficient to create large-scale alignments. Indeed, the direction imprinted by the mixing would then only be associated to each source independently. Rather, in the simplest formulation of this spinless-particle scenario, the magnetic field needs to be crossed by photons from different sources to reproduce such an alignment: one should thus consider magnetic fields encountered on the way; see, \textit{e.g.} Refs.~\cite{Hutsemekers:1998,Hutsemekers:2001,Hutsemekers:2005,Jain:2002vx}. The redshift-dependence of the effect might indicate that the mixing happens in different places along the line of sight and would thus require fairly extended \draftun{extragalactic} magnetic fields (enough to encompass angularly most of the sources of a given region).

\subsection{It is possible to produce an alignment in a toy model}\label{sec:toymodel}

		The amount of additional linear polarisation required to produce \draftun{in one region} an alignment similar to those observed from a random distribution of polarisations can in fact already be achieved in a toy model (where one assumes that the magnetic field is homogeneous), taking into account \draftun{current constraints on spinless particles}. Although this is but a mere step towards a realistic scenario, it is a useful one, that can be thought of as representing the average magnetic field inside a supercluster of galaxies. \draftun{In the case of the local supercluster (which has the advantage of being suitably extended angularly), reported magnetic field measurements have given field strengths of the order of $0.3~\mu$G with a coherence length of 10~Mpc~\cite{Vallee:2002,Vallee:2011}.}
\draftun{Such a field has already been considered in the literature to reproduce the alignments in the spinless-particle scenario: this was first done in Ref.~\cite{Jain:2002vx} with the field strength reported at the time~\cite{Vallee:1990}, and has been followed by Ref.~\cite{Das:2004qka} with the current one.}

		\draftun{Here, we simply apply the results discussed in Sec.~\ref{sec:mixing} to the alignment effect in visible light.} More complex magnetic field morphologies will be discussed in forthcoming chapters.

	\subsubsection{Alignments due to dichroism}

	To reproduce coherent alignments of the polarisations of light from quasars in this spinless-particle scenario, dichroism will be the main mechanism leading to the generation of a systematic amount of linear polarisation. As estimated in Ref.~\cite{Hutsemekers:2010fw}, this has to be at least 0.5\% and certainly not more than 2\% to explain \draftun{the observations}.	

\begin{figure}
			\centering
				\includegraphics[width=\textwidth]{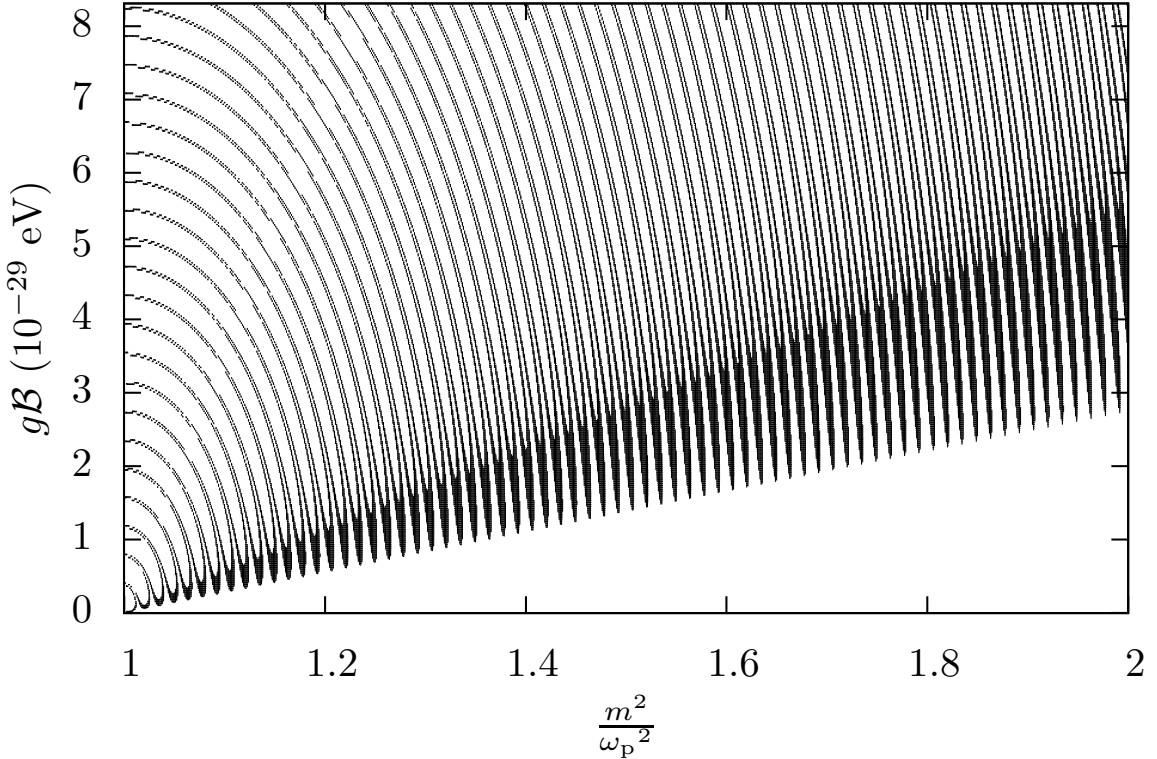}

			\caption{Same as Fig.~\ref{fig:japfan_unpol}, but only showing parameters such that the linear polarisation generated through pseudo\-scalar-photon mixing lies between 0.5\% and 2\%.
			}
			\label{fig:japfan_unpol_okalign}
		\end{figure}

	In order to present only this additional linear polarisation, and avoid the arbitrariness of the initial one, we first consider unpolarised light beams of optical frequency.\footnote{Note that even for partially linearly polarised light, which for quasars is at the 1\%-level as we discussed already, the unpolarised contribution will remain the dominant component.} We can isolate the part of the full parameter space (Fig.~\ref{fig:japfan_unpol}) able to reproduce the additional linear polarisation required, for a given travelled distance $z$ and for a fixed value of $\omega_{\mathrm{p}}$. We give the result in Fig.~\ref{fig:japfan_unpol_okalign}. Pseudo\-scalar masses smaller than $\omega_{\mathrm{p}}$ are not shown because the linear polarisation degree is an even function of ($m^2 - {\omega_{\mathrm{p}}}^2$), as discussed in Sec.~\ref{sec:mixing}.

		We see in Fig~\ref{fig:japfan_unpol_okalign} that the mixing effect can in principle be observable and produce enough linear polarisation, even in faint\hspace{1pt}---\hspace{1pt}but extended\hspace{1pt}---\hspace{1pt}magnetic fields ($\mathcal{B}=0.1~\mu$G and $g=10^{-11}$~GeV$^{-1}$ correspond to $g\mathcal{B} = 1.95\times10^{-29}$~eV), provided that $m$ and $\omega_{\mathrm{p}}$ are not too different.
		More precisely, the maximum amount of linear polarisation due to dichroism is only determined by the mixing angle $\theta$, and this implies that $\left(m^2 - {\omega_{\mathrm{p}}}^2\right)$ has to be roughly of the same order of magnitude as $g\mathcal{B}\omega$.
		The electron density, which determines the plasma frequency, is very small in superclusters, and even smaller in cosmic voids where only upper bounds exist~\cite{Deffayet:2001pc,Csaki:2001jk,Mortsell:2002dd,DeAngelis:2007yu,Mirizzi:2006zy}; the value typically considered in superclusters is $n_{\mathrm{e}} = 10^{-6}$~cm$^{-3}$~\cite{Vallee:1990,Vallee:2011,Das:2004qka,Burrage:2008ii}, corresponding to $\omega_{\mathrm{p}}=3.7\times10^{-14}$~eV.

		We therefore understand that this scenario, in which the mixing is supposed to take place in faint magnetic fields encountered on the way between the quasars and the observer, requires the existence of axion-like particles which are nearly massless. The effects of the mixing at optical frequencies would indeed be completely negligible otherwise.

\begin{figure}
			\centering
					\includegraphics[width=\textwidth]{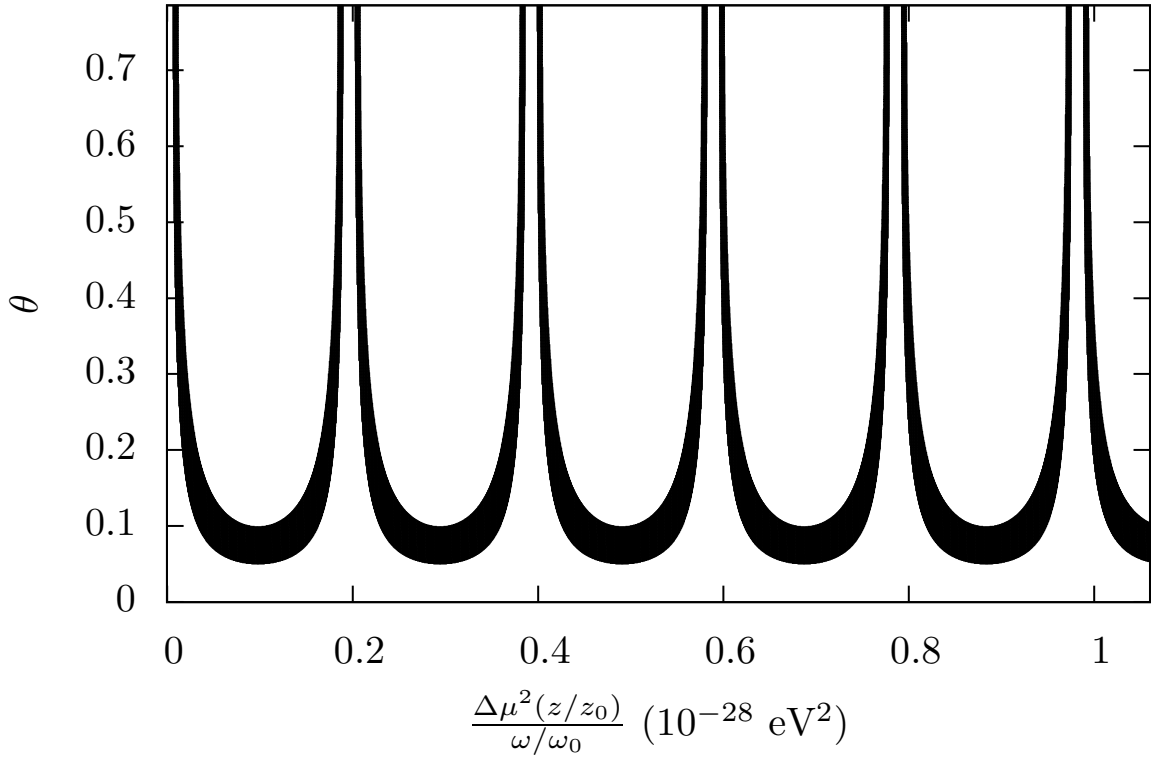}
			\caption{Same as Fig.~\ref{fig:rtheta_plin} but only displaying pairs of $\big(\frac{\Delta\mu^2}{\omega} z,\theta\big)$ such that the additional linear polarisation degree due to the mixing $p_{\mathrm{lin}}$ is in the interval $[0.005,0.02]$. For fixed $z=z_0$ and $\omega=\omega_0$, this is equivalent to a reparametrisation of Fig.~\ref{fig:japfan_unpol_okalign}.
			}
			\label{fig:rthetaplinok}
		\end{figure}

	Now, to estimate how efficient the mixing has to be, we can use the expression of the linear polarisation degree in terms of $\frac{\Delta\mu^2}{\omega} z$ and $\theta$ for unpolarised light given in Eq.~\eqref{eq:plinrtheta}, and only display parameters able to generate the correct amount of additional polarisation in this toy model; this leads to Fig.~\ref{fig:rthetaplinok}. We can also consider the average of the additional polarisation over one period in $z$, and impose that it lies between 0.005 and 0.02, giving an allowed range of values for the mixing angle:
	\be
		0.07\leq |\theta| \leq0.14.
		\label{eq:thetarange}
	\ee

	\subsubsection{A clear prediction from birefringence}

	For astronomical sources, processes leading to the production of circularly polarised light are rare. Generally, most quasars only emit partially linearly polarised light; their polarisation degree is typically around 1\%~\cite{Stockman:1984qz,Berriman:1990qz,Sluse:2005}. In the following, we suppose that the initial distribution of polarisation angles is random, so that the radiation can be described by random initial values for the Stokes parameters $q_0$ and $u_0$, with $p_{\mathrm{lin},0}=\sqrt{u_0^2+q_0^2}=0.01$, and we assume no initial circular polarisation: $v_0=0$.

	Now, while the mixing can generate enough linear polarisation to reproduce the effect via axion-like particles thanks to dichroism, we can expect that birefringence will lead to an observable amount of circular polarisation.
	This well-known property of the mixing~\cite{Maiani:1986md,Raffelt:1987im} has been extensively discussed in Sec.~\ref{sec:mixing} and is readily seen in Eqs.~\eqref{eq:Stokes_alternative}: a linearly polarised light beam (with non-zero $u_0$) will develop a circular polarisation as it propagates.\footnote{
This is also true for low-mass axion-like particles, even if the induced phase shift $\Phi$ drops quickly as the mass decreases: in the weak-mixing limit \cite{Cameron:1993mr},
	\begin{equation}
		\Phi = \theta^2 \left[ \frac{m^2 z}{2 \omega} - \sin\left(\frac{m^2 z}{2 \omega}\right) \right].
	\end{equation}
	In this astrophysical context, $\Phi$ is not small when the considered magnetic field regions are huge.
} From a technical point of view, an initial angle of $\frac{\pi}{4}$ with the direction of the external magnetic field leads to the maximal amount of generated circular polarisation; it corresponds to $u_0=p_{\mathrm{lin},0}$.

\begin{figure}
			\centering
					\includegraphics[width=\textwidth]{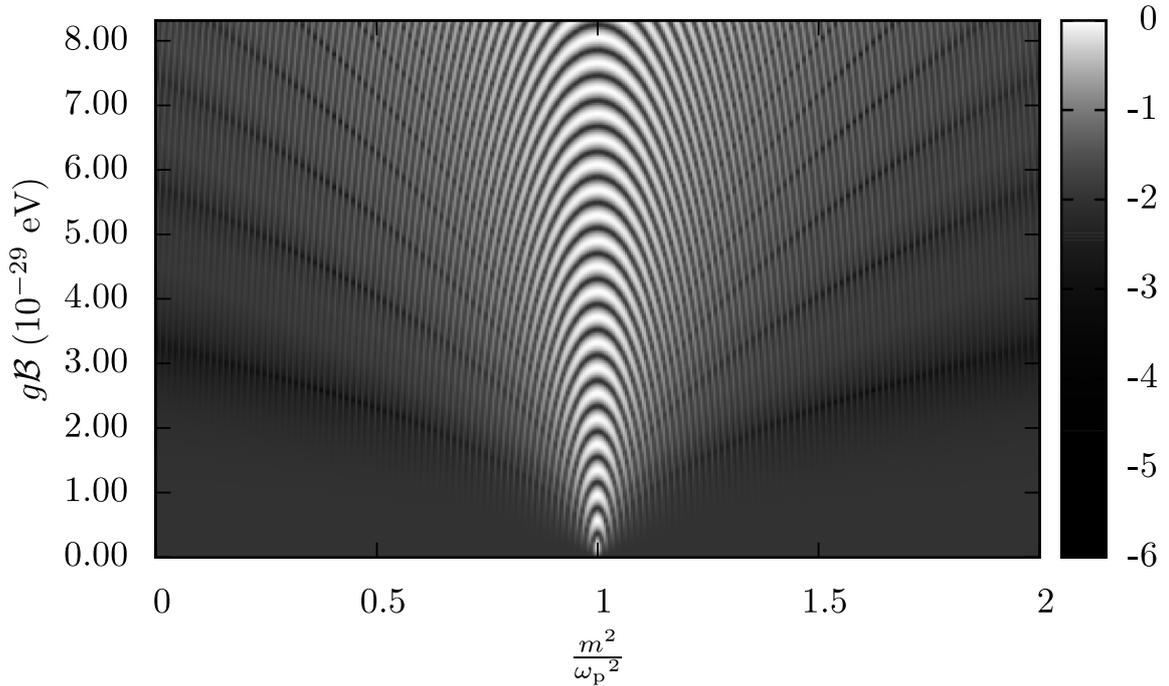}
			\caption{Linear polarisation degree after pseudoscalar-photon mixing in a transverse magnetic field region, in the case of initially polarised light beams of frequency $\omega=2.5$~eV with $u_0=0.01, q_0=0$. The plasma frequency and the size of the magnetic region are the same as in Fig.~\ref{fig:japfan_circ}. Note that the right-hand box gives the base-10 logarithm of the linear polarisation.
			}
			\label{fig:IPplin_pl}
		\end{figure}

	As we did for linear polarisation in the initially unpolarised case, we can calculate the circular polarisation $v$ predicted by pseudo\-scalar-photon mixing when light is described by plane waves with $\omega=2.5$~eV, for an initial linear polarisation of $1\%$; the case $u_0=p_{\mathrm{lin},0}=1\%$ can be obtained directly from Fig.~\ref{fig:japfan_circ}.\footnote{Note that the circular polarisation is an odd function of ($m^2 - {\omega_{\mathrm{p}}}^2$) as long as $v_0=0$.} The corresponding linear polarisation is shown in Fig.~\ref{fig:IPplin_pl}, and is of course very similar to that of the unpolarised case, which is the dominant contribution; in that figure, we observe minima that correspond to the extrema of $v$ in Fig.~\ref{fig:japfan_circ}.

	Figure~\ref{fig:japfan_circ} indicates that a large region of the parameter space leads to an observable circular polarisation. We can actually anticipate that this is a general result.
	Indeed, even for light coming from a single quasar, a number of regions with different uncorrelated magnetic fields will be encountered on the way towards us. It is thus impossible to avoid $u_0\neq0$ at the beginning of some of these regions.
	Therefore, according to this plane-wave treatment, if the alignment tendency is due to pseudoscalar-photon mixing, light from these quasars should be circularly polarised, with a circular polarisation of the order of the observed linear one, except if one assumes very specific distributions of magnetic field orientations along the line of sight for each quasar.
	While dichroism is an interesting way to produce linear polarisation and, in particular, to explain the observations concerning quasars, birefringence on the other hand should provide a very clear signature of the mixing.

\chapter{Circular polarisation in quasars}\label{chap:circular}

\section{Data and circular polarisation}\label{sec:prob}

	Following the predictions of axion-photon mixing, the circular polarisation of some quasars belonging to the sample~\cite{Hutsemekers:2005} has been recently accurately measured in visible light~\cite{Hutsemekers:2010fw}.
	The data consist of circular polarisation measurements using a ``Bessell V filter''~\cite{Sterken:1992ap,EFOSC2:2008}, together with a compilation of previous measurements~\cite{Landstreet:1972so,Stockman:1984qz,Moore:1981circ,*Valtaoja:1993circ,*Takalo:1993circ,*Impey:1995circ,*Tommasi:2001a,Tommasi:2001b}, most of them in white light (unfiltered).
	Out of \draftun{the reported} 21 V-filter measurements, all but two are compatible with no circular polarisation at the $3\sigma$ level.
	The two positive detections concern highly linearly polarised blazars (both with $p_{\mathrm{lin}}>22\%$), which could be intrinsically circularly polarised~\cite{Hutsemekers:2010fw}, as in the case of a highly linearly polarised BL Lac object ($p_{\mathrm{lin}}>26\%$), for which a non-zero circular polarisation in white light had also been previously observed~\cite{Tommasi:2001b,Hutsemekers:2010fw}.\footnote{What is seen from blazars (including BL Lac) is very different from the characteristics of other quasars; see, \textit{e.g.} Refs.~\cite{Schneider:2006bk,FiftyYearsQuasars:2012bl,Borguet:2009phd}. In particular, their optical emission is a featureless non-thermal continuum that is highly polarised and highly variable. Such properties can be explained if they are observed with a jet of \draftun{relativistic} particles almost along the line of sight; as for their radio emission, the optical emission would be mostly the result of \draftun{beamed} synchrotron radiation of ultra-relativistic electrons inside the jet, rather than the usual thermal emission from the accretion disc. This could also explain the origin of the observed circular polarisation in radio and in visible light; see, \textit{e.g.} Refs.~\cite{Hutsemekers:2010fw,Tinbergen:2003cp}.}
	These special cases set aside, the objects observed have a circular polarisation consistent with zero, be it in white light or as seen through a V filter: one infers for instance that the average of $|v|$ is $0.035\% \pm 0.016\%$, using V-filter data of 13 objects with $z>1$~\cite{Hutsemekers:2010fw}.

\begin{figure}
		\centering
		\includegraphics[width=\textwidth, trim = 0mm 4.5mm 0mm 0mm, clip]{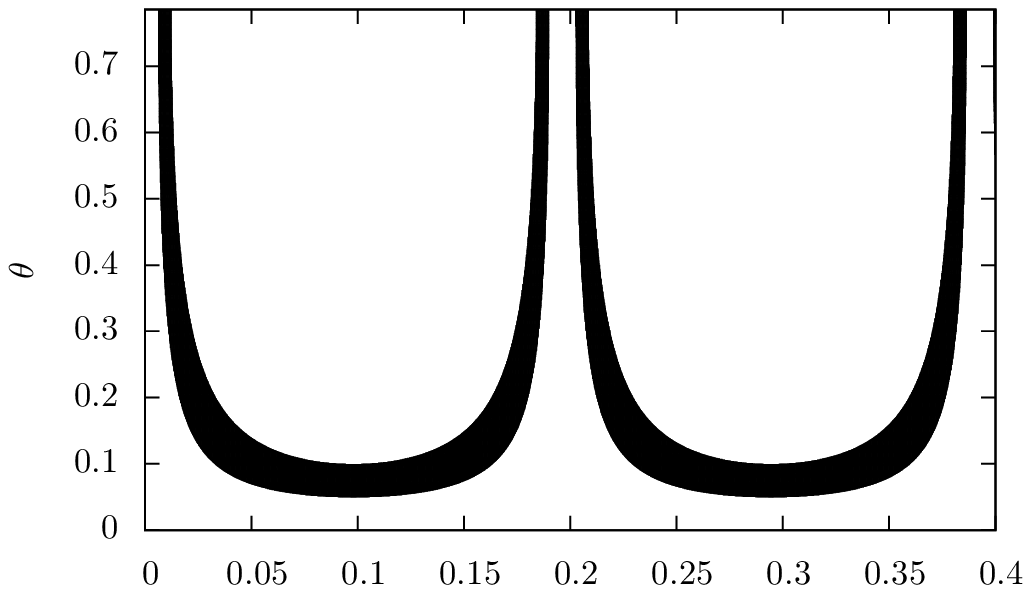}
		\includegraphics[width=\textwidth]{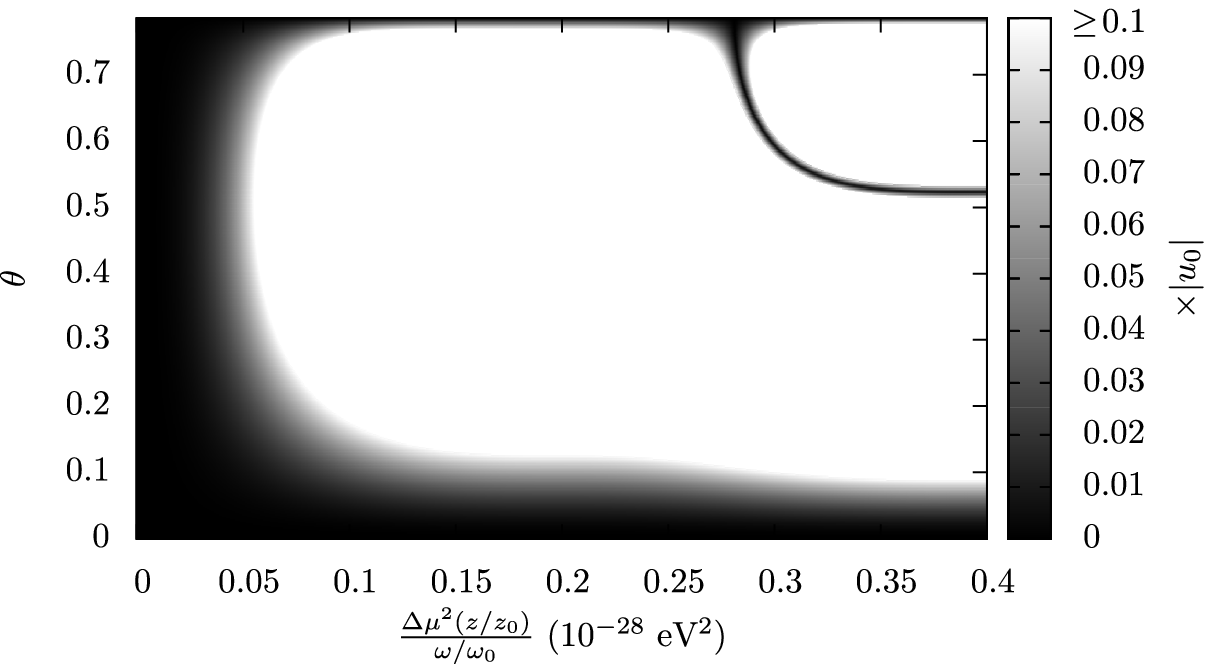}
		\caption{\emph{Bottom}: Amount of circular polarisation $|v|$ in the case $|u_0|=p_{\mathrm{lin},0}$ in the small region of Fig.~\ref{fig:rtheta_pcirc} where $v$ is small in modulus compared to the linear polarisation. \emph{Top}: Zoom on the corresponding region from Fig.~\ref{fig:rthetaplinok}, which shows the parameters able to produce the estimated amount of additional linear polarisation starting with initially unpolarised light\hspace{1pt}---\hspace{1pt}that remains essentially unchanged in the $|u_0|=p_{\mathrm{lin},0}$ case if we require $|v|$ to be small, as we explain.}
		\label{fig:probcirc}
\end{figure}

	This is clearly in contradiction with the results presented in the previous chapters for the mixing in the case of plane waves, in which we have stressed that $|v|$ is expected to be of the same order as $p_{\mathrm{lin}}$ \draftun{(\textit{i.e.} $\sim1\%$)} if we try to reproduce the linear polarisation data from quasars.
	If we go back to Fig.~\ref{fig:rtheta_pcirc}, we see that this is true for most of the parameters, except in a small region:
$\frac{\Delta\mu^2(z/z_0)}{(\omega/\omega_0)}\lesssim0.4\times10^{-28}$~eV$^{-2}$, and $|\theta|$ with values similar to the ones in Eq.~\eqref{eq:thetarange}; a zoom of this small area is pictured in Fig.~\ref{fig:probcirc}.

	Keeping an additional linear polarisation of the order of 1\% while suppressing the circular one then requires a considerable amount of \draftun{fine tuning} of the masses, or smaller regions of magnetic field. The latter seems excluded as the correlations of polarisations over huge distances require magnetic fields to be coherent in large regions (we shall discuss this in more details in Chap.~\ref{chap:moregenmagn}).

	To illustrate the need for fine tuning, we can give a pseudo\-scalar mass for which enough linear polarisation is created (using the fact that the maximum of $p_{\mathrm{lin}}$ is determined by $\theta$), and such that the circular polarisation is much smaller than the linear one (choosing a suitable $\Delta\mu^2$). We can consider $\theta$ as determined in the unpolarised case: the additional polarisation will indeed not be very different here, as by requiring no $v$, we essentially constrain $u$ not to change.
First, we write the pseudo\-scalar mass as a function of $\theta$ and of $\Delta\mu^2$ for a fixed value of $\omega_{\mathrm{p}}$:
	\be
		m = \sqrt{{\omega_{\mathrm{p}}}^2 + \Delta\mu^2 \cos\left(2\theta\right)},\quad\textrm{if }m > \omega_{\mathrm{p}};
	\ee
	\be
		m = \sqrt{{\omega_{\mathrm{p}}}^2 - \Delta\mu^2 \cos\left(2\theta\right)},\quad\textrm{if }m < \omega_{\mathrm{p}}.
	\ee
	For $z=z_0=10~$Mpc and $\omega= \omega_0=2.5$~eV, $u(0)=1\%$, and using $\theta = 0.1$, we then obtain that the only allowed ALP masses able to reproduce data would be such that:
	\be
		\frac{m}{\omega_{\mathrm{p}}} \in \left[0.99,1.01\right].\label{eq:finetuned}
	\ee
	This is a very fine-tuned situation, especially given that we have allowed $v$ to be as large as 0.1\% in this example.\footnote{If we require $v<0.01\%$, the range of allowed values for the mass shrinks to the interval $m \in \left[0.998,1.002\right]~\omega_{\mathrm{p}}$.} \draftun{Moreover, as} the plasma frequency is expected to vary along the light trajectory, \draftun{the constraint}~\eqref{eq:finetuned} cannot be maintained.
	These data thus strongly disfavour the ALP hypothesis under its simplest form.
	Note that neglecting the initial pseudoscalar flux, as was first done for simplicity, seems a good approximation, at least if one supposes a coupling only to photons.
	Indeed, reproducing alignments requires nearly massless particles (see Chap.~\ref{chap:alignments}), and, for such masses, the mixing is highly suppressed inside quasars at the energies we consider.\footnote{To present knowledge, $n_{\mathrm{e}}$ typically is huge in quasars~\cite{Peterson:2006ed,*FiftyYearsQuasars:2012ed,*Peterson:1997ed,*Osterbrock:1989ed,*Davidson:1979ed}, so that the mixing is vanishing despite much stronger magnetic field strengths (\textit{cf.} the Hillas plot for instance~\cite{Hillas:1984is}).
	The smallest value of $n_{\mathrm{e}}$ we could find ($2.5\times10^7$ times higher than what we use in superclusters) was used in Ref.~\cite{SanchezConde:2009wu} for some small ($\sim 0.01$~pc) regions of a given AGN, together with a magnetic field of 1.5 G.
	Related to dichroism, axion production is given by the loss of intensity: $I(0)-I(z)$; see Eqs.~\ref{eq:Stokes_alternative} for a single zone. To be conservative, we calculate the associated maximum, determined by $\sin^2 2\theta$ (as first derived in Ref.~\cite{Raffelt:1987im}), which is at best $\sim 4.5\times10^{-3}$ for a frequency redshifted to $\omega = 2.5$~eV.}
	\draftun{Still, this is not crucial for our conclusions, as obtained when we consider more general magnetic fields, \textit{e.g.} with a domain structure (see Chap.~\ref{chap:moregenmagn}), in which pseudoscalars are generated in the first domains anyway.}

	In the following, we are going to check whether this conclusion changes in more refined models.
	We first introduce a more rigorous treatment in which light from quasars is no longer described with plane waves but with wave packets that allow for a proper description of coherence, which may be crucial for circular polarisation.

\section{A wave-packet treatment}\label{sec:wptreatment}

	We now consider the possibility of reducing circular polarisation by considering wave packets, which automatically include the possibility of decoherence between waves of different frequencies. As circular polarisation is a matter of phase shifts, decoherence effects can significantly reduce it.
	There is also a natural observational reason for taking into account this effect: astronomers perform polarimetric measurements in given ranges of frequencies, with given filters.

	From the Lagrangian \eqref{eq:lagrangian}, the system of relevant equations is
	\be
		\left\{
			\begin{split}
				&(\square + {\omega_{\mathrm{p}}}^2) E(z,t) - g\mathcal{B} \partial^2_t\phi(z,t) = 0\\
				&(\square + m^2) \phi(z,t) + g\mathcal{B} E(z,t) = 0,
			\end{split}
		\label{eq:system}
		\right.
	\ee
	where we simplify the notation: from now on, $E\equiv E_{\parallel}$. Note that the solution for $E_{\perp}$ will simply be that for $E_{\parallel}$, with $g\mathcal{B}$ set to zero.

	We consider the case in which a wave packet is sent into a region of constant magnetic field $\mathcal{B}$, starting at $z=0$, and use wave packets in $\omega$:
	\be
		E (z,t) = \int_{-\infty}^{\infty} d\omega\    e^{- i \omega t} \widetilde{E} (z,\omega)
		\qquad\textrm{and}\qquad
		\phi (z,t) = \int_{-\infty}^{\infty} d\omega\    e^{- i \omega t} \widetilde{\phi} (z,\omega).
		\label{eq:decompsuromega}
	\ee
	Equations~\eqref{eq:system} have then to be satisfied by the integrands of \eqref{eq:decompsuromega} in each region, with $\mathcal{B}=0$ if $z\leq0$ (region I), and $\mathcal{B}\neq0$ if $z\geq0$ (region II); the solutions are given in Appendix~\ref{app:wp}. For the rest of the discussion, the incident packet in the first region has the initial shape
	\be
		\draft{\widetilde{{E}}_{\mathrm{i, I}}(z = 0,\omega) = \widetilde{{E}}_0\ e^{-\frac{a^2}{4}(\omega-\omega_0)^2},}
		\label{eq:gaussianpacket}
	\ee
	\draft{where $\widetilde{{E}}_0$ and $a$ are given below.}

	\subsection{Size of the wave packets}

	Continuum light coming from most quasars, at least in UV and visible wavelengths, is thermally emitted in the accretion disc. In order to obtain an estimate of the wave-packet size in this case, we can start with results for black-body radiation: we decompose the accretion disc into a concentric collection of black bodies of different temperatures at different radii~\cite{Shakura:1973bh,*Kochanek:2007ta}.\footnote{Strictly speaking, one would then have to average the results obtained for different black bodies over the range of frequencies actually observed.} For a black-body radiation of Wien wavelength $\lambda_{\mathrm{w}}$, estimates of the longitudinal coherence length $l_{\mathrm{c}}$ have been obtained in interferometry~\cite{Donges:1998}: $l_{\mathrm{c}} \simeq\lambda_{\mathrm{w}}\simeq \bar{\lambda}$, with $\bar{\lambda}$, the mean wavelength of the radiation.
	The relation that we use for the value of $a$ which enter Eq.~\eqref{eq:gaussianpacket} is then:
	\be
		a = \frac{2\sqrt{\ln(2)}}{\pi}\bar{\lambda},
	\ee
	and the initial full-width at half-maximum in position is:
	\be
		\Delta z\simeq\sqrt{2\ln(2)}a,
	\ee
	which is thus of the order of the wavelengths considered.

		\begin{figure}
			\centering
					\includegraphics[width=\textwidth, trim = 0mm 3mm 0mm 0mm, clip]{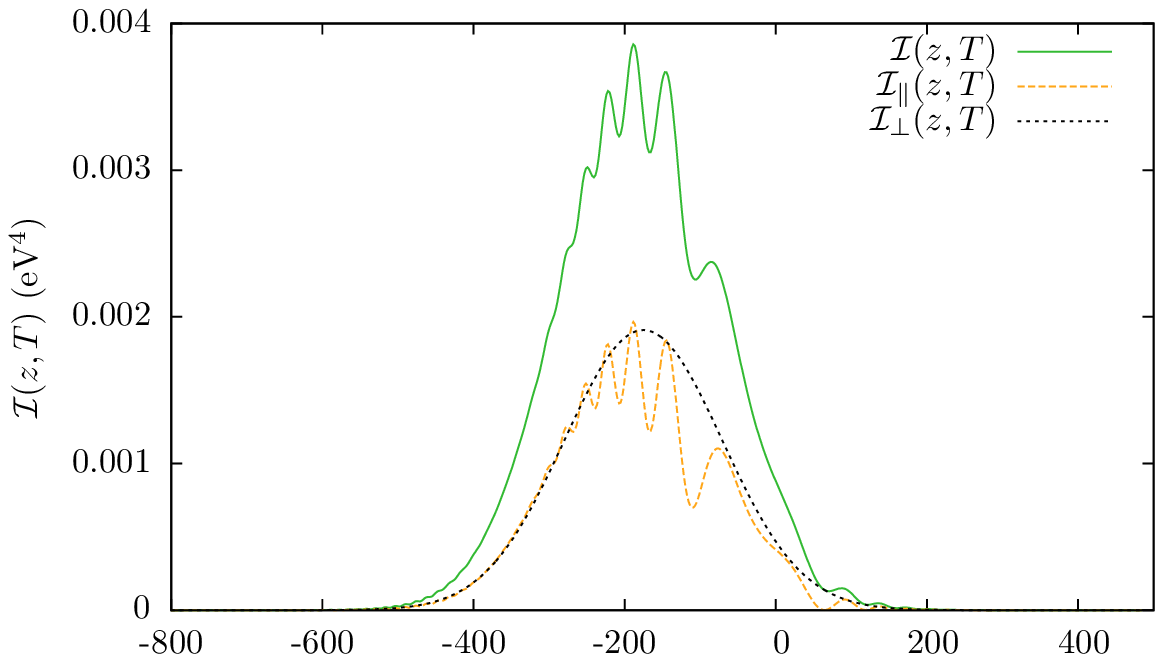}
					\includegraphics[width=\textwidth]{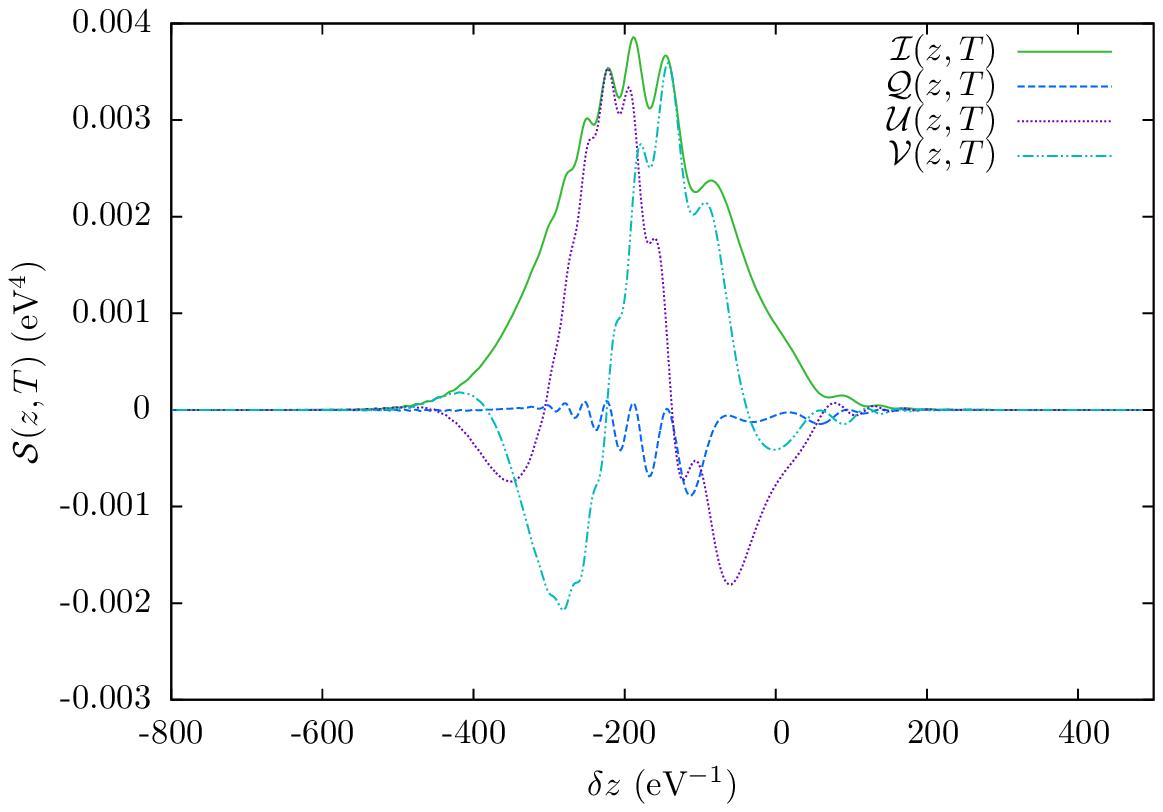}
			\caption{The shape of the wave packets at time $T=10~\textrm{Mpc}/c$ for a light beam with $u(0)=1$. The abscissa is $\delta z\equiv z-cT$, which is the shift in position with respect to a frame moving at the speed of light $c$; \textit{i.e.} here the origin is at 10~Mpc. \emph{Top}: we show the total intensity and the intensities for the polarisations parallel and perpendicular to the magnetic field, before integration. \emph{Bottom}: we show the contributions to the other Stokes parameters. We used $\omega_{\mathrm{p}}=3.7\times10^{-14}$~eV, $m=4\times10^{-14}$~eV, $\omega_0=2.5$~eV (\textit{i.e.} $\lambda_0=500$~nm), $a=1.34$~eV$^{-1}$, and $g\mathcal{B}=3\times10^{-29}$~eV.
			}
			\label{fig:packets}
		\end{figure}

	\subsection{Stokes parameters and partially polarised light}

	As wave packets go through the detector much faster than its time resolution, one has to integrate the packets over the exposure time $\Delta t$ to calculate the Stokes parameters. Let us now represent by $S$ any of the four Stokes parameters; with the notations of Eq.~\eqref{eq:StokesE}, the observed quantities are then
	\be
		{S}(z) = \langle \mathcal{S}(z,t) \rangle \equiv \frac{\Delta N}{\Delta t} {\int}_{t - \frac{\Delta t}{2}}^{t + \frac{\Delta t}{2}} dt\ \mathcal{S}(z,t),\label{eq:stokesint}
	\ee
	where $\Delta N/\Delta t$ is the number of packets during the interval $\Delta t$. As the normalisation does not matter, to simplify we choose $\Delta N/\Delta t = 1$~eV and
	\be
		\widetilde{{E}}_0=\frac{\sqrt{a}}{{(2\pi)}^{3/4}}\times 1~\textrm{eV}^{3/2};
	\ee
	this corresponds to packets with initial intensities of 1~eV$^4$.
	\draft{Note that calculating the propagation of these wave packets is a numerical challenge, as the computation of the Stokes parameters requires a spatial resolution better than 1~$\mu$m after a propagation over distances of the order of 10~Mpc; there is a factor of about $10^{30}$ between these scales.}\footnote{We use the Multiple-Precision Floating-point library with correct Rounding~\cite{mpfr}.}

	In Fig.~\ref{fig:packets}, we illustrate the packets after a propagation inside region II in quite a strong mixing case. For photon polarisations parallel to $\vec{\mathcal{B}}$ we see the effect of interferences within the packet, while there is only a spread for non-mixing photons. To \draftun{obtain finally} the values of the Stokes parameters at a given $z$, these quantities have to be integrated over $t$ using Eq.~\eqref{eq:stokesint}.\footnote{The integrands are in fact the functions that we see in Fig.~\ref{fig:packets} with $z$ replaced by $t$; indeed, being of the order of 10~Mpc, these only differ by a tiny quantity $\delta z$ (of the order of $1~\mu$m).}
	Note that the case illustrated is for 100\% polarised light, so that the obtained $U(z)$ and $V(z)$ are actually much larger than what the same conditions would give for typical quasars light.
	Now we need a correct description of what happens to initially unpolarised and partially linearly polarised light described with wave packets.

		To treat partially polarised light for the case at hand, we can make use of a useful property of the Stokes parameters in the case of fully polarised light beams, defined as in Eq.~\eqref{eq:fullypollight}.\footnote{That result, derived from the definitions of the Stokes parameters themselves, is actually quite general. It is automatically satisfied, provided that the two electric fields defining the linear polarisation basis $\vec{E}^{(x)}$ and $\vec{E}^{(y)}$ have their directions left unchanged, so that $\mE_{\mathrm{r}_x} = \cos(\varphi_0)\vec{E}^{(x)}\cdot\vec{e}_x$ and $\mE_{\mathrm{r}_y} = \sin(\varphi_0)\vec{E}^{(y)}\cdot\vec{e}_y$. In particular, this condition is verified for light beams mixing with axion-like particles in a single magnetic field region; see Sec.~\ref{sec:solution_onezone_polarisationbasis}.
		}
		For fixed external conditions, calculating the Stokes parameters~\eqref{eq:StokesE} of any such light beam in the $\varphi_0=\frac{\pi}{4}$ case gives us access to the following quantities:
		\be
			\begin{split}
				I_{\mathrm{pol}}(z;\varphi_0=\frac{\pi}{4}) &= \langle \frac{1}{2} \left({|E_{\parallel}|}^2 + {|E_{\perp}|}^2\right) \rangle \equiv C_1(z) \\
				Q_{\mathrm{pol}}(z;\varphi_0=\frac{\pi}{4}) &= \langle \frac{1}{2} \left({|E_{\parallel}|}^2 - {|E_{\perp}|}^2\right) \rangle \equiv C_2(z) \\
				U_{\mathrm{pol}}(z;\varphi_0=\frac{\pi}{4}) &= \langle \operatorname{Re}\{E_{\parallel}E^*_{\perp}\} \rangle \equiv C_U(z) \\
				V_{\mathrm{pol}}(z;\varphi_0=\frac{\pi}{4}) &= \langle \operatorname{Im}\{E_{\parallel}E^*_{\perp}\} \rangle \equiv C_V(z), 
                	\end{split}
			\label{eq:lesC}
		\ee
		which do evolve with $z$ in our case, due to the mixing with pseudo\-scalars.
		Using the definitions of these new quantities, it is then straightforward to show that the evolution of the Stokes parameters for any other light beam $\vmE_{\mathrm{r}}$, with initial angle $\varphi_0$, in the same conditions is determined by
		\be
			\left\{
				\begin{split}
        	        	        I_{\mathrm{pol}}(z;\varphi_0) &= C_1(z) - C_2(z) \cos(2\varphi_0)\\
        	        	        Q_{\mathrm{pol}}(z;\varphi_0) &= C_2(z) - C_1(z) \cos(2\varphi_0)\\
        	        	        U_{\mathrm{pol}}(z;\varphi_0) &= C_U(z) \sin(2\varphi_0)\\
					V_{\mathrm{pol}}(z;\varphi_0) &= C_V(z) \sin(2\varphi_0).
        	        	\end{split}
			\right.
				\label{eq:Stokesangle}
                \ee
		Now, as we know, unpolarised light can be thought of as the average over every possible initial angle $\varphi_0$, or equivalently as a sum at the Stokes-parameters level of the fully linearly polarised cases $\varphi_0=0$ and $\varphi_0=\frac{\pi}{2}$ for instance.
Applied to Eq.~\eqref{eq:Stokesangle}, such an averaging then gives the result in the case of an unpolarised light beam:
		\be
			\left\{
				\begin{split}
		        	        I_{\mathrm{unpol}}(z) &= C_1(z)\\
		        	        Q_{\mathrm{unpol}}(z) &= C_2(z)\\
		        	        U_{\mathrm{unpol}}(z) &=0\\
					V_{\mathrm{unpol}}(z) &=0.
	                	\end{split}
			\right.
                \ee

		The evolution of any Stokes parameter $S$ of a light beam characterised by an initial partial linear polarisation, defined by a given value of $p_{\mathrm{lin},0}$ and a value of $\varphi_0$, then follows:
		\be
			S_{\mathrm{partial}}(z;\varphi_0;p_{\mathrm{lin},0})
			=
			p_{\mathrm{lin},0}\ S_{\mathrm{pol}}(z;\varphi_0) + \frac{1}{2\pi} {\int}_0^{2\pi} (1 - p_{\mathrm{lin},0})\ S_{\mathrm{pol}}(z;\varphi) \ d\varphi.
			\label{eq:Stokesanglepartialexplicit}
		\ee
		This finally leads to a natural generalisation of the fully polarised case:
		\be
			\left\{
				\begin{split}
		        	        I_{\mathrm{partial}}(z;\varphi_0;p_{\mathrm{lin},0}) &= C_1(z) - p_{\mathrm{lin},0}\big[C_2(z) \cos(2\varphi_0)\big]\\
		        	        Q_{\mathrm{partial}}(z;\varphi_0;p_{\mathrm{lin},0}) &= C_2(z) - p_{\mathrm{lin},0}\big[C_1(z) \cos(2\varphi_0)\big]\\
		        	        U_{\mathrm{partial}}(z;\varphi_0;p_{\mathrm{lin},0}) &= p_{\mathrm{lin},0}\big[C_U(z) \sin(2\varphi_0)\big]\\
					V_{\mathrm{partial}}(z;\varphi_0;p_{\mathrm{lin},0}) &= p_{\mathrm{lin},0}\big[C_V(z) \sin(2\varphi_0)\big];
	                	\end{split}
			\right.
			\label{eq:Stokesanglepartial}
                \ee
		\textit{i.e.} it is sufficient to calculate the coefficients defined in Eqs.~\eqref{eq:lesC} once to derive the evolution of the Stokes parameters for any\hspace{1pt}---\hspace{1pt}linearly polarised (either fully or partially) or unpolarised\hspace{1pt}---\hspace{1pt}initial beam travelling the same distance in the same external conditions.

		\begin{figure}
			\centering
					\hspace{3cm}\includegraphics[width=0.7\textwidth]{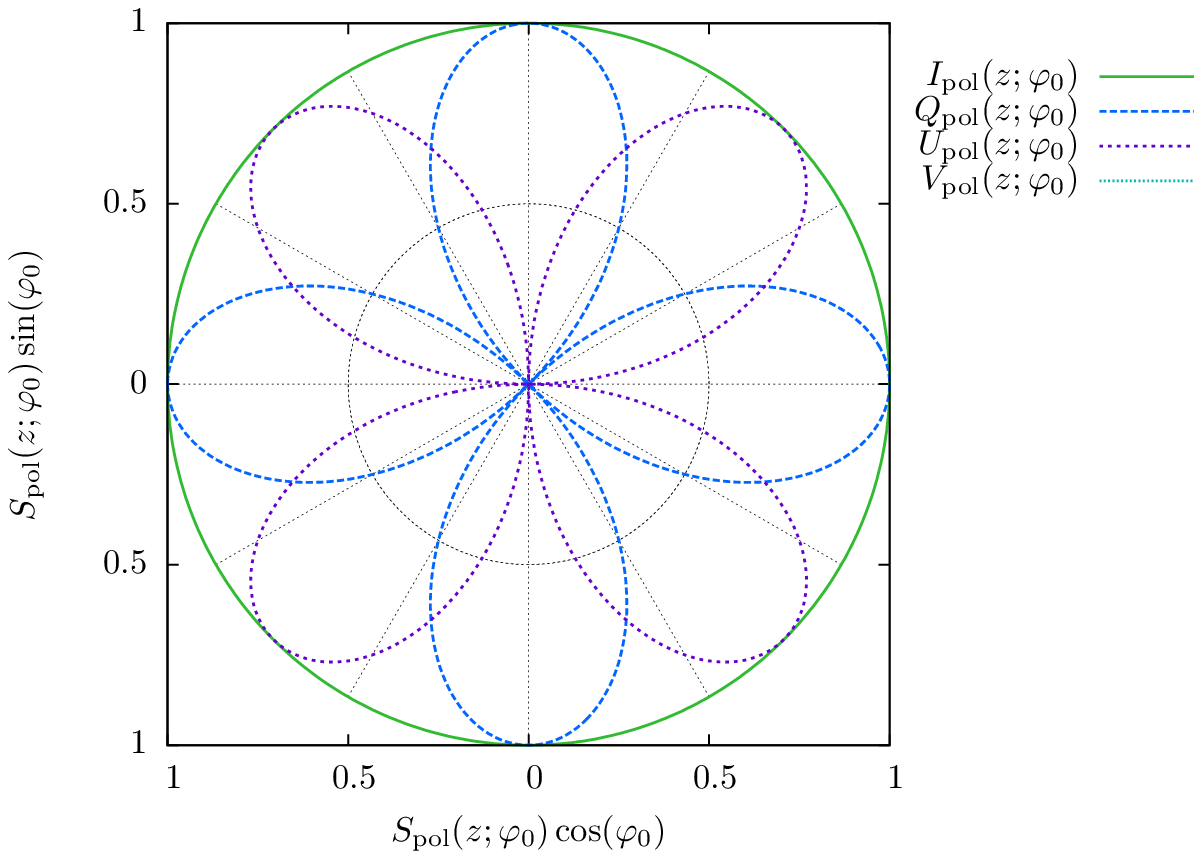}

					\includegraphics[width=0.55278\textwidth]{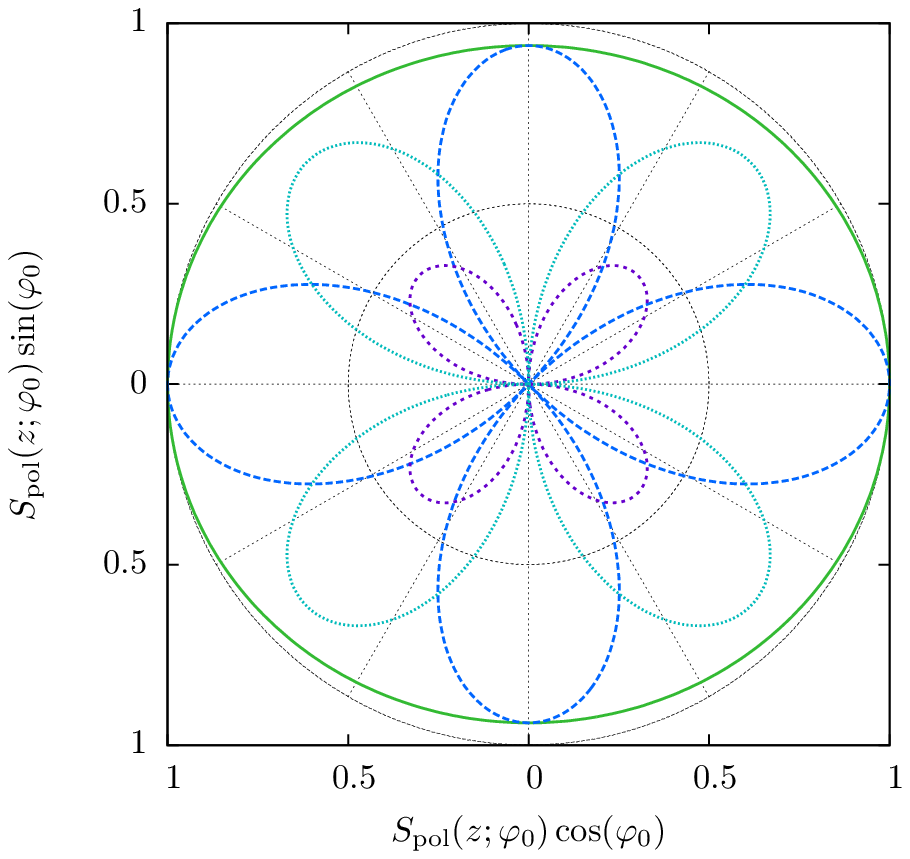}\includegraphics[width=0.45722\textwidth]{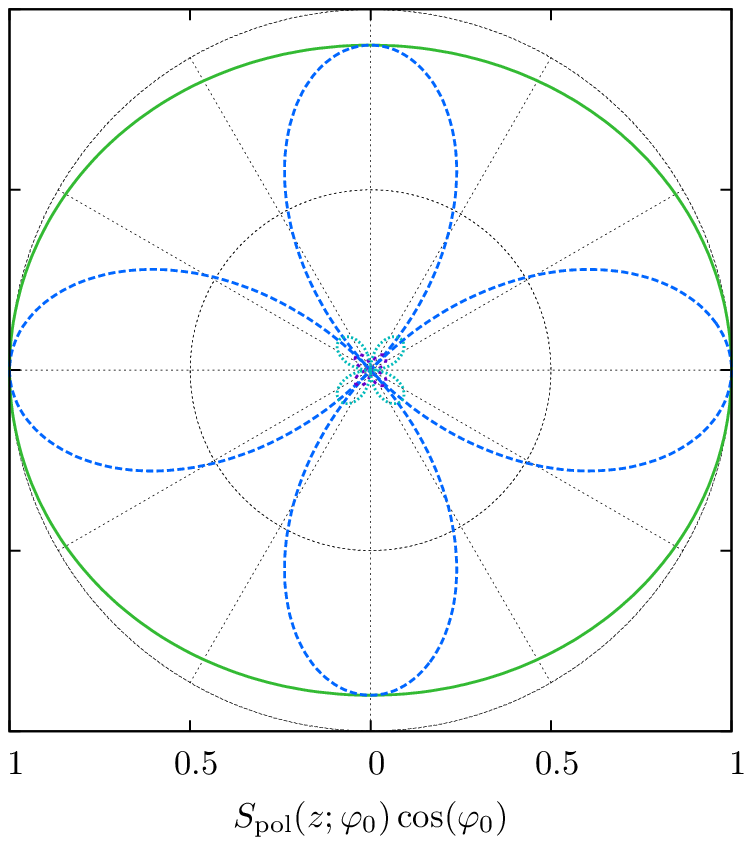}
			\caption{Stereographic views of each of the Stokes parameters \draft{in eV$^4$} before (\emph{top}), and after a 10~Mpc propagation inside a magnetic field with plane waves (\emph{bottom left}), and with wave packets (\emph{bottom right}), for initially 100\% linearly polarised light. The distance of the curves to the origin gives the value of the parameters. To enable direct comparisons, the angular coordinate in the three figures is the initial angle, $\varphi_0$. The direction of the magnetic field is the one given by $\varphi_0=\pm\frac{\pi}{2}$. This relatively strong mixing case is shown for $m=4.5\times10^{-14}$~eV, $\omega_{\mathrm{p}}=3.7\times10^{-14}$~eV, $g\mathcal{B}=5\times10^{-29}$~eV, $\omega_0=2.5$~eV, and $a = 1.34$~eV$^{-1}$.
			}
			\label{fig:polarstokes}
		\end{figure}

	\subsection{Results for white light}\label{section:results1zone}

	We now present the results of the mixing of photons with axion-like particles in a wave-packet formalism. We shall argue that these packets can be used to describe white light, with no photometric filter. The photomultipliers used to perform the white-light measurements of polarisation have a broad spectral-response range (from 185 to 930~nm for the ones used in Ref.~\cite{Landstreet:1972so}), which is indeed similar to the width of our wave packets.\footnote{Note however that there is an atmospheric cutoff for wavelengths below 330~nm~\cite{Patat:2011atm}.}

	Note that current upper limits on the pseudo\-scalar coupling are $g=10^{-11}$~GeV$^{-1}$. Together with $\mathcal{B}=0.3~\mu$G, this means that $g\mathcal{B}\lesssim6\times10^{-29}$~eV.

	In Fig.~\ref{fig:polarstokes}, we first illustrate the different Stokes parameters at a given distance, for each initial value of the angle $\varphi_0$. For a given angle, the distance between the origin and each $S(z;\varphi_0)$ curve is the value of this Stokes parameter. Now, for wave packets, we obtain that the circular polarisation $V(z)$ is strongly reduced with respect to the plane-wave prediction; notice that $U(z)$ is also affected in the same way, as it is also very sensitive to phase effects.
This can be understood if one goes back to Fig.~\ref{fig:packets}: we see that these two quantities change sign within the packet itself, due to the extremely frequency-dependent character of the birefringent effect induced by the pseudo\-scalars.
Also note that, whereas for plane waves the Stokes parameters obey $I^2(z)=Q^2(z)+U^2(z)+V^2(z)$ for any $\varphi_0$, this is no longer true in the wave-packet case, even for light initially fully linearly polarised.

\begin{figure}
			\centering
					\includegraphics[width=\textwidth]{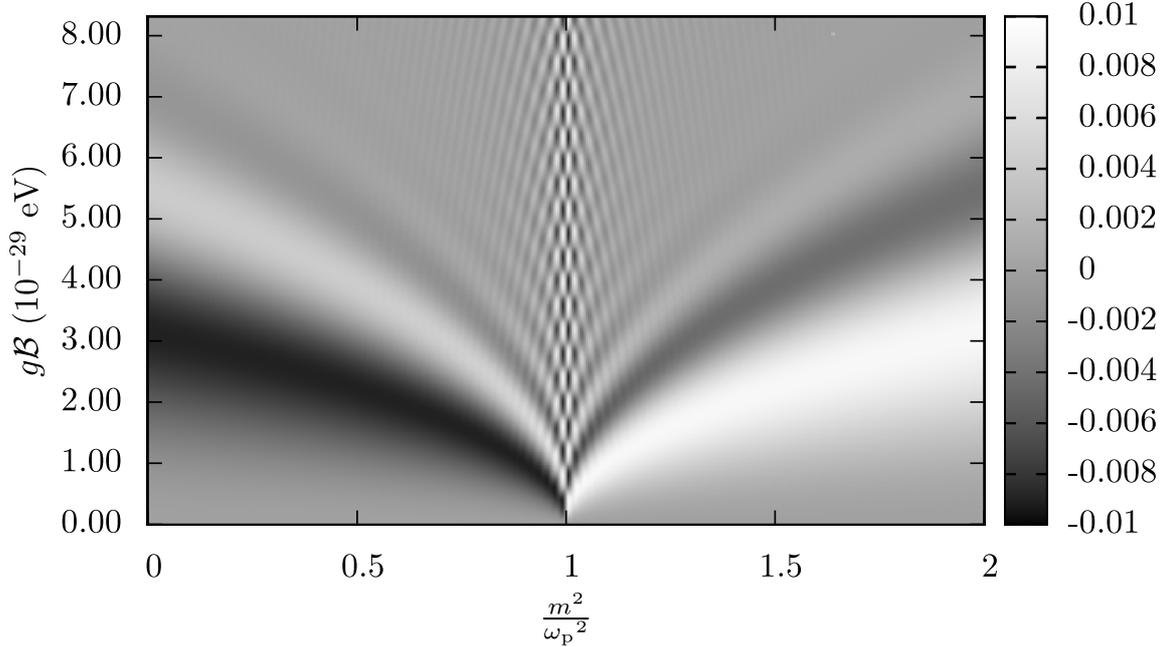}
			\caption{Same (circular polarisation) as Fig.~\ref{fig:japfan_circ} but with light described by wave packets; this is in the case $u_0=0.01$ and $q_0=v_0=0$. We have used $\omega_0=2.5$~eV and $a=1.34~$eV$^{-1}$.
			}
			\label{fig:WPpcirc}
		\end{figure}

\begin{figure}
			\centering
					\includegraphics[width=\textwidth]{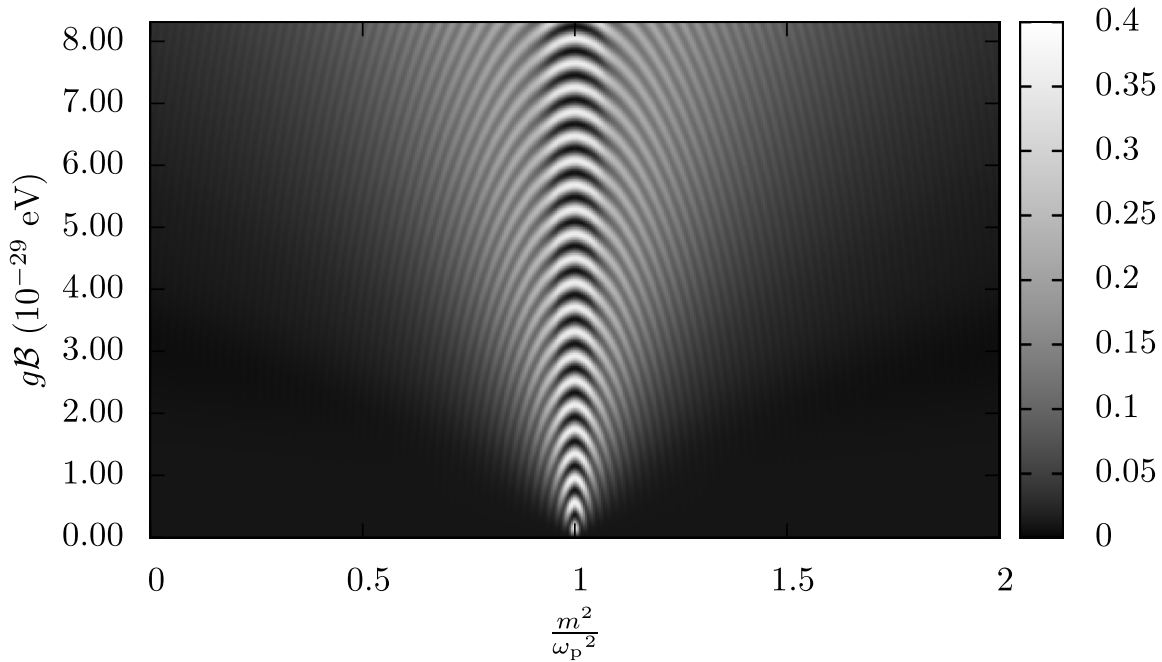}
			\caption{Same (linear polarisation) as Fig.~\ref{fig:IPplin_pl} but with light described by wave packets. We have used $\omega_0=2.5$~eV and $a=1.34~$eV$^{-1}$. Note that the right-hand box gives the base-10 logarithm of the linear polarisation.
			}
			\label{fig:WPplin}
		\end{figure}

	For partially polarised light the expected amount of circular polarisation will of course be even smaller.
	This is shown in Figs.~\ref{fig:WPpcirc} and~\ref{fig:WPplin}, which are the wave-packet results analogous to those of the plane-wave case (Figs.~\ref{fig:japfan_circ} and~\ref{fig:IPplin_pl}). We obtain a large suppression of circular polarisation for most of the parameters. Let us emphasise that this is in the case leading to the highest amount of $v$; \textit{i.e.} $u(0)=p_{\mathrm{lin},0}=0.01$. Besides, we notice that the maximum linear polarisation attainable for some of the parameters is smaller than in the plane-wave case, and that the contrast is not as sharp: this is related to the loss (and averaging) of $u$ that happens in this case, as also seen in Fig.~\ref{fig:polarstokes}, and to the averaging of $q$.

\begin{figure}
			\centering
					\includegraphics[width=\textwidth]{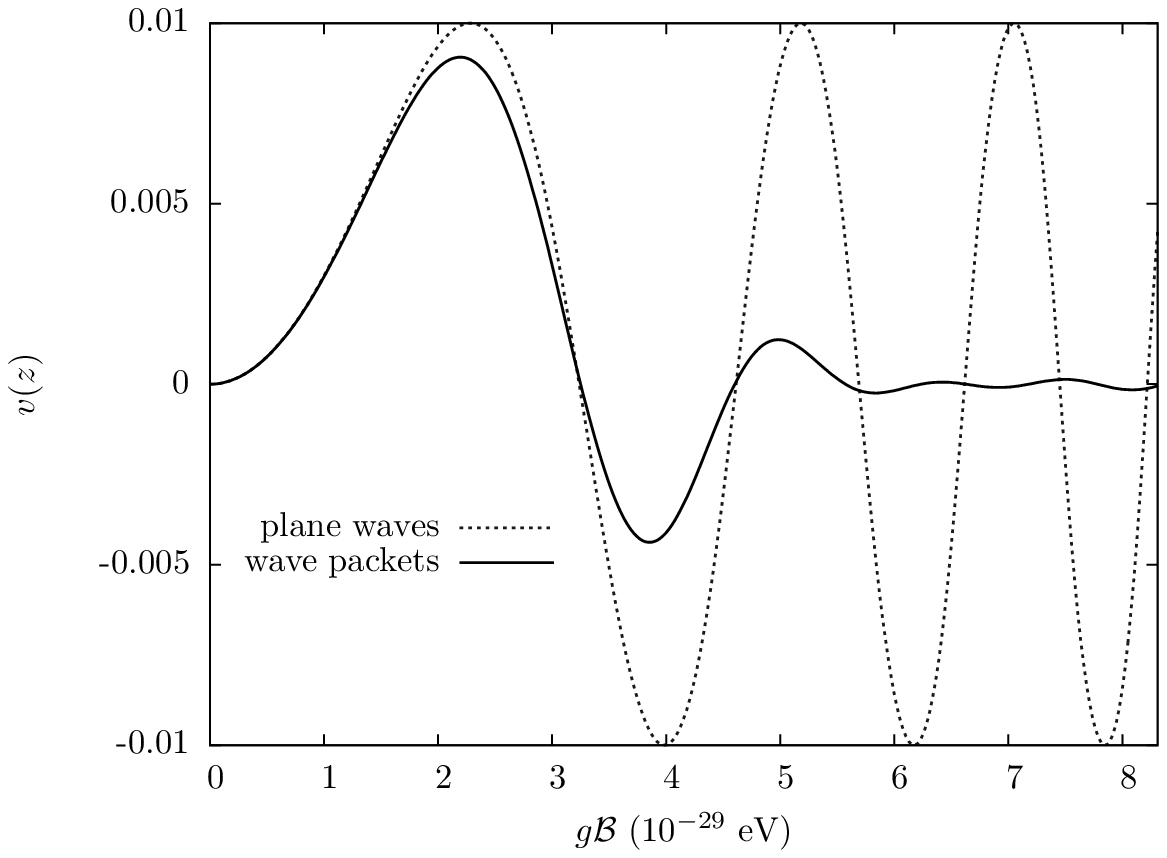}
			\caption{Comparison of results obtained with plane waves and with wave packets, for the same parameters. This is a cut, respectively of Fig.~\ref{fig:japfan_circ} with $u_0=0.01$ and of Fig.~\ref{fig:WPpcirc}, for the pseudo\-scalar mass $m=4.5\times10^{-14}$~eV.
			}
			\label{fig:pcirc_compared}
		\end{figure}

	In Fig.~\ref{fig:pcirc_compared}, we also directly compare the two descriptions for different values of the coupling; one can see that, for very small $g\mathcal{B}$, the results are similar, and that the suppression is more efficient at larger values of $g\mathcal{B}$.

	Finally, we have generalised our calculations to the case where the packets are initially described by the frequency distribution~\eqref{eq:gaussianpacket}, somewhere in the first region, at some $\tilde{z}\ll0$. The first region can then represent a cosmic void, where $\omega_{\mathrm{p}}$ can also typically have a smaller value; this allows the packet to propagate a long time, which makes it spread, before it enters the second region. We have checked that the results we have presented above hold in this case as well (even if the first region is taken to be one gigaparsec long). This confirms that the main mechanism that reduces the circular polarisation is not related to the separation and the spread of photon packets of different polarisation, but rather because of phase shifts within the packets that mix.
	This can be understood as $v(\omega)$ can change sign within the packet, averaging to zero, while $p_{\mathrm{lin}}(\omega) = \sqrt{q^2(\omega) + u^2(\omega)}$ cannot, keeping an alignment possible.

\begin{figure}
			\centering
					\includegraphics[width=0.7\textwidth]{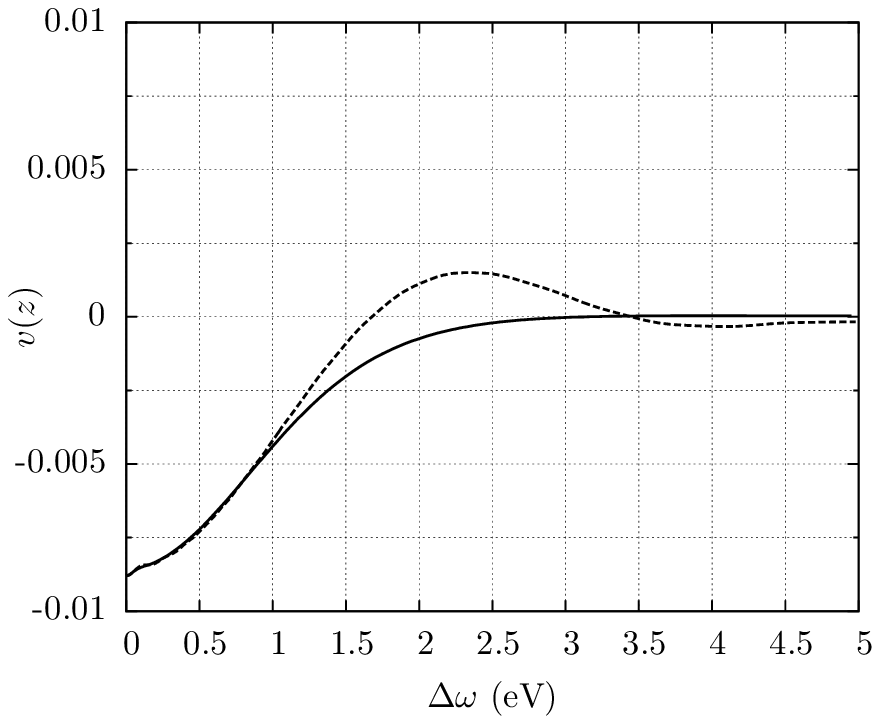}
					\includegraphics[width=0.7\textwidth]{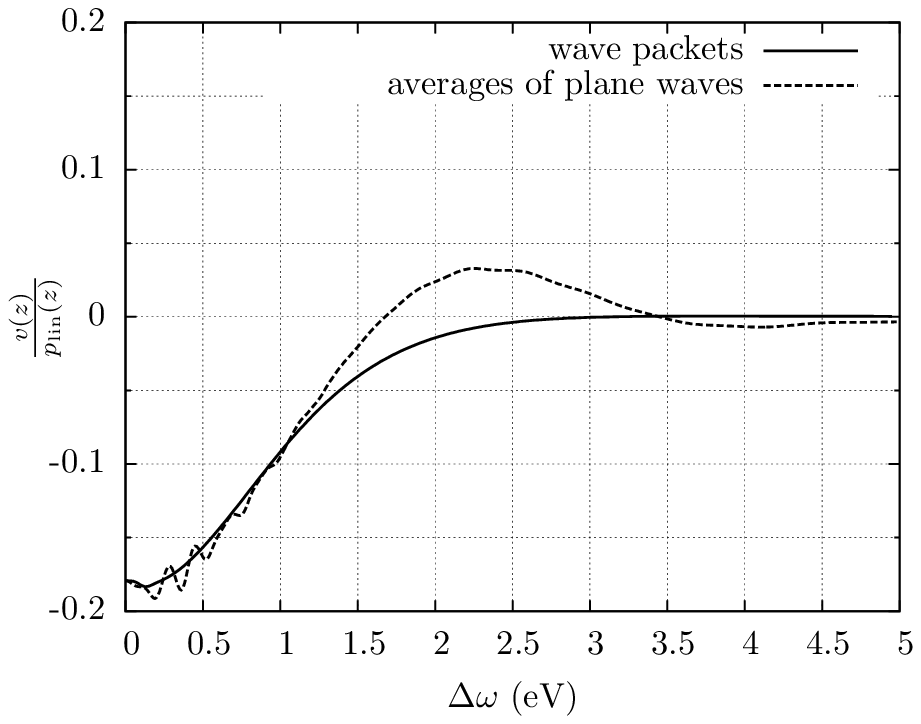}
			\caption{Comparing averaging methods for wave packets (with $a=\frac{4\sqrt{\ln(2)}}{\Delta\omega}$) and averages of plane waves: in both cases, the absolute values of $v$ (\emph{top}) and of $\frac{v}{p_{\mathrm{lin}}}$ (\emph{bottom}) are reduced with increasing $\Delta \omega$ with respect to the monochromatic case (\textit{i.e.} $\Delta \omega = 0$). Here, we used $g\mathcal{B} = 6\times10^{-29}$~eV, and $\Delta \omega$ is centered around $\omega_0 = 2.5$~eV; the other parameters are the same as in Fig.~\ref{fig:pcirc_compared}.
			}
			\label{fig:av_vs_pckts}
		\end{figure}

	A simpler approach~\cite{Stodolsky:1998tc} is to use direct averages of the plane-wave Stokes parameters of Eq.~\eqref{eq:Stokes_alternative} over frequency, instead of wave packets. This will give the same qualitative results, as illustrated in Fig.~\ref{fig:av_vs_pckts}, where quantities are plotted against $\Delta \omega$, the bandwidth over which each averaging is performed. For the averages of plane waves, we have used the analytical formulas~\eqref{eq:Stokes_alternative} averaged over a step profile in $\omega$, centered around $\omega_0$ and of width $\Delta \omega$. For the Gaussian wave packets of Eq.~\eqref{eq:gaussianpacket}, on the other hand, we chose $\Delta \omega$ to represent the full-width-at-half-maximum in $\omega$ of each initial packet (\textit {i.e}, $a=4\sqrt{\ln(2)}{\left(\Delta\omega\right)}^{-1}$). In either case, the larger the band of frequencies over which the averaging is done, the smaller the absolute value of the circular polarisation and its relative importance compared to the linear polarisation. This holds whatever the details of the averaging. Similar results have also been obtained in different contexts (chameleons~\cite{Burrage:2008ii}, and high-energy gamma sources~\cite{Bassan:2010ya}).

	Therefore, as far as white-light data are concerned, phenomenological implications of axion-like particles mixing with photons can be reconciled with circular polarisation measurements~\cite{Payez:2009kc,Payez:2009vi,Payez:2010xb}.
	
	\subsection{Results for Bessell V filter}\label{sec:vfilterKO}

	Most of the recent circular polarisation data of Ref.~\cite{Hutsemekers:2010fw} were taken using a Bessell broadband V filter~\cite{Sterken:1992ap,EFOSC2:2008}. This filter is centered around $\lambda = 547.6$~nm and the associated full-width-at-half-maximum is 113.2~nm. To mimic this cut in frequencies, one can convolve wave packets with the spectral response of the filter, or proceed to averages of plane-wave results over $\omega$ using \draftun{its} frequency profile.

	We then find that, even though it is a broadband filter, the typical values of the astrophysical parameters are such that the circular polarisation does not change sufficiently over this bandwidth to be strongly reduced when averaging over $\omega$. This is illustrated in Fig.~\ref{fig:av_vs_pckts} for small values of $\Delta \omega$ ($\approx 0.5$~eV). The circular polarisation degree is slightly smaller than in the monochromatic case, but the effect is certainly not sufficient to reconcile the \draftun{spinless-particle scenario} with the data. Except for very specific choices of parameters, the axion-like particle parameters able to create an alignment will also predict a sizeable amount of circular polarisation.
	If axions were at work, given the\hspace{1pt}---\hspace{1pt}somehow narrow\hspace{1pt}---\hspace{1pt}bandwidth of the broadband V filter, circular polarisation should have been observed.

	\chapter{Mixing in a more general magnetic field}\label{chap:moregenmagn}

	Among 21 quasars with circular polarisation measured in the V filter, 18 are located in the same direction of the sky, towards the regions of alignments A0 and A1, that we have presented in Sec.~\ref{sec:descriptionalignment}.
	In this chapter, we are going to focus on the regions where these V-filter data have been taken from and check the sensitivity of our conclusions to changes in the magnetic field morphology. 
	The location of these regions actually points towards the center of the Virgo supercluster~\cite{deVaucouleurs:1953lsc,*deVaucouleurs:1975lsc,*Tully:2002en,*Hu:2006lsc}, which is our local supercluster, shortened as ``LSC.''

	\section{Models for supercluster magnetic fields}

	\subsection{Motivation}

	The LSC magnetic field is essentially the last relevant magnetic field encountered by extragalactic photons coming towards us, given the energies and the masses determined by the problem at hand. For this reason, the axion-like particle explanation of quasar data will be ruled out if the influence of this field creates too much circular polarisation, as any $v$ created there should have been detected.

	This contrasts with what one can expect from photons of higher energies; see, \textit{e.g.} Ref.~\cite{Simet:2007sa}. In our case, the influence of our galactic magnetic field can be neglected because the mixing with nearly massless pseudoscalars would be inefficient at the optical frequencies at which the alignments have been detected (we use the values for $\mathcal{B}$ and $\omega_{\mathrm{p}}$ considered in Ref.~\cite{Bassan:2010ya}).
	Moreover, it is even smaller at the galactic latitudes at which the data have been obtained, as the field strength decreases exponentially in the direction transverse to the galactic plane~\cite{Giovannini:2003yn}.
	Finally, if the mixing was to happen inside the galaxy, one would have expected the galactic stars angularly close to the quasars to be similarly affected. It is not the case: their observed polarisation is much lower, as shown in Ref.~\cite{Payez:2010xb}.

		\subsection{Structure}\label{sec:struct}

	\begin{figure}
			\centering
			\includegraphics[width=0.65\textwidth]{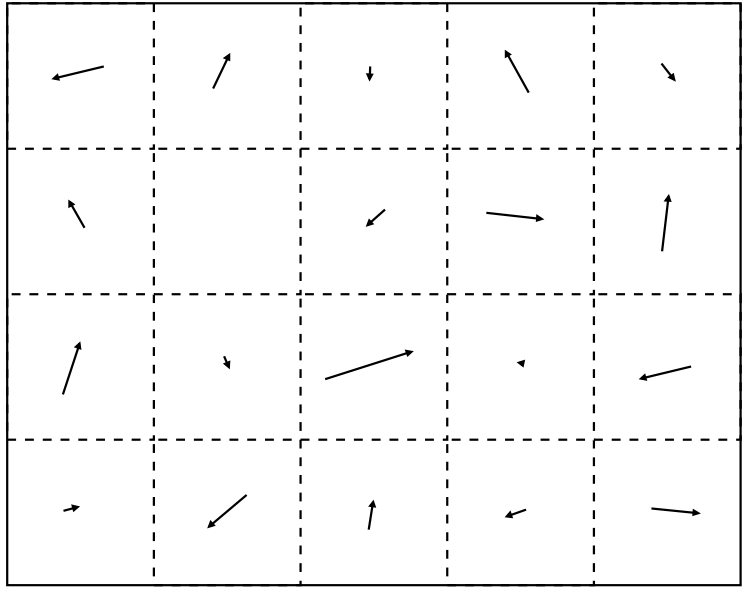}
			\caption{Sketch of a slice of the assumed morphology in domains for the transverse magnetic field in superclusters, projected onto the sky.}
			\label{fig:magnfield_patchy}
	\end{figure}

			On the observational side, what is typically assumed in the literature for magnetic fields inside clusters or superclusters is a domain structure~\cite{Vallee:2002,Vallee:2011,Xu:2005rb}, possibly with an additional weaker background field coherent over several domains. One usually considers a collection of domains of a given size, and lets the magnetic field direction change from one domain to the other while keeping the same field strength $|\vec{{B}}_{\mathrm{domain}}|$, which averages to a smaller value at the supercluster scale. Note that it is also what is typically considered to discuss the propagation of cosmic rays, and what is obtained from structure formation; see for instance Ref.~\cite{Dolag:2005a,*Dolag:2003ft} for results about the LSC. 
			Finally, the same kind of model is usually used when one discusses magnetic fields beyond supercluster scales; see for instance Ref.~\cite{Mirizzi:2006zy}. 

			Additionally, we perform general tridimensional rotations of the magnetic field direction between domains; it will thus pick up a longitudinal component most of the time. As emphasised in Chap.~\ref{chap:mixing}, only the projection of the magnetic field onto the transverse plane is relevant for ALP-photon mixing. Hence, we allow values of the transverse magnetic field strength $\mathcal{B}$ much smaller than $|\vec{B}_{\mathrm{domain}}|$. Taking a slice of such a magnetic field and projecting it onto the sky, we get something similar to what is depicted in Fig.~\ref{fig:magnfield_patchy}.
	
			In such a configuration, one sees different fields when looking in different directions, once everything is put together. Because of this, and because the exact magnetic structure is not known inside the Virgo supercluster, we assume that the effect of these irregularities is equivalent to considering that light coming from different objects essentially passes through random fields.

		\subsection{Field strength}\label{sec:Bstrength}

		Often in the astrophysics literature, extended magnetic fields are classified as galactic or extragalactic. Let us note that this is not precise enough for our purpose as the \draft{adjective} ``extragalactic'' encompasses quite different environments, with different magnetic fields: we can at least distinguish between clusters of galaxies, superclusters, and, finally, anything beyond these structures.

		Magnetic fields in galaxy clusters are well established~\cite{Clarke:2000bz,Giovannini:2003yn,Vallee:2011} and have typical strengths of several $\mu$G, although they can be as large as 40$~\mu$G. Our knowledge of these fields is sometimes refined enough that not only the field strength and the coherence lengths are known but also hints about the orientation are available~\cite{Pfrommer:2009hn}. In our case, even though the regions A0 and A1 are roughly centered on the Virgo cluster, most of the quasars are seen through a larger domain, and we need to model the magnetic field at the scale of the Virgo supercluster.

		There is evidence for a sizeable magnetic field in the surroundings of galaxy clusters\hspace{1pt}---\hspace{1pt}\draftun{near the Coma cluster, for instance, field strengths in the range $0.2$--$0.4~\mu$G have been detected on scales that can be as large as $\sim4$~Mpc~\cite{Kronberg:2007wa}}\hspace{1pt}---\hspace{1pt}and some results are also available for superclusters, see \textit{e.g.} Refs.~\cite{Widrow:2002ud,Giovannini:2003yn,Vallee:2011} for reviews.
		For instance, it has been inferred from Faraday rotation measures in radio wavelengths~\cite{Vallee:2002,Vallee:2011} that the structure of the magnetic field within the local supercluster plane (centered on the Virgo cluster, the direction we focus on) can be described as a collection of $\sim2~\mu$G magnetic field zones coherent over $\sim100$~kpc, adding up to the supercluster scale. If one interprets these data with a magnetic field coherent over the supercluster scale, or assumes much larger coherence lengths, one then obtains field strengths about 5--10 times smaller; such a field was used thus far for illustration.
		A similar larger-scale magnetic field is also considered in Ref.~\cite{Xu:2005rb}, which reports values for typical magnetic field strengths in the Hercules and in the Perseus-Pisces superclusters of the order of $0.3\pm0.1~\mu$G (over 800~kpc) or $0.4\pm0.2~\mu$G (over 400~kpc) from rotation measures.

		For completeness, let us simply say that little is known about cosmological-scale magnetic fields, except that they would be much weaker: a lower observational bound~\cite{Neronov:2010zz} indicates that they are $\geq 3\times10^{-16}$~G, while upper bounds~\cite{barrow:1997csm,*blasi:1999csm} give field strengths of $\lesssim10^{-9}$~G.

\section{Mixing formalism in a more general field}\label{sec:cells}

	Consider several regions with different magnetic fields, their direction and strength changing from one region to another.
	The states that define our polarisation basis no longer satisfy Eq.~\eqref{eq:polbasisoneregion} as they propagate in such a case; the evolution of their polarisations is now more general and we must consider Eq.~\eqref{eq:egen} instead. As a direct consequence of this, we can no longer solve the mixing with axion-like particles for both of these states at once. The benefit of using such a basis is still substantial however, as we can reconstruct any other polarisation state once we derive these basis solutions.

	First of all, let us work out axion-photon mixing in a arbitrarily oriented transverse magnetic field $\vec{\mathcal{B}} = \mathcal{B}\cos\delta~\vec{e}_1 + \mathcal{B}\sin\delta~\vec{e}_2$, where $\vec{e}_1$ and $\vec{e}_2$ define the orthonormal basis that we will use throughout to keep track of an absolute direction and to define Stokes parameters.

	We approximate $(\omega^2 + {\partial_z}^2) \simeq 2\omega\left(\omega + i\partial_z\right)$ in the equations of motion for the fields, as the masses we use are indeed much smaller than the photon energies entering the problem.\footnote{This simplification is similar to what is done in Ref.~\cite{Raffelt:1987im}.} Inside a region, the system of equations reads
	    \be
		    \Bigg[\Big(\omega + i\frac{\partial}{\partial z}\Big) -
	                                \left(
	                                \begin{array}{ccc}
	                                 \frac{{\omega_{\mathrm{p}}}^2}{2\omega} & 0            & \frac{-g\mathcal{B}\cos\delta}{2}\\
	                                0            &  \frac{{\omega_{\mathrm{p}}}^2}{2\omega} & \frac{-g\mathcal{B}\sin\delta}{2}\\
	                                \frac{-g\mathcal{B}\cos\delta}{2}            & \frac{-g\mathcal{B}\sin\delta}{2}  & \frac{m^2}{2\omega}
	                                \end{array} \right)\Bigg] \left(\!\! \begin{array}{c}A_{1}(z) \\ A_{2}(z) \\\phi(z)\end{array}  \!\!\right) = 0.\label{eq:eom_genplanewaves}
	    \ee
	We introduce $a_1(z)$, $a_2(z)$, and $\chi(z)$ such that we remove the $e^{i\omega z}$-dependence of the solutions:
	\be
	\left(\!\! \begin{array}{c}A_{1}(z) \\ A_{2}(z) \\\phi(z)\end{array}  \!\!\right) = \left(\!\! \begin{array}{c}a_{1}(z) \\ a_{2}(z) \\\ \chi(z)\end{array}  \!\!\right)e^{i\omega z},
	\ee
	and then rotate by $(\frac{\pi}{2}-\delta)$ to an appropriate basis $(\vec{e_{\perp}}, \vec{e_{\parallel}})$, such that $\vec{\mathcal{B}} = (0, \mathcal{B})$. Solving the equations in a way similar to the one used in Sec.~\ref{sec:solution_onezone_polarisationbasis}, and going back to the $(\vec{e_{1}}, \vec{e_{2}})$ basis, we finally obtain
	\be
		\left(\!\! \begin{array}{c}a_{1}(z) \\ a_{2}(z) \\\ \chi(z)\end{array}  \!\!\right) =
		\left(
		\begin{array}{ccc}
		K_{11}	&K_{12}	&K_{13}\\
		K_{12}	&K_{22}	&K_{23}\\
		K_{13}	&K_{23}	&K_{33}
		\end{array} \right)
		\left(\!\! \begin{array}{c}a_{1}(0) \\ a_{2}(0) \\\ \chi(0)\end{array}  \!\!\right),
	\ee
	with
	\be
		\left\{
		\begin{array}{llllll}
		K_{11} = \sin^2\delta\ e^{-i\frac{{\omega_{\mathrm{p}}}^2}{2\omega} z} + \cos^2\delta \left({(\textrm{c}\theta)}^2\ e^{-i \frac{{\mu_{C}}^2}{2\omega}z} + {(\textrm{s}\theta)}^2\ e^{-i \frac{{\mu_{D}}^2}{2\omega}z}\right)\\
		K_{22} = \cos^2\delta\ e^{-i\frac{{\omega_{\mathrm{p}}}^2}{2\omega} z} + \sin^2\delta \left({(\textrm{c}\theta)}^2\ e^{-i \frac{{\mu_{C}}^2}{2\omega}z} + {(\textrm{s}\theta)}^2\ e^{-i \frac{{\mu_{D}}^2}{2\omega}z}\right)\\
		K_{12} = -\sin\delta\cos\delta\ e^{-i\frac{{\omega_{\mathrm{p}}}^2}{2\omega} z} + \cos\delta\sin\delta \left({(\textrm{c}\theta)}^2\ e^{-i \frac{{\mu_{C}}^2}{2\omega}z} + {(\textrm{s}\theta)}^2\ e^{-i \frac{{\mu_{D}}^2}{2\omega}z}\right)\\
		K_{13} = \cos\delta\ \frac{\sin(2\theta)}{2}\left(e^{-i \frac{{\mu_{C}}^2}{2\omega}z} - e^{-i \frac{{\mu_{D}}^2}{2\omega}z}\right)\\
		K_{23} = \sin\delta\ \frac{\sin(2\theta)}{2}\left(e^{-i \frac{{\mu_{C}}^2}{2\omega}z} - e^{-i \frac{{\mu_{D}}^2}{2\omega}z}\right)\\
		K_{33} = {(\textrm{s}\theta)}^2\ e^{-i \frac{{\mu_{C}}^2}{2\omega}z} + {(\textrm{c}\theta)}^2\ e^{-i \frac{{\mu_{D}}^2}{2\omega}z},\\
		\end{array}
		\right.
	\ee
	where $\mu_C$ and $\mu_D$ are respectively $\mu_+$ and $\mu_-$ when $\omega_{\mathrm{p}}>m$, and the other way around when $m>\omega_{\mathrm{p}}$.

	When we consider light travelling through different magnetic field domains, we use this result inside each domain, ensuring the continuity of the fields at the boundaries and neglecting reflected waves, which have an amplitude of the order of $\frac{\Delta\mu^2}{\omega^2}\lesssim10^{-27}$ times the incident ones.
	Indeed, as already used to obtain Eqs.~\eqref{eq:eom_genplanewaves}, for the different states, the masses (which are roughly of the same order) are all negligible compared to $\omega$ in our problem; their momenta are therefore all essentially equal: for any two states $i$ and $j$, $k_i \sim k_j \sim \omega$ to an excellent approximation.\footnote{Of course, for the mixing, the difference is crucial.} This is also true if $i$ and $j$ label the momenta inside regions of different transverse magnetic field strengths, as $g\mathcal{B}$ and $\omega_{\mathrm{p}}$ can change from one region to the other but remain much smaller than $\omega$. In such a situation, we understand that reflected waves can be neglected, as in a simple quantum mechanical problem with 2 regions (see, \textit{e.g.} Ref.~\cite{Cohen-Tannoudji:1977}): if we write the momenta in regions $i$ and $j$ as $k_i$ and $k_j$ respectively, the reflected amplitude in such a case is given by the incident one times
	\be
		\frac{k_i - k_j}{k_i+k_j}=\frac{{k_i}^2 - {k_j}^2}{{(k_i + k_j)}^2}\sim\frac{\Delta {m_{ji}}^2}{4\omega^2},
	\ee
	where $\Delta {m_{ji}}^2$ is the difference of the square of the masses, to be compared to $\omega^2$.\footnote{The very fact that we can neglect the reflected waves actually implies that the continuity conditions for the derivatives of the fields at the boundaries can be neglected.}

	As stated earlier, for the domain-structure model, the magnetic field between domains is not only rotated in the transverse plane, as usually found in the literature, \textit{e.g.} Ref.~\cite{Mirizzi:2006zy}: instead we allow it to undergo the most general tridimensional rotation.
	\draft{When we allow an additional underlying field, we keep it in the $\vec{e}_2$ direction throughout. For each domain, we then only consider the transverse part of the associated external magnetic field and determine the angle $\delta$.}
	\draft{It is only at the end of the last domain that we compute the Stokes parameters after using Eqs.~\eqref{eq:fullypollight} and~\eqref{eq:unpollight}.}

\section{Simulations}

	As far as pseudoscalar-photon mixing is concerned in the external conditions of a supercluster, light seen through a V filter can essentially be considered as monochromatic, of frequency $\omega = 2.25$~eV. 
	We have checked that the results we obtain remain indeed stable if we use frequency profiles similar to that of the V filter or if we simply consider monochromatic waves as done in the following simulations.\footnote{While it does not make any difference for the viability of the spinless-particle scenario, note that in the next chapter, we shall use exactly the V-filter profile in our computations.}

	We now perform our simulations in more general magnetic field configurations for the Virgo supercluster.

	\subsection{Simulations in the uniform-field scenario}

	We know since Sec.~\ref{sec:toymodel} that the mixing inside a uniform magnetic field can produce coherent orientations with respect to the magnetic field direction; however, the existence of axion-like particles responsible for an alignment also implies that circular polarisation is produced and contradicts data as discussed in~Sec.~\ref{sec:vfilterKO}.
	As was done for the linear polarisation data in Sec.~\ref{sec:descriptionalignment}, let us nonetheless simply check in a $(q, u)$ space what such an alignment of linear polarisation looks like, if we start from a random distribution of polarisations.

	\begin{figure}
			\centering
			\includegraphics[width=0.65\textwidth]{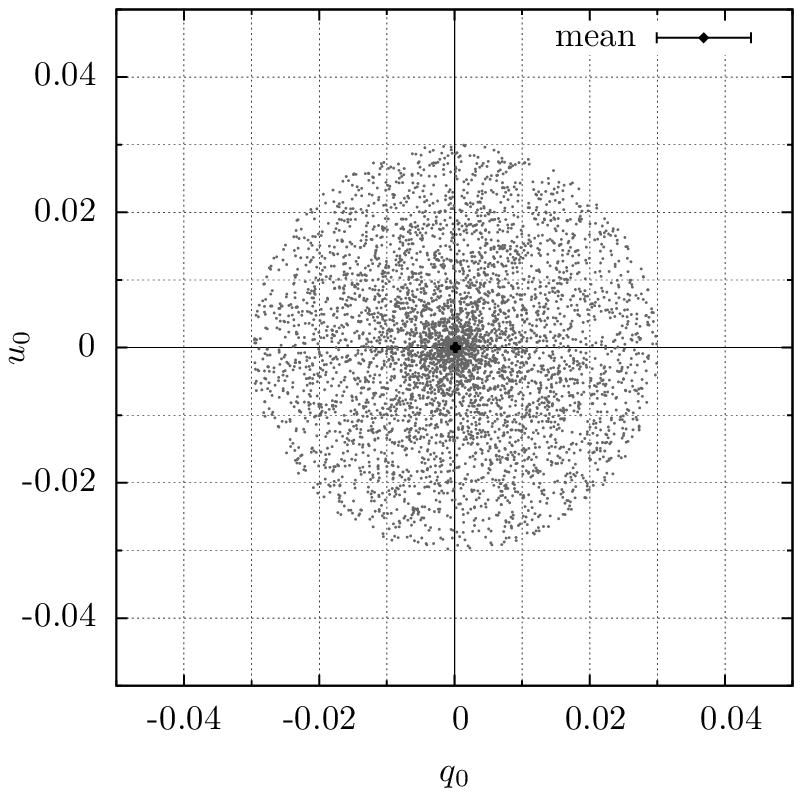}
			\includegraphics[width=0.65\textwidth]{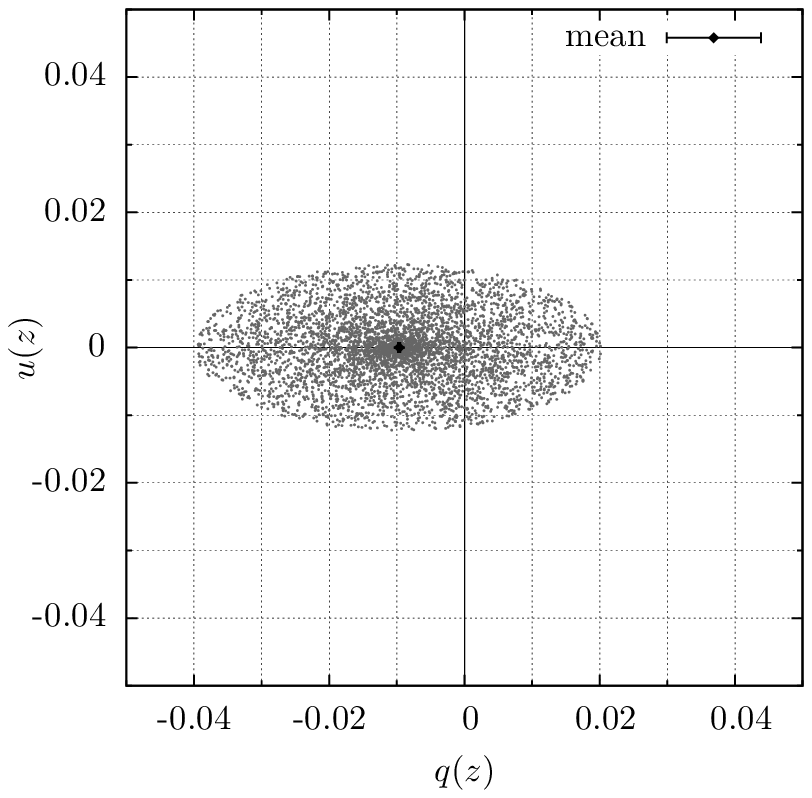}
			\caption{5000 beams are generated. \emph{Top}: Initial distribution in the $(q,u)$ space. \emph{Bottom}: The associated distribution after effects induced by axion-photon mixing in the uniform case. The parameters used here are $\omega_0=2.25$~eV, $\omega_{\mathrm{p}}=3.7\times10^{-14}$~eV, $m=4.5\times10^{-14}$~eV, $g = 3.5\times10^{-12}$~GeV$^{-1}$, $\mathcal{B}=|\vec{B}_{\mathrm{uniform}}|=0.3~\mu$G, and $z=10$~Mpc.
			}
			\label{fig:q_u_uniform}
	\end{figure}

	To mimic a random distribution of initial quasar polarisations, we first generate partially polarised light beams with $p_{\mathrm{lin}}$ randomly taken between 0 and 3\%, and with polarisation angles independently randomly generated. In Fig.~\ref{fig:q_u_uniform}, on top, we plot this initial distribution of light beams, each random realisation being displayed using its Stokes parameters $q$ and $u$.
	In the bottom panel, we show what this distribution becomes, due to axion-photon mixing inside the 10~Mpc uniform magnetic field. As expected, we see that there is indeed a departure from a random distribution acquired through the mixing, corresponding to an asymmetry in the $(q, u)$ space; see the discussion made in Sec.~\ref{sec:descriptionalignment}. The fact that the asymmetry appears along one of the axes is only due to our specific choice for the basis; only $p_{\mathrm{lin}}$ is a physical quantity, independent of the choices made by the observer.
	More quantitatively, the means that we obtain for $q$ and $u$ in this example lead to a value of $p_{\mathrm{lin}}=0.01$ after axion-photon mixing, while they were compatible with zero initially.\footnote{Note that the loss $u_0-u(z)$ that the see between the top and bottom panels directly illustrates the generation of $v(z)$ which contradicts data.}

	\subsection{Simulations in the domain-structure scenario}
	\subsubsection{Pure randomness}
	As already mentioned, making the magnetic field vary in a domain structure may suppress $v$: as we have seen in Sec.~\ref{sec:prob}, in small-enough magnetic-field regions, the induced circular polarisation can be smaller than the linear one.
Nevertheless, circular polarisation is not the main problem in this picture.

Indeed, it is obvious that such a field will not help create an alignment: if the magnetic field can be thought of as small domains with magnetic field directions distributed in a random way from one to the other, this will be the case along the line of sight, but also transversally. Then, two objects which are angularly separated will pass through two different magnetic field configurations. There is thus no way to create an alignment, as there is no preferred direction in this problem that is common to all quasars.

	\subsubsection{With an underlying uniform field}

	We can go further and use a more refined model, where there is at least some correlation between domains rather than a complete randomness. To do this, we sum the magnetic fields of the uniform and of the domain-structure models, using results presented in Sec~\ref{sec:cells}. We thus have magnetic fields with a domain-structure on top of a fainter field, this one being coherent over the LSC scale: this would lead to some semi-randomness between domains on the LSC scale, as discussed in Ref.~\cite{Vallee:2002}.
	\begin{figure}
			\centering
			\includegraphics[width=0.65\textwidth]{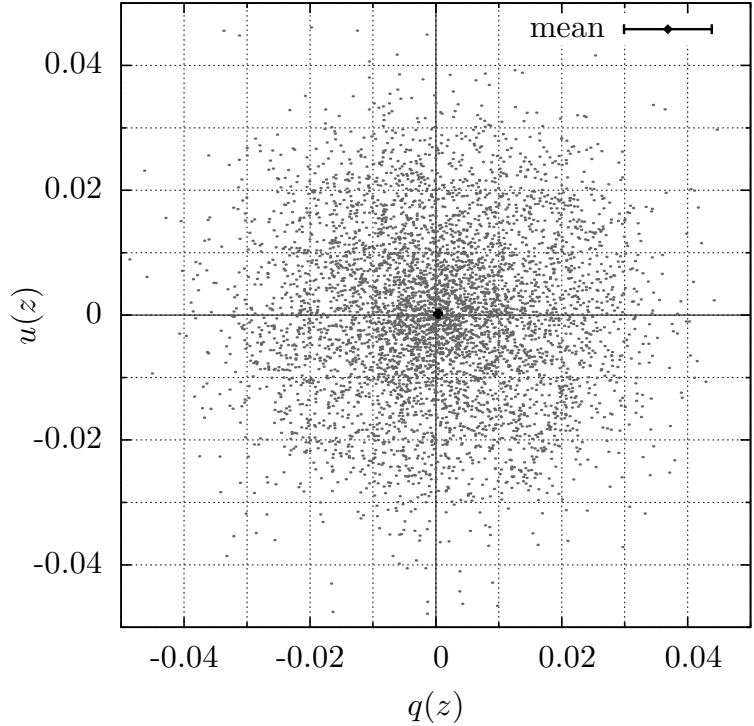}
			\caption{
			Same as Fig.~\ref{fig:q_u_uniform} (bottom panel) but after axion-photon mixing in the domain-structure case with an underlying uniform field. The parameters used here are the same as before, except $m=1.85\times10^{-13}$~eV, and the magnetic fields of course. For the uniform magnetic field, we use $|\vec{B}_{\mathrm{uniform}}|=0.3~\mu$G over 10~Mpc; for the randomly oriented part of the magnetic field, we use $|\vec{B}_{\mathrm{domain}}| = 2~\mu$G in a hundred $100~$kpc-long domains.
			}
			\label{fig:q_u_patchyplusundl}
		\end{figure}

	In this case, to keep a final linear polarisation of the order of 1\%, we have to consider either smaller values of the coupling, or bigger values of $|m^2 - {\omega_{\mathrm{p}}}^2|$ than in the uniform case because the magnetic field is stronger (see Eq.~\eqref{eq:thetamix}).

	For this reason, and because the field strength of the uniform component is $\approx7$ times smaller than that of the randomly oriented one, it is not surprising that there is no obvious departure from isotropy due to axion-photon mixing in this case. Indeed, the effect induced by the uniform component of the magnetic field is then strongly suppressed: therefore, an alignment cannot be achieved.
	For the example we present in Fig.~\ref{fig:q_u_patchyplusundl}, we obtain that the means of $q$ and $u$ are compatible with zero at the 1$\sigma$-level, namely hardly any improvement with respect to the initial distribution, and certainly not enough additional linear polarisation to be able to explain the observed alignments. 
	We further checked that if the relative intensities of the random and background fields are chosen to produce an alignment, then the circular polarisation is again always too high.
	Finally, we have also checked that our results are stable with respect to fluctuations of a factor of 2 (up or down) of the parameters, among which the plasma frequency, the domain sizes, and the magnetic field strength.

	To conclude, if this second possibility turns out to be a satisfactory model of the magnetic field of the local supercluster, not only would there be some circular polarisation, but axion-like particles will be unable to create coherent orientations in that field. Note that these results are general and do not apply only to the LSC magnetic field.
	In particular, considering another magnetic field to produce alignments would require even larger coherence scales: it should indeed be located beyond the LSC and still be large enough for light from angularly distant quasars to pass through a field giving the same preferred direction. Moreover, even if such a magnetic field exists, one should make sure that no circular polarisation is generated once photons eventually pass through the LSC.

\chapter{New constraints on axion-like particles}\label{chap:constraints}

	If spinless particles cannot explain the large-scale alignments of quasar polarisations, these data can be exploited to constrain the parameters~\cite{Payez:2012vf,Payez:2011mk}. We shall not discuss here the coherent alignment effect further, but instead derive new constraints for low-mass ALPs.
	Similar polarimetric constraints on these particles, based on linear polarisation, have been discussed in the literature in various contexts.
	For instance, the authors of Ref.~\cite{Gill:2011yp} consider the influence of the mixing in magnetic white dwarfs on polarisation, and give $g\lesssim10^{-11}$~GeV$^{-1}$ for $m\lesssim 10^{-7}$~eV.\footnote{Should the strongest magnetic field strengths ever reported for two magnetic white dwarfs be confirmed, the whole astrophysical region discussed in the introduction of this thesis might then be excluded, as the limit for nearly massless ALPs would then become $g\lesssim5\times10^{-13}$~GeV$^{-1}$ according to Ref.~\cite{Gill:2011yp}.}
	Another recent example is Ref.~\cite{Horns:2012pp}, in which the authors consider the effect of the mixing in a cosmological-scale magnetic field of strength $B_{\mathrm{csm}}$ on the polarisation angle of ultraviolet photons, and obtain $g\left(\frac{B_{\mathrm{csm}}}{1~\textrm{nG}}\right)\lesssim 10^{-11}$~GeV$^{-1}$ for $m\lesssim 10^{-15}$~eV.

	Being as conservative as possible and using present quasar polarisation data in visible light, we show in this chapter that one can derive new constraints on ALPs using supercluster magnetic fields, and narrow down the parameter-space region of astrophysical interest.
	The change of polarisation induced by the mixing is indeed a very specific prediction, and this is especially true for circular polarisation. Depending on the mass and coupling of the ALPs, the mixing with light in external magnetic fields can be very efficient and contradict observations, as there is indeed little room for a modification of polarisation.

	Again, we are going to consider objects located behind the Virgo supercluster, as we have information on $p_{\mathrm{lin}}$ and $p_{\mathrm{circ}}$ in this direction. We then consider pseudoscalar-photon mixing in the magnetic field of that supercluster, and check for which parameters the generated polarisation would remain consistent with the observations.

	\section{Optical polarisation data}

		\subsection{Properties of the full sample}

			In the sample of 355 quasar polarisations discussed in Chap.~\ref{chap:alignments}, only objects with a polarisation degree $p_{\mathrm{lin}} \ge 0.6\%$ and $p_{\mathrm{lin}} / \sigma_{p_{\mathrm{lin}}} \ge 2$ were considered. In the present chapter, we extend the sample to include low-polarisation objects with $p_{\mathrm{lin}} < 0.6\%$ measured with uncertainties $\sigma_{p_{\mathrm{lin}}} \le 0.3\%$. Since $p_{\mathrm{lin}}$ is positive definite, it is biased at low signal-to-noise ratio. A reasonably good estimator of the true (debiased) polarisation degree is computed using $\tilde{p}_{\mathrm{lin}} = (p^2_{{\mathrm{lin}}} - \sigma^2_{p_{\mathrm{lin}}})^{\frac{1}{2}}$ when $p_{\mathrm{lin}} > \sigma_{p_{\mathrm{lin}}}$, and $\tilde{p}_{\mathrm{lin}} = 0$ when $p_{\mathrm{lin}} \le \sigma_{p_{\mathrm{lin}}}$~\cite{Simmons:1985ub}. In the rest of this chapter, we only consider $\tilde{p}_{\mathrm{lin}}$ when we discuss the linear polarisation. As for circular polarisation in visible light, which has rarely been measured, our starting point is the compilation presented in Ref.~\cite{Hutsemekers:2010fw} together with new V-filter data, that we have discussed in Sec.~\ref{sec:prob}.

		\subsection{Subsample and criteria used to obtain constraints}\label{sec:criteria}

			\begin{figure}
				\centering
				\includegraphics[width=.8\textwidth]{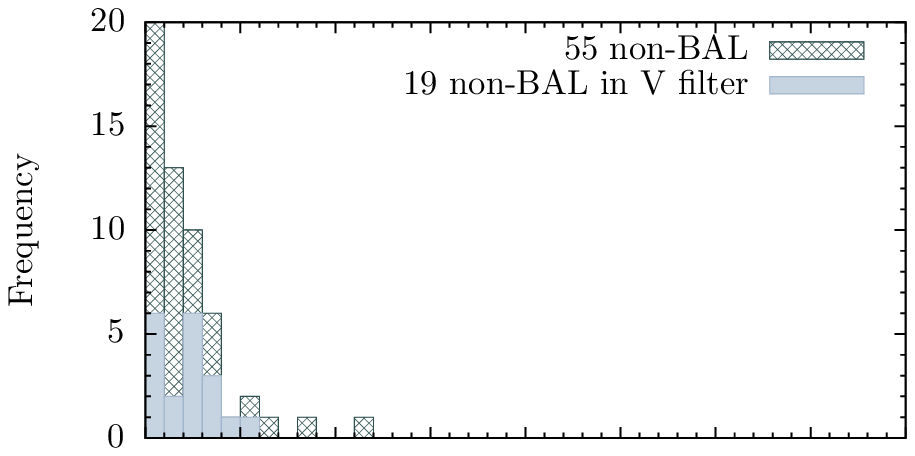}
				\includegraphics[width=.8\textwidth]{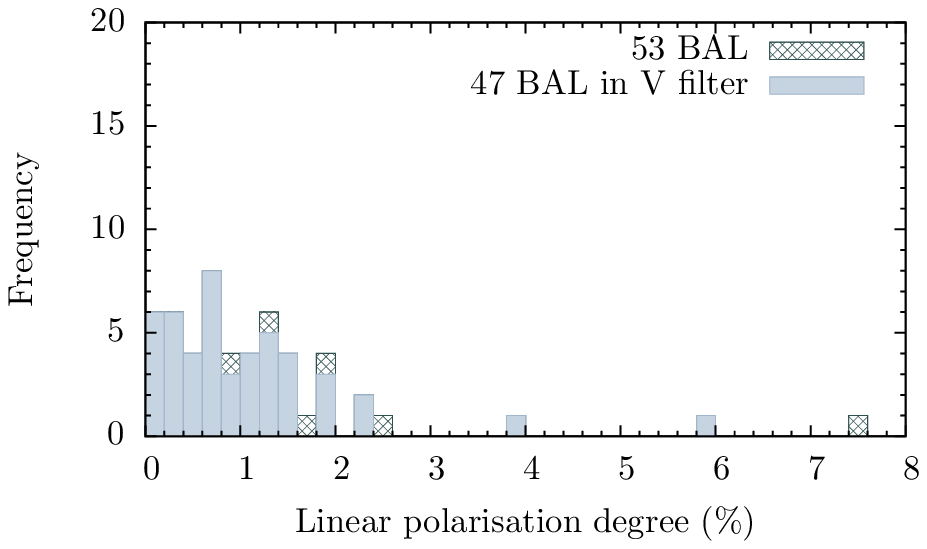}
				\caption{Comparison of the distributions of debiased linear polarisation between non-BAL quasars (\emph{top}) and BAL quasars (\emph{bottom}), for objects taken in the direction of region A1. Clearly, BAL quasars are often more linearly polarised than quasars without broad absorption lines.}
				\label{fig:balnonbal}
			\end{figure}

			Since our goal is to derive constraints from the polarisation induced by axion-like particles, we restrict our sample to spectroscopically defined classes of quasars known to have the smallest intrinsic polarisations.

			We therefore discard the following objects from the sample of quasars with linear polarisation measurements: known radio-loud quasars (\textit{i.e.} classified as such in the NASA/IPAC Extragalactic Database (NED)~\cite{NED} or objects from Refs.~\cite{Impey:1990qz,Impey:1991qz,Wills:1992qz,Visvanathan:1998qz}), near-infrared (2MASS) selected objects, in particular those from Ref.~\cite{Smith:2002qz}, and Broad-Absorption-Line (BAL) quasars (classified as such in NED). A comparison of non-BAL measurements to the polarisation distribution of BAL quasars is shown in Fig.~\ref{fig:balnonbal}, demonstrating that differences are significant; they are also preserved by any mechanism possibly affecting light on the line-of-sight~\cite{Hutsemekers:2001}.

			Given that the observation of null circular polarisation is a much more stringent constraint for axion-like particles in V filter than in white light, we first restrict our sample to these measurements. Then, again, as we restrict ourselves to classes of quasars with the smallest polarisations, we then discard the two spectroscopically identified BL Lac objects, more prone to be intrinsically strongly polarised.

			With these criteria, the reduced sample then mostly consists of radio-quiet / optically selected non-BAL quasars. Such objects are known to be linearly polarised at most at the 1\% level~\cite{Stockman:1984qz,Berriman:1990qz}.

\begin{figure}
				\centering
				\includegraphics[width=.8\textwidth]{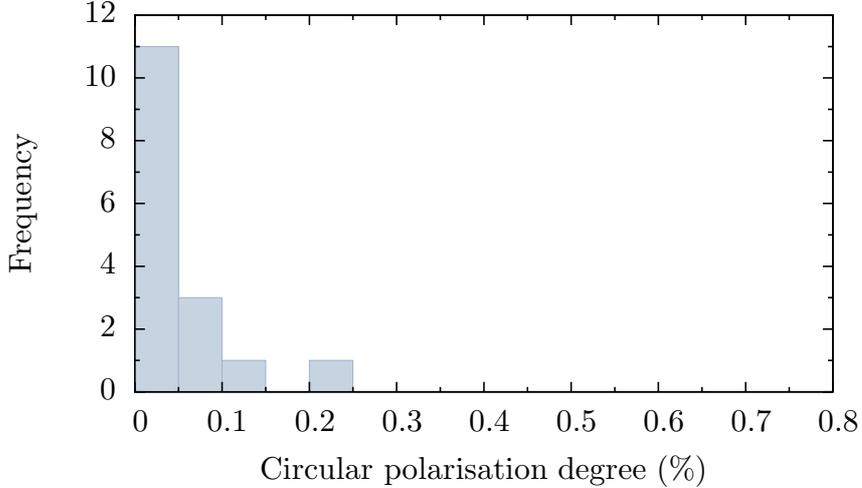}
				\caption{Distribution of the degree of circular polarisation for the central values of the 16 quasars in the direction of region A1, measured in the V filter. Note in particular the different scale, compared to the linear polarisation shown in Fig.~\ref{fig:balnonbal}.}
				\label{fig:pcirc}
			\end{figure}

			We then focus on quasars located in the direction in which most of the V-filter circular polarisation data were taken, \textit{i.e.} towards the center of the (local) Virgo supercluster. Keeping only objects with right ascensions between 168\textdegree{} and 218\textdegree, we are left with a final sample of 55 quasars with measured linear polarisation in white light or in V filter and of 16 quasars with measured circular polarisation in V filter (all compatible with zero at $3\sigma$). The distribution of the circular polarisation degree $p_{\mathrm{circ}}$ associated with the central values of the 16 circular polarisation measurements is shown in Fig.~\ref{fig:pcirc}.

			We know that quasars intrinsically emit polarised light: otherwise we would simply not see a difference in the distribution of degrees of linear polarisation depending on the spectroscopic type. However, we cannot access the initial distribution of polarisation. For this reason, in order to avoid any overestimation of the final polarisation generated by the mixing, we work with initially unpolarised light and conservatively allow the observed linear polarisation of the quasars to be only due to the interaction with ALPs.
			Under this hypothesis, one can impose that the linear polarisation generated by the ALP-photon mixing model does not exceed the observed one.

			The idea is therefore to calculate \draftun{for various values of the parameters} the polarisation predicted by the mixing with ALPs inside external magnetic fields encountered on the way to Earth, and to compare it with data.
			For both the linear and the circular polarisations, we thus compute the probability that the polarisation ($p^{\textrm{th}}$) due to ALPs is smaller than the observed one ($p^{\mathrm{obs}}$) as:
			\be
			P = N (p^{\mathrm{obs}} \geq p^{\textrm{th}}) / N_{\mathrm{total}},\label{eq:proba}
			\ee
			by taking the ratio of the number of measurements in bins with a polarisation higher than $p^{\textrm{th}}$, to the total number of measurements $N_{\mathrm{total}}$.

	\section{Minimal constraints}

	\subsection{Model of the intrasupercluster medium}\label{sec:modelfluct}

			We use the domain-structure magnetic field from Sec.~\ref{sec:struct}, and additionally consider fluctuations from one domain to the other of both the electron density and of the domain size.
			More precisely, we let these quantities fluctuate by 50\% up or down.
			Note that, as we allow for fluctuations of domain sizes, only the total distance $z_{\mathrm{tot}}$ is well-defined from one magnetic field realisation to another, not the number of domains, as we stop adding domains as soon as we reach $z_{\mathrm{tot}}$.
			Let us also remind that we allow for 3D rotations of the magnetic field from a domain to the other, so that the transverse magnetic field strength $\mathcal{B}$ which enters in the mixing can take values much smaller than $|\vec{B}_{\mathrm{domain}}|$.

			Finally, note that the size of the supercluster ($\sim$ 20~Mpc) is small enough to neglect cosmological redshift. As the mixing is more efficient at higher frequencies (see Eq.~\eqref{eq:thetamix}), redshift effects could only increase the polarisation produced in any case relevant to this study and make our bounds stronger.
			Moreover, as we only take into account the influence of the last magnetic field region, we again underestimate the amount of polarisation due to ALP-photon mixing.
			We thus stress that our constraints are conservative.

	\subsection{Method}\label{sec:method}

		Our constraints on ALPs are derived iteratively for different points in the $(m,g)$ parameter space: a given couple of parameters corresponds to a given hypothetical axion-like particle. For fixed ALP parameters, we generate random configurations, defined as the set of transverse field strengths and directions, of domain sizes, and of electron densities.
		For a given value of the frequency $\omega$, we have given the solutions for the two electric fields of the radiation after a propagation in each domain in such a configuration in Sec.~\ref{sec:cells}.
		We use initially unpolarised light beams (as discussed in Sec.~\ref{sec:criteria}), calculate the four Stokes parameters at the end of the last magnetic field domain, and compare the final polarisation with data.

		Now, as explained in Chap.~\ref{chap:circular}, the expected circular polarisation due to ALP-photon mixing can be reduced when one considers wave packets, or averages over frequencies. This effect, though typically small in V filter, is taken into account when deriving our limits.

\begin{figure}
	\centering
	\includegraphics[width=.68\textwidth]{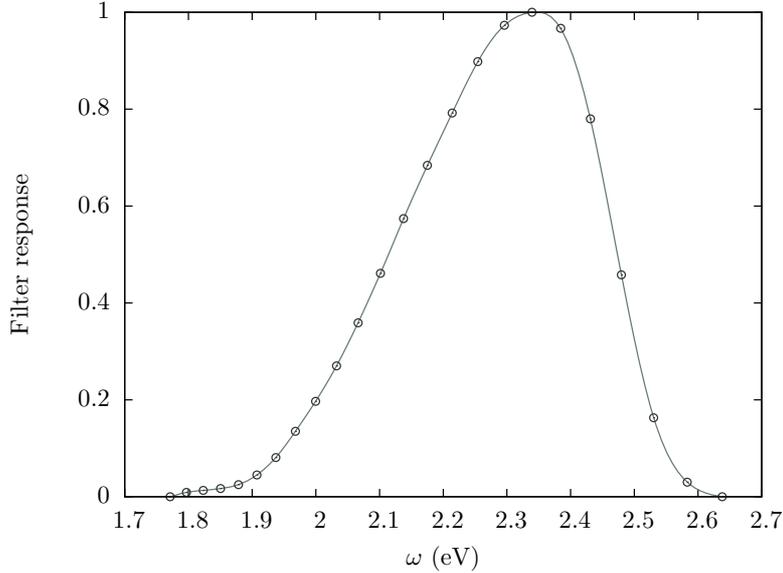}
	\caption{Interpolation of the filter response of the Bessell V filter as a function of $\omega$; we use the values reported in Ref.~\cite{Bessell:1990hs} and calculate the interpolated profile, using splines~\cite{gsl}, in our integration routine.}
	\label{fig:Vfilter_interpolation}
\end{figure}

		To implement it, for a single light beam, we completely generate a random configuration. We then integrate over $\omega$ the monochromatic results for the Stokes parameters at the end of the last zone of magnetic field, with the spectral response of the V filter as a weighting function (Fig.~\ref{fig:Vfilter_interpolation}). This corresponds to one iteration: it gives the values of the Stokes parameters through the V filter for a given source at the end of one configuration.

		After that, we compare the generated linear and circular polarisations with the data histograms\footnote{The results are not affected by the presence of white-light data for linear polarisation.}, associate to each polarisation a probability given by Eq.~\eqref{eq:proba}, respectively $P^{(\mathrm{lin})}$ and $P^{(\mathrm{circ})}$, and obtain the final probability $\mathcal{P}$ to be compatible with data for this particular configuration by multiplying these two individual probabilities. We then repeat the procedure with a new random configuration.

		We do this many times (5000 times) to minimise the influence of statistical fluctuations, and average the value of $\mathcal{P}$ over the configurations as we do not have access to the actual configuration for each measurement. We thus integrate it out and give the average probability for a given ALP $(m,g)$ not to exceed the observed polarisation in a random configuration. Then we start over for a new couple $(m,g)$.

		Finally, we summarise this information by saying that parameters are ``excluded at 1$\sigma$'', when the average probability for this particle to produce too much polarisation compared with data is 68.3\%; and similarly for 2$\sigma$ (95.5\%) and 3$\sigma$ (99.7\%).
		In Appendix~\ref{app:distributions}, we show two theoretical distributions of polarisation, associated with points excluded at 2$\sigma$ and 3$\sigma$; let us stress that this is only for illustration, as the technique described above to obtain constraints does not directly involve such distributions.

	\subsection{Results}

		\begin{figure}[h]
			\centering
			\includegraphics[width=\textwidth]{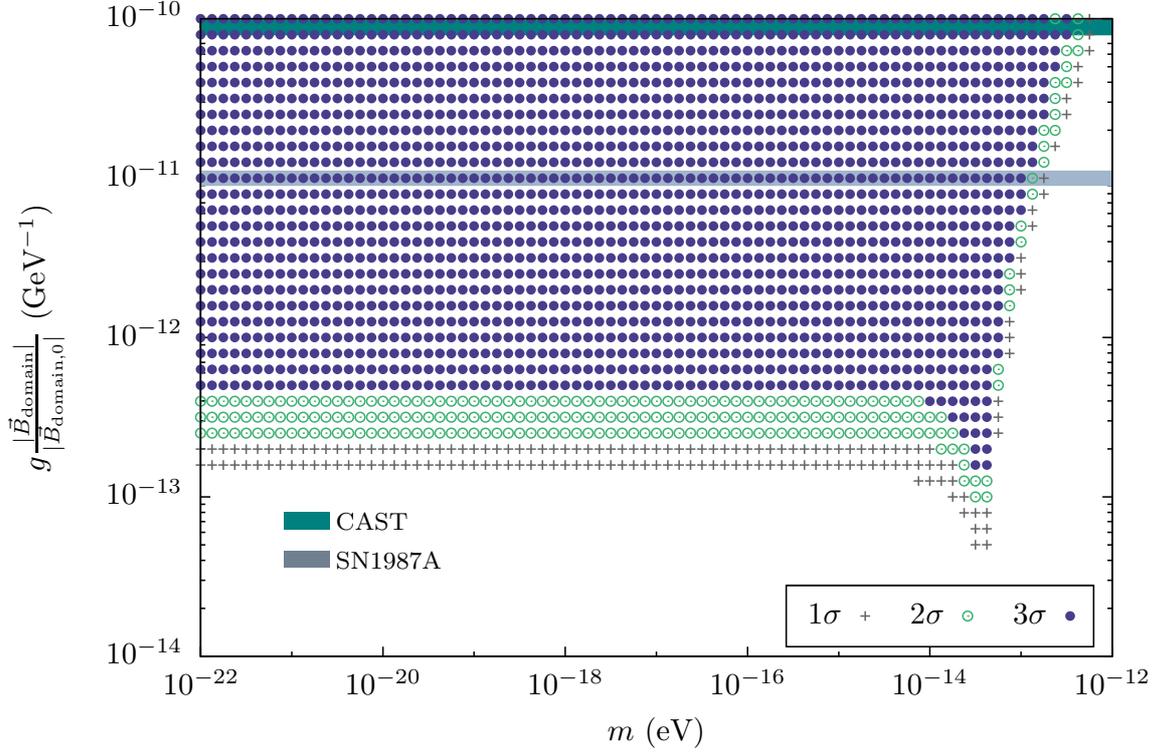}
			\caption{Exclusion plot. The total distance considered here is $z_{\mathrm{tot}}=10$~Mpc; the average domain size is 100~kpc; the norm of the total magnetic field in each domain is $|\vec{B}_{\mathrm{domain,0}}|=2~\mu$G; the average electron density is $n_{\mathrm{e}}=10^{-6}$~cm$^{-3}$ (corresponding to a plasma frequency $\omega_{\mathrm{p}}=3.7\times10^{-14}$~eV).
			}
			\label{fig:exclusion}
		\end{figure}

		The results we obtain are shown in Fig.~\ref{fig:exclusion}. The upper limits obtained by CAST and from the energy-loss considerations associated with SN1987A are also shown. 

		Let us first stress a very important point: as there is only one physical magnetic field scale in the problem, namely the total field strength inside a domain $|\vec{B}_{\mathrm{domain}}|$, this exclusion plot remains the same for any other value of the magnetic field inside the domains. Indeed, as stressed in Sec.~\ref{sec:mixing}, only the product of the magnetic field with the coupling constant $g$ enters the equations. Therefore, while there are uncertainties about the magnetic field strength, as discussed in Sec.~\ref{sec:Bstrength}, the exclusion limit on $g$ can be rescaled once our knowledge of the magnetic field inside the supercluster improves.

		In Fig.~\ref{fig:exclusion}, we observe two features: a dip and a plateau for small ALP masses. Both can be understood from the expression of the mixing angle, Eq.~\eqref{eq:thetamix}, and of the oscillating phase, Eq.~\eqref{eq:deltamu2z_omega}. The position of the dip is determined by $\omega_{\mathrm{p}}$; for ALP masses very close to the plasma frequency, the mixing is maximal, $\theta\approx\pi/4$, and so is its effect on (linear) polarisation. The value of the bound then depends on the phase, \textit{i.e.} on $\mathcal{B}$ and on the distance travelled, and is thus model-dependent.
		As for the plateau, for ALP masses much smaller than the plasma frequency, the mixing angle is essentially independent of $m$:
		\be
			\theta\stackrel{(m\lll\omega_{\mathrm{p}})}{=}\frac{1}{2} \textrm{atan}\left(\frac{2g\mathcal{B}\omega}{- {\omega_{\mathrm{p}}}^2}\right),
		\ee
		and the oscillations due to the phase~\eqref{eq:deltamu2z_omega} can be averaged.
		Indeed, for a travelled distance inside a domain $L\approx100~$kpc, $\mathcal{B}\lesssim2~\mu$G, and $\omega_{\mathrm{p}}\approx3.7\times10^{-14}~$eV, one obtains that the oscillations of the Stokes parameters are of short wavelength compared with $L$ if $g\mathcal{B}\omega\ll {\omega_{\mathrm{p}}}^2$, \textit{i.e.} when $g\ll10^{-11}~$GeV$^{-1}$.
		Hence, in that case, the plateau is rather stable and the exclusion band can be extended to massless ALPs. Furthermore, the fact that higher values of the coupling $g$ are more constrained is natural, as the mixing is more efficient and, thus, more polarisation is produced.
		
		Here, for the average electron density inside the supercluster, we have used $n_{\mathrm{e}} = 10^{-6}$~cm$^{-3}$ (which gives a plasma frequency $\omega_{\mathrm{p}}=3.7\times10^{-14}$~eV), as in previous works related to ALP-photon mixing in similar conditions, \textit{e.g.} Refs.~\cite{Das:2004qka,Burrage:2008ii}.
		Nonetheless, the properties of the intrasupercluster medium are not well-known, including the value of the average electron density.
		Searches for gas in the Shapley supercluster~\cite{Molnar:1998sc} have bounded $n_{\mathrm{e}}$ to be less than $5\times10^{-6}$~cm$^{-3}$, and a subsequent tentative detection, for the local supercluster~\cite{Boughn:1999sc}, gave $n_{\mathrm{e}}\approx2.5\times10^{-6}$~cm$^{-3}$. More recent gas-dynamics and N-body simulations of the local supercluster are also available in the direction we are looking at (\textit{i.e.} at galactic latitudes $b_{\mathrm{gal}}>30\degr$, extending to distances of about 30~Mpc around the LSC center). They lead to values for the plasma frequency between $\omega_{\mathrm{p}}\approx2.3\times10^{-14}$~eV and $\omega_{\mathrm{p}}\approx5.5\times10^{-14}$~eV for gas overdensities between $\delta_{\mathrm{g}}=1$ and $\delta_{\mathrm{g}}=10$, corresponding to filamentary structures such as superclusters~\cite{Kravtsov:2002ac}; the authors define the overdensity as
		\be
			\delta_{\mathrm{g}} = \frac{\rho_{\mathrm{g}}}{\bar{\rho_{\mathrm{g}}}},
		\ee
		where $\rho_{\mathrm{g}}$ and $\bar{\rho_{\mathrm{g}}}$ are respectively the gas density and its average value.

		Now, if we take an extreme case and allow the intrasupercluster electron density to be one order of magnitude larger on average than what we have considered in Fig.~\ref{fig:exclusion} (\textit{i.e.} allow fluctuations, see Sec.~\ref{sec:modelfluct}, up to values as large as $n_{\mathrm{e}}=15\times10^{-6}$~cm$^{-3}$), we have checked that, while our limits would then change, they would still improve the current bounds on ALPs. For smaller values of the electron density $n_{\mathrm{e}}<10^{-6}$~cm$^{-3}$, our limits are stable, and would in fact be slightly more stringent.

		We have checked the stability of our constraints under the change of the total magnetic field size $z_{\mathrm{tot}}$. For total magnetic field sizes of 5~Mpc and 20~Mpc, the exclusion plot obtained essentially does not change its shape compared to Fig.~\ref{fig:exclusion}, but is shifted along the $y$-axis. For instance, the $2\sigma$-limit we obtain for nearly massless ALPs is $g\lesssim2.5\times10^{-13}$~GeV$^{-1}$ for $z_{\mathrm{tot}}=10$~Mpc, and becomes $g\lesssim3.2\times10^{-13}$~GeV$^{-1}$ and $g\lesssim2\times10^{-13}$~GeV$^{-1}$ for $z_{\mathrm{tot}}=5$~Mpc and $z_{\mathrm{tot}}=20$~Mpc respectively. Our limits improve the best bounds to date.

		We have also checked that our constraints are stable: a) under the addition of a uniform background field (typically weaker: $|\vec{B}_{\mathrm{uniform}}|\sim0.4~\mu$G), which would allow some correlation between domains over the supercluster scale;\footnote{\draft{As shown in Appendix~\ref{app:constraints}, the limits we obtain in that case} are indistinguishable from those of Fig.~\ref{fig:exclusion}\hspace{1pt}---\hspace{1pt}but then, of course, one cannot rescale the limits derived for $|\vec{B}_{\mathrm{domain}}|$ and $|\vec{B}_{\mathrm{uniform}}|$ changed independently.} b) if we use the histograms of linear polarisation for non-BAL quasars measured in white light and V filter, or exclusively in V filter to define our probability; c) if we include the measurements of circular polarisation for the two BL Lac objects; d) if we do not allow for fluctuations of the size of the domains and of the electron density between domains; e) if we use slightly different statistical criteria, as we have done in Ref.~\cite{Payez:2011mk}.

		Note that the limits obtained here are much stronger (by about a factor 50 on $|\vec{B}_{\mathrm{domain}}|$) than those one would obtain using a constant supercluster magnetic field, without any fluctuations; \draft{see Appendix~\ref{app:constraints}}. However, in that case, the limits become very dependent on the assumption one makes about the initial polarisation, as assuming an unpolarised emission from quasars excludes the creation of circular polarisation.

\begin{introccl}

	\chapter{Conclusion}

\section{What have we learned?}

	We have investigated what was so far the best hope to explain the existence of correlations of optical quasar polarisations over cosmological scales, and have shown that this scenario\hspace{1pt}---\hspace{1pt}involving the mixing with new light spinless particles\hspace{1pt}---\hspace{1pt}is highly disfavoured by current data.

	One of the major issues for that spinless-particle scenario is that the phenomenology expected from the mixing is in severe contradiction with the reported measurements of vanishing circular polarisation for objects from the quasar sample: the creation of a sizeable amount of circular polarisation via birefringence is indeed one of its key predictions, along with the modification of linear polarisation via dichroism that would have generated the alignments in faint extended magnetic fields.
	For white unfiltered light, we have shown that this was not necessarily an issue as, compared to the simpler monochromatic description, the predicted amount is reduced when averaging over the frequency, as in the wave-packet treatment that we have proposed.
	However, as we have discussed, the problem is rather that no circular polarisation has been found when observing through a Bessell broadband V filter, which is not broad enough to average the circular polarisation to almost zero.

	We have carefully verified the stability of our conclusion via analytical and numerical studies, including a wave-packet treatment, the use of more general magnetic field morphologies, as well as the influence of fluctuations among domains.
	In general, there is either too much circular polarisation, too much linear polarisation, or no alignment: the production of alignments of linear polarisation, while keeping the circular polarisation small, fails already at the qualitative level.
	This is what we obtain when we take into account constraints from good-quality linear and circular polarisation measurements, current estimates of magnetic field strengths and structures from the literature, and existing limits on pseudoscalar-photon couplings.

	We have then extended previous proposals to use polarisation data to constrain the parameters of axion-like particles by using a realistic magnetic field and an average over frequencies. We have presented constraints derived simultaneously from linear and circular polarisation measurements, sticking to a very conservative approach where we allowed the observed polarisation of non-BAL quasars to be entirely due to the mixing with these particles.
	Using reported properties of the intrasupercluster medium, we have shown that current bounds on nearly massless pseudoscalars are improved.
	Indeed, for a large portion of the parameter space region of astrophysical interest, the polarisation produced by the mixing with such particles in this medium would be too large.

	We have kept a conservative method throughout, so that the constraints that we have presented are the minimal ones; we have also checked that they are robust. The weak point of the method comes of course from the observational uncertainties on the electron density and on the magnetic field. As we emphasised, our constraints can be straightforwardly rescaled for any other value of the magnetic field strength, once this knowledge is improved. For the time being, one might also consider a scenario in which the maximum transverse field strength in one domain $|\vec{B}_{\mathrm{domain}}|$ is at most equal to the average magnetic field strength reported over the supercluster scale, namely 5 to 10 times weaker. Doing so would still improve current bounds on ALPs.

	Throughout, we have focused our discussion on the pseudoscalar case but, as we emphasised in the introduction of this manuscript, similar results would also hold for scalar particles. 

	\section{Outlooks}

	Both for light axion-like particles and for the large-scale alignment effect, it would certainly be worth investigating what polarisation has to tell us at other wavelengths.

\subsection{Pursuing the search for axion-like particles}
	First of all, measuring X-ray polarisation is highly promising, in particular for the search for axion-like particles.
	These particles are of course still interesting: a part of the astrophysical window that we have discussed in the introduction remains indeed intact ($m\sim 10^{-13}$--$10^{-10}$~eV and $g\lesssim10^{-11}~$GeV$^{-1}$) and could be explored with photons of higher energies, which might lead to a detection.
	Moreover, if no signal from these elusive particles is found with polarimetry, our method could in principle be adapted to set new constraints, replacing the supercluster by galaxy clusters for instance.
	X-rays are needed for this as an effect in visible light inside clusters would be suppressed except at the resonance: the plasma frequency is indeed much higher in such an environment.
	As electron densities and magnetic fields are better determined inside galaxy clusters, this would lead to more reliable constraints, could confirm our bound, and also extend it to higher values of the mass.
	Along with other authors, see for instance Refs.~\cite{Bassan:2010ya,Mena:2011xj,Gill:2011yp}, we therefore stress that the polarisation properties of very-high-energy photons is certainly a promising tool to search for ALPs.

\subsection{Towards an understanding of the alignment effect}

	Now, regarding the existence of large-scale alignments of quasar polarisations in visible light, no satisfactory explanation seems to be available to date in the literature. 

	\subsubsection*{Is there a similar effect in radio waves?}

	In fact, one of the reasons that have severely disfavoured most of the alternatives to the spinless-particle scenario is the reported absence of similar alignments in radio waves~\cite{Joshi:2007yf}.
	Recently however, the authors of Ref.~\cite{Tiwari:2012rr} have claimed that they have found extremely significant alignments at these wavelengths (reporting 5$\sigma$) for different cuts.\footnote{They also mentioned that one should determine whether the mixing with ALPs in magnetic fields might be related to their result. We discuss this in Appendix~\ref{app:radio} and explain why it cannot be a signature of such particles.}
	The situation in radio waves should therefore definitely be clarified with a detailed study; if this effect is confirmed, it would be extremely interesting to know both its general characteristics and its behaviour with redshift as this would bring valuable information to understand what causes such alignments of polarisation. 

	\subsubsection*{Is there an alignment for type-2 quasars?}

	Active galactic nuclei are understood as all related to the same physical mechanism: the accretion of matter onto a supermassive black hole. It is believed that the different classes of AGN correspond to the same kind of objects with various central luminosities, and their other individual properties would then reflect the fact that we observe them with different viewing angles~\cite{Antonucci:1993sg,Urry:1995mg}.

	In particular, type-1 quasars are high-luminosity AGN thought to be observed face on, so that most of the light emitted from their center goes straight towards us, with only little absorption or scattering by the dust torus. This would be compatible with their small linear polarisation degrees ($\sim 1\%$) and the high luminosity observed. On the other hand, type-2 quasars are thought to be observed with the dust torus edge on. While being intrinsically just as luminous as their type-1 counterparts, their optical continuum is heavily obscured as their central region is hidden from us; a significant part of the flux that we observe from them comes in the form of light scattered on dust, hence their higher linear polarisation degrees ($\sim 10\%$).

	In fact, the latest sample of good-quality quasar polarisations used to characterise the alignment effect contains only measurements of type-1 quasars~\cite{Hutsemekers:2005}. This is both because \textit{bona-fide} type-2 quasars have only been identified quite recently (see, \textit{e.g.} Refs.~\cite{Zakamska:2003nj,Zakamska:2004gv} and references therein) and because, during the dedicated observational campaigns, objects have been selected according to their apparent magnitude in the optical continuum, bright sources being preferred~\cite{Hutsemekers:2001}.

	As type-2 quasars are intrinsically identical to type-1 quasars in the unification model, it would be very instructive to know if the polarisation of visible light from type-2 quasars also appears aligned over large scales.
	Indeed, a mechanism acting on quasars themselves and leading to coherent orientation of quasar axes would be expected to affect these objects independently of their orientation with respect to the line of sight: if this is the case, the polarisation vectors should be aligned for type-1 and type-2 alike, irrespectively of their polarisation degree.
	The same could be said about mechanisms that would act during the propagation of light and change only the polarisation angle independently of the polarisation degree; in that case, this would actually hold true even if type-1 and type-2 quasars turn out to be intrinsically different.
	On the contrary, in mechanisms leading to a small systematic additional linear polarisation for each source, as we had in the spinless-particle scenario, one would not expect the polarisation of type-2 objects to be aligned.

	\subsubsection*{Evidence for a large quasar group towards Virgo}

	Finally, we note that an extremely large elongated structure involving quasars has been identified very recently at $z\sim1.3$ towards Virgo~\cite{Clowes:2012pn}. The authors report that its longest extension is larger than 1~Gpc.

	While this might very well have no connection whatsoever with the effect we have discussed, the existence of such structures is nonetheless very interesting, as one of the surprising properties of the alignment effect is that it is observed over huge scales, among seemingly unrelated quasars. Its location inside one of the most significant regions of alignments is also quite striking, and it would certainly be worth investigating whether the polarisations of the objects in this large quasar group display the same kind of alignments as those observed in the rest of region A1.

	Note however that this group of quasars, with right ascensions essentially between 160$\degr$ and 170$\degr$, is situated on the outskirts of that region; compare with Fig.~\ref{fig:align_nonlocal}. In particular, it does not overlap with region A1+, from where comes most of the significance of the effect, in the center of region A1.

\end{introccl}

\begin{appendices}

\chapter{Notations and conventions}\label{app:notations}

	\section*{Units}
	In this thesis, we make use of the so-called ``system of natural units,'' where
            \be
            \left\{
                \begin{array}{ll}
                    \textrm{the speed of light in vacuum}\ c=1,\\
                    \textrm{the reduced Planck constant}\ \hbar=1;
                \end{array} \right.
            \ee
	namely, we choose to express velocities in units of $c$, and actions and angular momenta in units of $\hbar$. As we set $c=\hbar=1$ everywhere, this let us choose in what units we want to express other physical quantities; indeed, we then have
        \be
		{[\textrm{energy}]} = {[\textrm{mass}]} = {[\textrm{time}]}^{-1} = {[\textrm{length}]}^{-1} = \dots
	\ee
	We also use the Heaviside--Lorentz system of electromagnetic units throughout, meaning for instance that our definition of the fine-structure constant is
	\be
		\alpha \equiv \frac{e^2}{4\pi} \approx \frac{1}{137}.
	\ee
	In particular, this implies that the conversion of one Tesla in {eV}$^{2}$ is as follows:
	\begin{align*}
		\textrm{1 eV}^2e^{-1}(\hbar c^2)^{-1} 
		&\approx
			\frac{(1.6\times10^{-19}\ \textrm{kg}\ \textrm{m}^2\ \textrm{s}^{-2})^2}
			{(1.6\times10^{-19}\ \textrm{C})\times(10^{-34}\ \textrm{kg}\ \textrm{m}^2\ \textrm{s}^{-1})\times{(3\times10^8\ \textrm{m}\ \textrm{s}^{-1})}^2}\\
		&\approx 1.78\times10^{-2}\ \textrm{T},
	\end{align*}
	leading to
	\be
	\textrm{1 T }\approx \frac{1}{1.78\times10^{-2} \times \sqrt{\frac{4\pi}{137}}}\ \textrm{eV}^2.
	\ee

	\section*{Special relativity}
	The convention we follow for the components of the metric tensor in Minkowski space is
        \be
            (\eta_{\mu \nu}) = \left(
                \begin{array}{cccc}
                1 & 0 & 0 & 0 \\
                0 & -1 & 0 & 0 \\
                0 & 0 & -1 & 0 \\
                0 & 0 & 0 & -1
                \end{array} \right);
        \ee
	Greek indices represent space-time components: $\mu, \nu = 0,1,2,3$ or $t,x,y,z$. 

	A given contravariant vector $u$ has components
	\be
		u^{\mu} = (u^0, u^1, u^2, u^3) = (u^0, \vec{u}),
	\ee
	with $\vec{u}$ its spatial part. Now, using the convention that repeated indices are summed over, the components of the associated covariant vector are then obtained via:
	\be
		u_{\mu} = \eta_{\mu\nu}u^{\nu} = (u^0, -\vec{u}),
	\ee
	and the scalar product of two vectors is defined as:
	\be
		u_{\mu}v^{\mu}=\eta_{\mu\nu}u^{\nu}v^{\mu}.
	\ee

	Given the coordinates $x^{\mu}$ in Minkowski space, one introduces the operators
	\be
		\partial_{\alpha} \equiv \frac{\partial}{\partial x^{\alpha}} = \Big(\frac{\partial}{\partial x^0},\vec{\nabla}\Big)
		\qquad \textrm{and} \qquad
		\partial^{\alpha} \equiv \frac{\partial}{\partial x_{\alpha}} = \Big(\frac{\partial}{\partial x^0},-\vec{\nabla}\Big),
	\ee
	as well as the d'Alembertian
	\be
		\square\equiv\partial^{\mu}\partial_{\mu}=\frac{\partial^2}{\partial t^2}-\nabla^2.
	\ee

	\section*{Electrodynamics}
	In the framework of special relativity, one introduces the 4-vector potential of electrodynamics, of components:
	\be
		A^{\mu} = (A^0, \vec{A}),
	\ee
	with $A^0$ and $\vec{A}$ being respectively the usual scalar and 3-vector potentials. Using this, the electromagnetic field strength tensor $F^{\mu\nu}$ is then defined as
	\be
		F^{\mu\nu} = \partial^{\mu}A^{\nu} - \partial^{\nu}A^{\mu},
	\ee
	and its dual is
	\be
		\widetilde{F}^{\mu\nu}\equiv\frac{1}{2}\ \epsilon^{\mu\nu\rho\sigma}F_{\rho\sigma},
	\ee
	with $\epsilon^{\alpha\beta\rho\sigma}$, the totally antisymmetric tensor.

\chapter{Stokes parameters as intensities}\label{app:stokes}

	Here we recall that the Stokes parameters, written in a given \draftun{arbitrary} basis $(\vec{e}_x, \vec{e}_y)$ as
		\be
                \left\{
                    \begin{split}
                        I(z) &= \langle\mathcal{I}(z,t)\rangle \ \!= \langle \mE_{\mathrm{r}_y}\mEs_{\mathrm{r}_y} + \mE_{\mathrm{r}_x}\mEs_{\mathrm{r}_x}\rangle\\
                        Q(z) &= \langle\mathcal{Q}(z,t)\rangle = \langle \mE_{\mathrm{r}_y}\mEs_{\mathrm{r}_y} - \mE_{\mathrm{r}_x}\mEs_{\mathrm{r}_x}\rangle\\
                        U(z) &= \langle\mathcal{U}(z,t)\rangle = \langle \mE_{\mathrm{r}_x}\mEs_{\mathrm{r}_y} + \mEs_{\mathrm{r}_x}\mE_{\mathrm{r}_y}\rangle\\
                        V(z) &= \langle\mathcal{V}(z,t)\rangle = \langle i(\mE_{\mathrm{r}_x}\mEs_{\mathrm{r}_y} - \mEs_{\mathrm{r}_x}\mE_{\mathrm{r}_y}) \rangle,
                    \end{split} \right.
                    \tag{\ref{eq:StokesE}}
                \ee
	can be built from intensities; this is of course obvious for $I$ which is, as announced, the total intensity, as $\langle \mE_{\mathrm{r}_x} \mEs_{\mathrm{r}_x} \rangle$ and $\langle \mE_{\mathrm{r}_y} \mEs_{\mathrm{r}_y} \rangle$ give the intensities in both directions.

	On the other hand, $Q$ measures the difference between these intensities, and is related to the linear polarisation: it indeed tells if the electric field lies preferentially in the $x$ or in the $y$ direction. It is maximal (resp. minimal) for a light beam fully linearly polarised along the $y$ axis (resp. $x$ axis). Note that it is equal to zero for circularly polarised light, but also for linearly polarised light with a polarisation angle $\varphi_0=\frac{\pi}{4}$. From this last observation, we can say two things: that the value of $Q$ depends on the orthogonal basis we choose, and that we need an additional parameter to fully describe linear polarisation.

	This would be $U$, which has in fact the same structure as $Q$. This can be shown introducing
	\be
		\mE_{{\mathrm{r}_{\pm}}} = \vmE_{\mathrm{r}}\cdot\vec{e}_{\pm}; \qquad \vec{e}_{\pm} = \frac{\vec{e}_x \pm \vec{e}_y}{\sqrt{2}},
	\ee
	where $(\vec{e}_-, \vec{e}_+)$ define another orthogonal basis in the transverse plane, rotated by a $\frac{\pi}{4}$ angle compared to $(\vec{e}_x, \vec{e}_y)$. Then, as
	\be
		\mE_{\mathrm{r}_y} = \frac{1}{\sqrt{2}}\left(\mE_{\mathrm{r}_+} - \mE_{\mathrm{r}_-}\right) \textrm{\quad and \quad} \mE_{\mathrm{r}_x} = \frac{1}{\sqrt{2}}\left(\mE_{\mathrm{r}_+} + \mE_{\mathrm{r}_-}\right),
	\ee
	we indeed find that, in this new basis, $U$ corresponds to
	\be
		U(z) = \langle \mE_{\mathrm{r}_x}\mEs_{\mathrm{r}_y} + \mEs_{\mathrm{r}_x}\mE_{\mathrm{r}_y}\rangle = \langle \mE_{\mathrm{r}_+}\mEs_{\mathrm{r}_+} - \mE_{\mathrm{r}_-}\mEs_{\mathrm{r}_-} \rangle,
	\ee
	which has exactly the same form that $Q$ has in the $(\vec{e}_x, \vec{e}_y)$ basis. The Stokes parameter $U$ therefore compares intensities in two orthogonal directions, which are rotated by a $\frac{\pi}{4}$ angle with respect to the $x$ and $y$ axes.
	\begin{figure}
		\centering
		\includegraphics[width=0.35\textwidth]{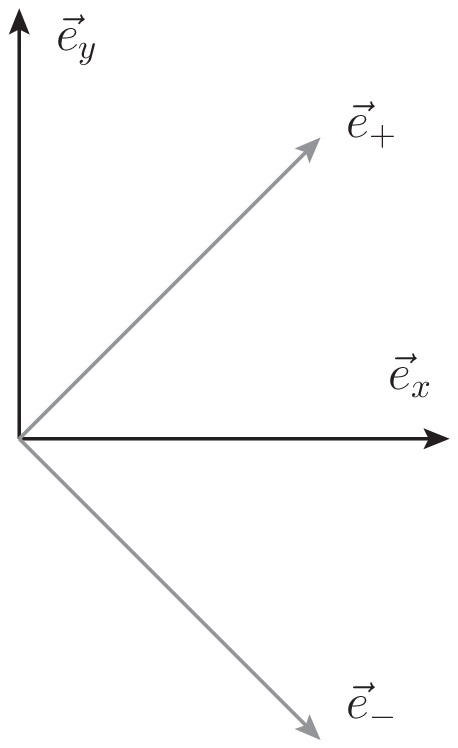}
	\end{figure}

	As we discussed, the individual values of the Stokes parameters describing linear polarisation are basis-dependent. In another basis $(\vec{e}_{\bar{x}},\vec{e}_{\bar{y}})$, in which the electric field has components:
	\be
		\mE_{\mathrm{r}_{\bar{x}}} = \mE_{\mathrm{r}_{x}}\cos\delta + \mE_{\mathrm{r}_{y}} \sin\delta
		\quad\textrm{and}\quad
		\mE_{\mathrm{r}_{\bar{y}}} = \mE_{\mathrm{r}_{y}} \cos\delta - \mE_{\mathrm{r}_{x}} \sin\delta
		\quad(\delta\in\mathbb{R}),
	\ee
	the values calculated for $\bar{Q}$ and $\bar{U}$ obey
	\be
	\left\{
	\begin{split}
		\bar{Q} &= Q \cos\left(2\delta\right) - U \sin\left(2\delta\right)\\
		\bar{U} &= U \cos\left(2\delta\right) + Q \sin\left(2\delta\right),
	\end{split}
	\right.
	\ee
	with $Q$ and $U$, the corresponding parameters obtained in the $(\vec{e}_{x},\vec{e}_{y})$ basis. From this, it follows that the degree of linear polarisation, defined as
	\be
		p_{\mathrm{lin}} = \frac{\sqrt{Q^2 + U^2}}{I} = \frac{\sqrt{\bar{Q}^2 + \bar{U}^2}}{I}
		\tag{\ref{eq:poldegree}}
	\ee
	is independent of the choice of axes; it also follows that the polarisation angle $\varphi$, which can be written as
	\be
		\varphi = \frac{1}{2}\atan\left(\frac{U}{Q}\right),\tag{\ref{eq:varphi}}
	\ee
	is defined modulo $\pi$\hspace{1pt}---\hspace{1pt}in fact, it just gives the orientation of the polarisation plane with respect to \draftun{a chosen} direction.

	Finally, we can show that measuring $V$, the Stokes parameter describing circular polarisation, can be reduced to a linear polarisation measurement with the use of retarders. In the simple monochromatic case for instance, one can first use a quarter-wave plate to induce a relative $\frac{\pi}{2}$ phase shift between $\mE_{\mathrm{r}_x}$ and $\mE_{\mathrm{r}_y}$. At the end of the quarter-wave plate, $\mE_{\mathrm{r}_x}$ is then essentially replaced by $i\mE_{\mathrm{r}_x}$, as one can redefine the electric field to absorb any additional global phase. Then, comparing intensities in the ``$+$'' and ``$-$'' directions (as done by the Stokes parameter $U$) after the retarder, we recover the value of $V$ for the initial light beam. Indeed, we then have:
	\be
		\langle (i\mE_{\mathrm{r}_x})\mEs_{\mathrm{r}_y} + {(i\mE_{\mathrm{r}_x})}^*\mE_{\mathrm{r}_y}\rangle = \langle i(\mE_{\mathrm{r}_x}\mEs_{\mathrm{r}_y} - \mEs_{\mathrm{r}_x}\mE_{\mathrm{r}_y}) \rangle.
	\ee

\chapter{Details of the wave-packet treatment}\label{app:wp}

	Here, we solve the system of Eq.~\eqref{eq:system} in the case of the step-like magnetic field presented in Sec.~\ref{sec:wptreatment}, using the decomposition of Eq.~\eqref{eq:decompsuromega}.
	Note that, as in the plane-wave case, we rephase $\phi(z,t)$ and use the gauge condition $A^0=0$, so that $\widetilde{E}(z,\omega)=i\omega\widetilde{A}(z,\omega)$.

	The solutions in the first region are
	\bea
		E_{\mathrm{I}} (z,t) &=& \int_{-\infty}^{\infty} d\omega\   e^{- i \omega t} i\omega\big[\widetilde{A}_{\mathrm{i, I}}(z = 0,\omega) e^{i k_E z} + \widetilde{A}_{\mathrm{r, I}}(z = 0,\omega) e^{- i k_E z}\big],\\
		\phi_{\mathrm{I}} (z,t) &=& \int_{-\infty}^{\infty} d\omega\   e^{- i \omega t} \big[\widetilde{\phi}_{\mathrm{i, I}}(z = 0,\omega) e^{i k_{\phi} z} \ + \widetilde{\phi}_{\mathrm{r, I}}(z = 0,\omega) e^{- i k_{\phi} z}\big],
	\eea
with the dispersion relations $k_E = \sqrt{\omega^2 - {\omega_{\mathrm{p}}}^2}$ and $k_{\phi} =\sqrt{\omega^2 - m^2}$. Here we have already used the fact that we will always consider amplitudes centered around $\omega_0$, with $\omega_0\gg\omega_{\mathrm{p}}, m$, and decreasing sufficiently quickly with $\omega$ for the contributions from $\omega\leq\omega_{\mathrm{p}}, m$ to be negligible. Similarly, the solutions in the second region read
        \begin{multline}
		E_{\mathrm{II}} (z,t) = \int_{-\infty}^{\infty} d\omega\   e^{- i \omega t}   i \omega  \big[ \widetilde{A}_{\mathrm{i, II}}(z = 0,\omega) \big((\textrm{c}\theta)^2 e^{i k_C z} + (\textrm{s}\theta)^2 e^{i k_D z}\big)\\
+  \widetilde{\phi}_{\mathrm{i, II}}(z = 0,\omega) \frac{\sin(2\theta)}{2}\big(e^{i k_C z} - e^{i k_D z}\big)\big],
	        \label{eq:EIIxtevolution}
	\end{multline}
        \begin{multline}
		\phi_{\mathrm{II}} (z,t) = \int_{-\infty}^{\infty} d\omega\    e^{- i \omega t} \big[ \widetilde{A}_{\mathrm{i, II}}(z = 0,\omega)  \frac{\sin(2\theta)}{2}\big(e^{i k_C z} - e^{i k_D z}\big)\\
+  \widetilde{\phi}_{\mathrm{i, II}}(z = 0,\omega) \big((\textrm{s}\theta)^2 e^{i k_C z}  +  (\textrm{c}\theta)^2 e^{i k_D z}\big)\big],
	        \label{eq:phiIIxtevolution}
	\end{multline}
	where $k_C$ and $k_D$ are respectively $k_+=\sqrt{\omega^2-{\mu_+}^2}$ and $k_-=\sqrt{\omega^2-{\mu_-}^2}$ when $\omega_{\mathrm{p}}>m$, and the other way around when $m>\omega_{\mathrm{p}}$.
	As we are focusing on the small mixing case in particular, remember that the heaviest eigenmode of propagation is then mostly made of the heaviest state among photons and pseudo\-scalars, and conversely for the lightest one, as we discussed with plane waves.
\bigskip

	The amplitudes $\widetilde{A}_{\mathrm{i, I}}(0,\omega)$, $\widetilde{A}_{\mathrm{r, I}}(0,\omega)$, $\widetilde{\phi}_{\mathrm{i, I}}(0,\omega)$, $\widetilde{\phi}_{\mathrm{r, I}}(0,\omega)$, $\widetilde{A}_{\mathrm{i, II}}(0,\omega)$ and $\widetilde{\phi}_{\mathrm{i, II}}(0,\omega)$ are determined by initial and boundary conditions. They correspond to incident ($\mathrm{i}$) or reflected ($\mathrm{r}$) amplitudes that appear as light goes from region I into the potential barrier.
To simplify our discussion, we now work in the case where there is no incident pseudo\-scalar flux in region I, namely ${\phi}_{\mathrm{i, I}}(0,\omega) = 0$.

	The continuity requirements on $E(z,t)$ and $\phi(z,t)$, and on their first derivatives with respect to $z$, at $z=0$ then lead to relations where the only free parameter left is $\widetilde{A}_{\mathrm{i, I}}(z=0,\omega)$. For completeness, they are
        \begin{multline}
		\widetilde{A}_{\mathrm{i, II}} (z=0, \omega) = \\  2k_E \Big[   \frac{k_C (\textrm{s}\theta)^2 + k_D (\textrm{c}\theta)^2 + k_{\phi}}{k_E k_{\phi} + k_C k_D + k_E \big( k_C(\textrm{s}\theta)^2 + k_D (\textrm{c}\theta)^2 \big) + k_{\phi} \big( k_C (\textrm{c}\theta)^2 + k_D (\textrm{s}\theta)^2 \big)}   \Big] \widetilde{A}_{\mathrm{i, I}}(0,\omega)\\
		\equiv \mathscr{V}\ \widetilde{A}_{\mathrm{i, I}}(0,\omega),
        	\label{eq:AiIIcoeff}
	\end{multline}
        \begin{multline}
		\widetilde{\phi}_{\mathrm{i, II}}(z=0,\omega) = \widetilde{\phi}_{\mathrm{r, I}}(z=0,\omega) = \\
 \Big[   \frac{k_E \big(k_C - k_D\big) \sin(2\theta)}{k_E k_{\phi} + k_C k_D + k_E \big( k_C(\textrm{s}\theta)^2 + k_D (\textrm{c}\theta)^2 \big) + k_{\phi} \big( k_C (\textrm{c}\theta)^2 + k_D (\textrm{s}\theta)^2 \big)}   \Big] \widetilde{A}_{\mathrm{i, I}}(0,\omega)\\
	\equiv \mathscr{W}\ \widetilde{A}_{\mathrm{i, I}}(0,\omega),
	        \label{eq:phiiIIcoeff}
	\end{multline}
        \begin{multline}
		\widetilde{A}_{\mathrm{r, I}}(z = 0,\omega) = \left(\mathscr{V} - 1\right) \widetilde{A}_{i,I}(0,\omega).\\
        \end{multline}

	In the case of an incident Gaussian wave packet
	\be
		\draft{\widetilde{{E}}_{\mathrm{i, I}}(z = 0,\omega) = \widetilde{{E}}_0\ e^{-\frac{a^2}{4}(\omega-\omega_0)^2},}
	\ee
	we obtain the following result for $E_{\mathrm{II}}(z,t)$ (which is the only amplitude entering the expressions of the Stokes parameters in the second region):
	\begin{multline}
		 E_{\mathrm{II}}(z,t) = \int_{-\infty}^{\infty} d\omega\ 
		 \draft{\widetilde{{E}}_0}\Big[\frac{\mathscr{V} - i \mathscr{W}}{4} \exp(-\frac{a^2}{4}(\omega-\omega_0)^2 + i(k_Cz - \omega t + 2\theta)) \\
		 + \frac{\mathscr{V} +  i \mathscr{W}}{4} \exp(-\frac{a^2}{4}(\omega-\omega_0)^2 + i(k_Cz - \omega t - 2\theta)) \\
		 + \frac{\mathscr{V}}{2} \exp(-\frac{a^2}{4}(\omega-\omega_0)^2 + i(k_Cz - \omega t)) \\
		 - \frac{\mathscr{V} - i \mathscr{W}}{4} \exp(-\frac{a^2}{4}(\omega-\omega_0)^2 + i(k_Dz - \omega t + 2\theta)) \\
		 - \frac{\mathscr{V} + i \mathscr{W}}{4} \exp(-\frac{a^2}{4}(\omega-\omega_0)^2 + i(k_Dz - \omega t - 2\theta)) \\
		 + \frac{\mathscr{V}}{2} \exp(-\frac{a^2}{4}(\omega-\omega_0)^2 + i(k_Dz - \omega t))];\label{Eparallel_generalexplicit}
	\end{multline}
	where $\mathscr{V}$, $\mathscr{W}$, $k_C$, $k_D$ and $\theta$ are functions of $\omega$. 
	Note that if $g\mathcal{B}$ is set to zero, this reduces to
	\be
		E_{\mathrm{II}}(z,t;g\mathcal{B}=0)=E_{\perp}(z,t) = \int_{-\infty}^{\infty} d\omega\ \draft{\widetilde{{E}}_0}\exp(-\frac{a^2}{4}(\omega-\omega_0)^2 + i(k_Ez - \omega t)).
		\label{E_gbzero}
	\ee
We finally Taylor expand the coefficients and the arguments of the exponentials around $\omega_0$ (up to the second order) to carry out the integrals~\eqref{Eparallel_generalexplicit} and \eqref{E_gbzero} analytically to better than 1\% for the case at hand (as was checked by estimating the contribution of the next order).

\chapter{Theoretical distributions}\label{app:distributions}

	We show distributions for the linear and the circular polarisations obtained after 5000 different magnetic field realisations for ALP parameters that are excluded according to the method that we propose in Chap.~\ref{chap:constraints}. We reproduce what is obtained for two points taken from Fig.~\ref{fig:exclusion}, both for an ALP mass $m=10^{-20}$~eV, but for different values of the coupling.

\begin{figure}
			\centering
			\includegraphics[height=5.8125cm]{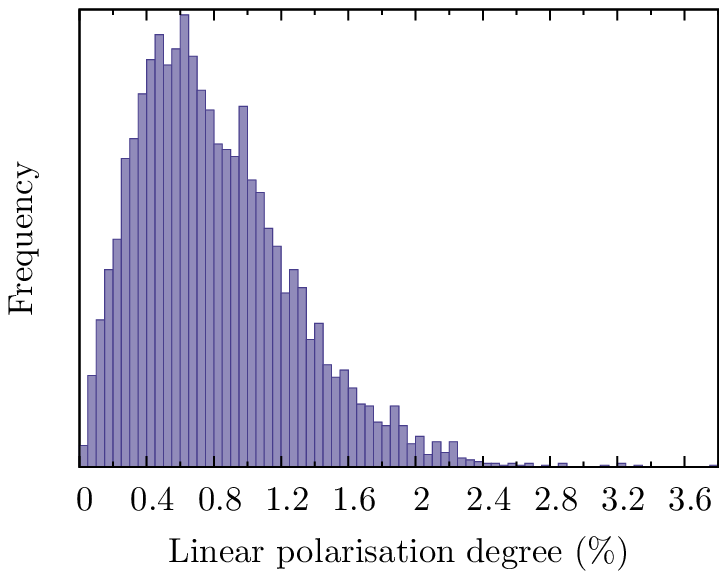}
			\includegraphics[height=5.8125cm]{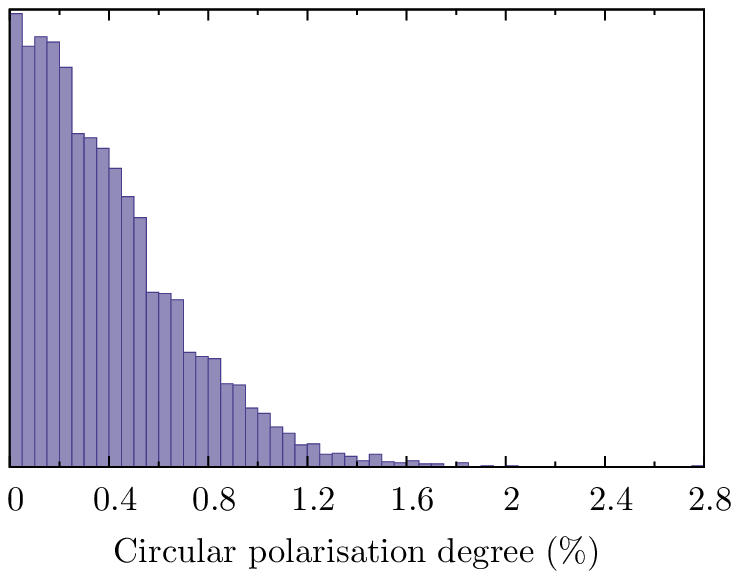}
			\caption{Theoretical distributions of polarisation for 5000 randomly generated configurations for initially unpolarised light. Here, $m=10^{-20}$~eV and $g=2.5\times10^{-13}$~GeV$^{-1}$; parameters and fluctuations are the same as in Fig.~\ref{fig:exclusion}.}
			\label{fig:2s}
		\end{figure}

\begin{figure}
			\centering
			\includegraphics[height=5.8125cm]{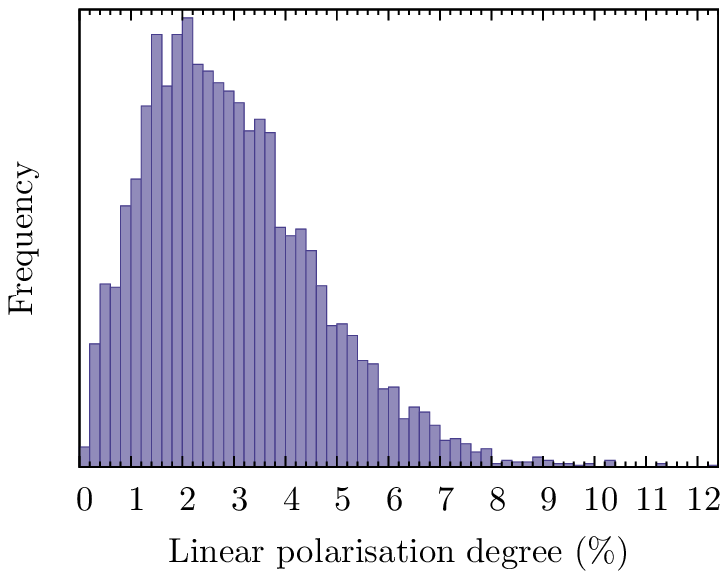}
			\includegraphics[height=5.8125cm]{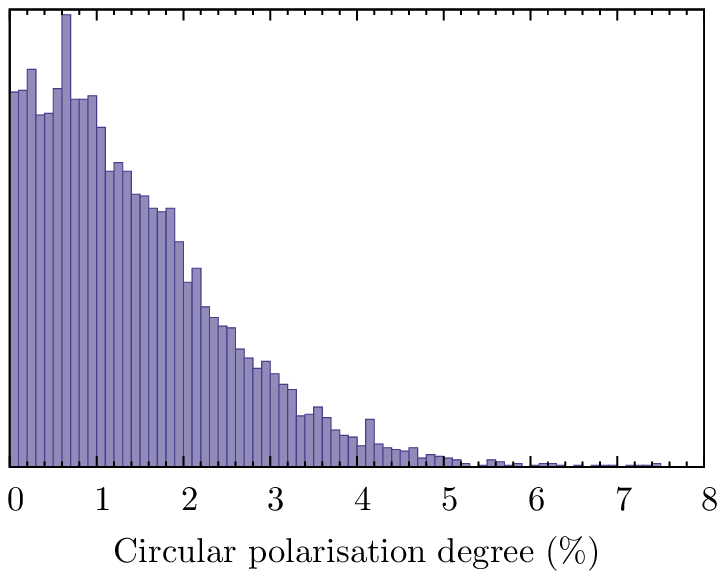}
			\caption{Theoretical distributions of polarisation for 5000 randomly generated configurations for initially unpolarised light. Here, $m=10^{-20}$~eV and $g=5\times10^{-13}$~GeV$^{-1}$; parameters and fluctuations are the same as in Fig.~\ref{fig:exclusion}.}
			\label{fig:3s}
		\end{figure}

	In Fig.~\ref{fig:2s}, we show the theoretical distribution of polarisations obtained for initially unpolarised light for a coupling $g=2.5\times10^{-13}$~GeV$^{-1}$, which is the lower bound of parameters excluded at 2$\sigma$ for nearly massless ALPs.
	In Fig.~\ref{fig:3s}, we show the same for a coupling $g=5\times10^{-13}$~GeV$^{-1}$, \textit{i.e.} the 3$\sigma$ lower bound for nearly massless ALPs.

	In both case, we see that ALP-photon mixing produces much more polarisation than observed (Figs.~\ref{fig:balnonbal} and~\ref{fig:pcirc}); in particular, the circular polarisation is very large. For values of the coupling close to the current bound from SN1987A, the linear and circular polarisations go above 50\% of polarisation in some configurations while we would expect at most around 1\% for the linear polarisation and a very tiny circular polarisation.

\chapter{\draft{Constraints in other magnetic fields}}\label{app:constraints}

	We illustrate the constraints that we obtain when we use other morphologies for the magnetic field of the local supercluster.

\begin{figure}[h!]
			\centering
			\includegraphics[width=\textwidth]{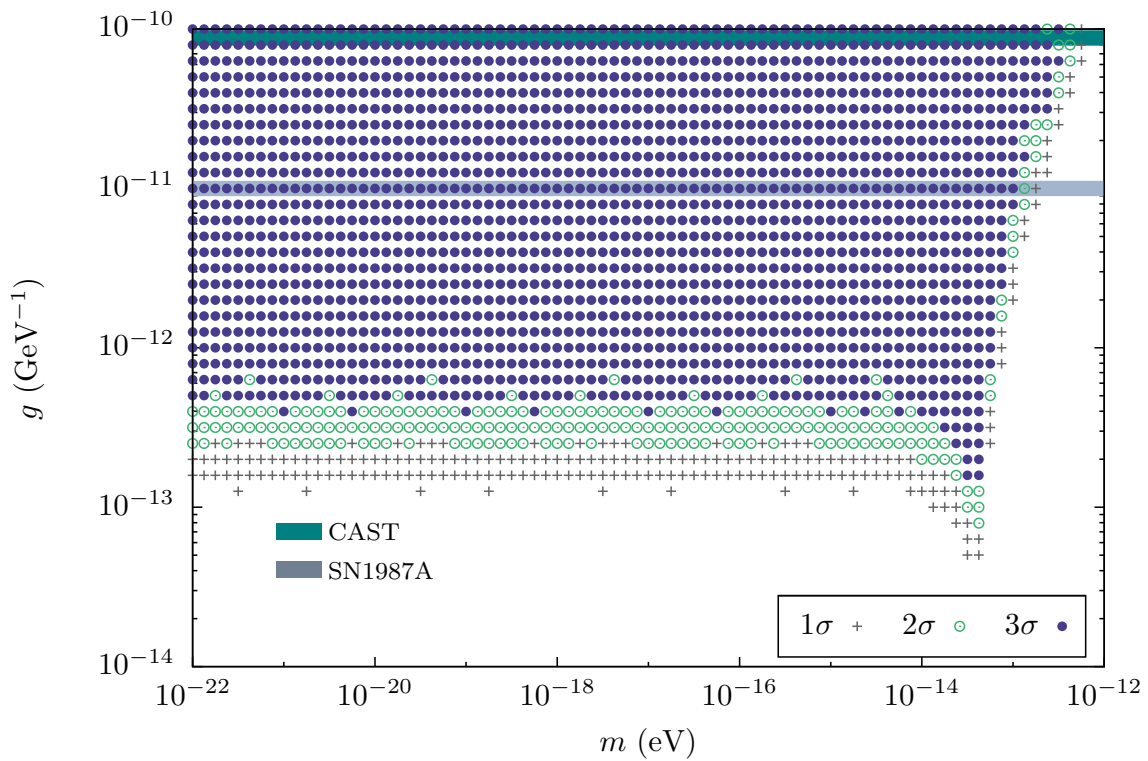}
			\caption{Exclusion plot. Same as Fig.~\ref{fig:exclusion}, but we also add a uniform transverse contribution of 0.3~$\mu$G among domains to the random field of $|\vec{B}_{\mathrm{domain,0}}|=2~\mu$G here. Our constraints remain stable.}
			\label{fig:const_patchyetfond}
		\end{figure}

	As stated in Chap.~\ref{chap:constraints}, including a weaker underlying uniform field, that can allow for correlations among domains of magnetic field, does not change our results. This is shown in Fig.~\ref{fig:const_patchyetfond} (note that we show low-statistics runs in this appendix).

\begin{figure}[h!]
			\centering
			\includegraphics[width=\textwidth]{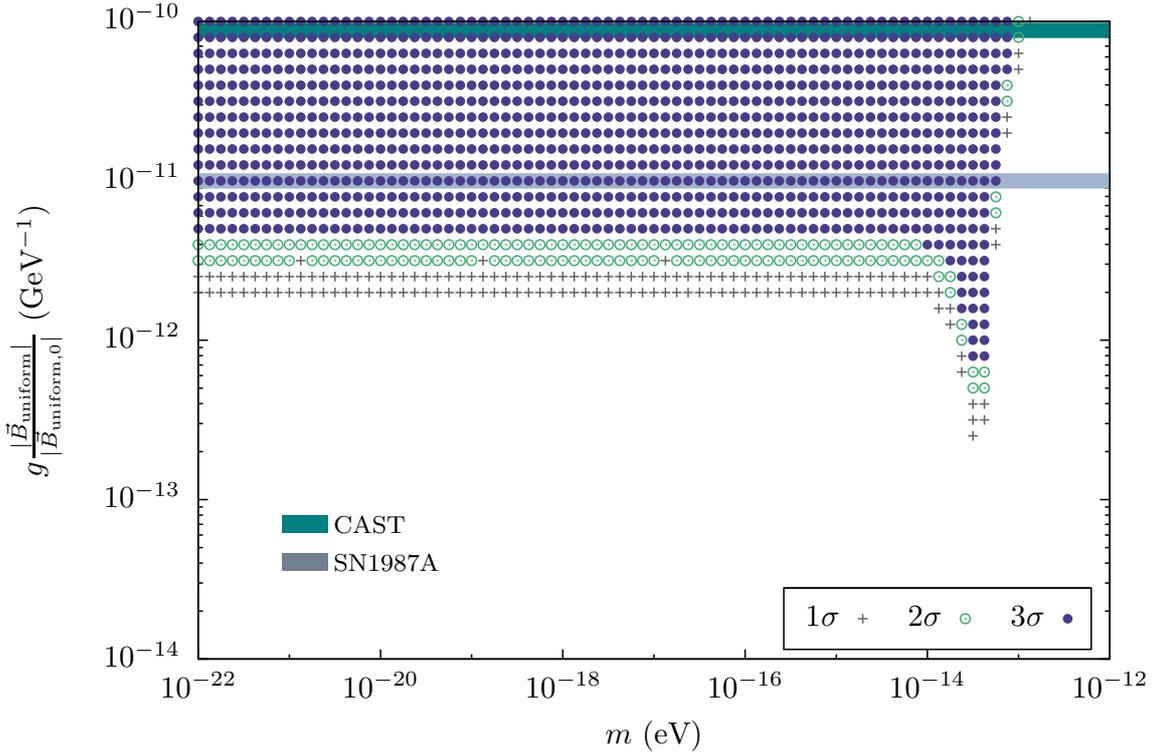}
			\caption{Exclusion plot. The total distance considered is $z_{\mathrm{tot}}=10$~Mpc; the average domain size is 100~kpc; the transverse magnetic field strength here is $\mathcal{B}=|\vec{B}_{\mathrm{uniform,0}}|=0.3~\mu$G, with constant direction throughout; the average electron density is $n_{\mathrm{e}}=10^{-6}$~cm$^{-3}$. This shows the importance of the constraint from circular polarisation, and from the fluctuating field.}
			\label{fig:constr_fondonlyfluct}
		\end{figure}

\begin{figure}[h!]
			\centering
			\includegraphics[width=\textwidth]{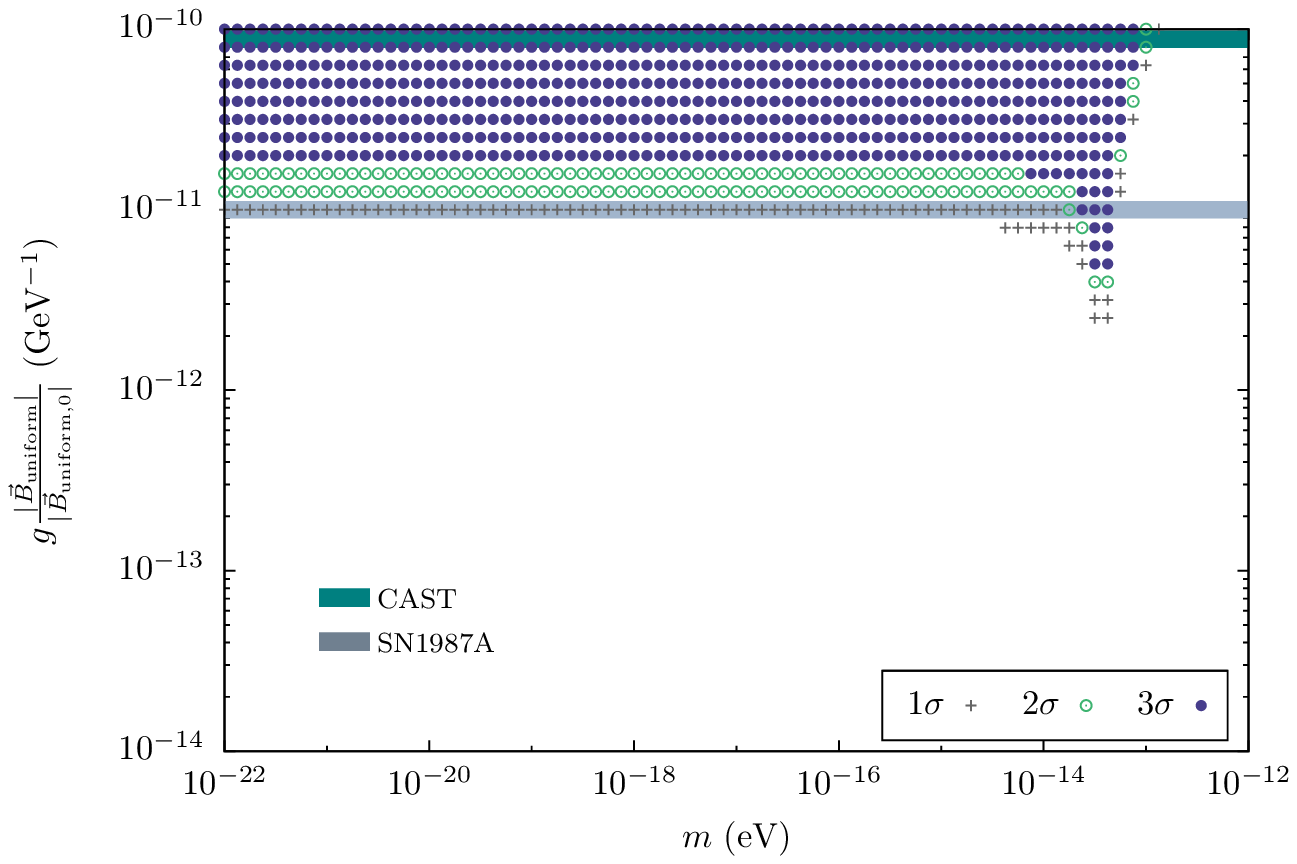}
			\caption{Exclusion plot. Same as Fig.~\ref{fig:constr_fondonlyfluct}, but without any fluctuations. This shows the importance of the fluctuations on $n_{\mathrm{e}}$.}
			\label{fig:constr_fondonlypasfluct}
		\end{figure}

	For completeness, we also show the constraints that we obtain when we apply our method in the uniform-field scenario:
	\begin{itemize}
		\item[-] in Fig.~\ref{fig:constr_fondonlyfluct}, where we allow for fluctuations of $n_{\mathrm{e}}$ among $\sim100$-kpc domains.
		\item[-] in Fig.~\ref{fig:constr_fondonlypasfluct}, without any fluctuations.
	\end{itemize}

	The problem in these cases is that the constraints obtained are highly dependent on the assumption made on the initial polarisation of quasars. For the domain-structure magnetic field, we have conservatively allowed the initial quasar polarisation to be zero, to avoid any overestimation of the final polarisation due to the mixing with axion-like particles. This is in fact too restrictive in the case of a uniform field as the mixing in a single zone will not produce circular polarisation if we start from unpolarised light, while a deviation from uniformity would lead to the generation of circular polarisation in that case.
	This is illustrated by the fact that the constraints are weaker than what we have for $|\vec{B}_{\mathrm{domain}}|=0.3~\mu$G in Fig.~\ref{fig:exclusion}, despite the coherence length being a hundred times smaller and the presence of tridimensional rotations allowing for smaller values of the transverse magnetic field.

\chapter{No radio alignments from ALPs}\label{app:radio}

		If there are similar alignment effects in different energy domains over cosmological scales, it is quite natural to think that they have the same origin.
		In the following, we emphasise~\cite{Payez:2012rc} that an observation of alignments in radio wavelengths as reported in Ref.~\cite{Tiwari:2012rr} cannot be explained by the mixing of photons with spinless particles in external magnetic fields. On the one hand, this is not expected in the (already excluded) scenario associated with alignments in visible light; on the other hand, if the mixing was efficient enough in radio waves, then there would be a strong contradiction with polarisation data in visible light.

		Let us consider the emission of light from a given quasar, and focus on two values of the frequency: $\omega_1$ and $\omega_2$, which will be redshifted as light propagates. Let us choose $\omega_1$ in such a way that it will be observed as visible light once it reaches us, $\omega_V = 2.5$~eV corresponding to 500~nm; while $\omega_2$ will be redshifted in the radio band, $\omega_R = 3.474\times10^{-5}$~eV corresponding to the observations at 8.4~GHz. For these two beams, the external conditions will be the same as they propagate towards us, so that we have at all times:\footnote{Here, we neglect the tiny difference of group velocity which is formally caused by the non-zero plasma frequency. Equivalently, we can suppose that the external conditions do not change on the time scale which separates the two wave fronts.}
		\be
		\frac{\tan\left(2\theta(\omega_2)\right)}{\tan\left(2\theta(\omega_1)\right)} = \frac{\omega_R}{\omega_V}=1.4\times10^{-5},\label{eq:compare_theta}
		\ee
		as these beams are redshifted in the same way.
		From Eq.~\eqref{eq:compare_theta} and the discussion above, it is already clear that the effect due to pseudoscalar-photon mixing in radio waves is inefficient compared to the one in visible light.

		As discussed in~Chap.~\ref{chap:alignments}, in order to produce an additional polarisation similar to the one needed in optical wavelengths, $\theta(\omega_1)=0.1$ was a typical value.
		Now, to determine the corresponding value for $\omega_2$, we can approximate $\tan\left(x\right)\approx x$ in Eq.~\eqref{eq:compare_theta}. Doing so, we introduce a relative error slightly bigger than 1\% for $\tan\left(2\theta(\omega_1)\right)$, but it allows us to continue the discussion in the general case.\footnote{One can also check this result using directly $\omega_V$ and $\omega_R$, and choosing values for the parameters $g$, $m$, $\mathcal{B}$, and $\omega_{\mathrm{p}}$ such that $\theta(\omega_V)=0.1$. With the same parameters, one can then calculate $\theta(\omega_R)$.}
		We then finally obtain that, while $\theta(\omega_1)=0.1$, the mixing angle corresponding to the other light beam is as small as $\theta(\omega_2)=1.4\times10^{-6}$ under the same external conditions.

		In order to give a quantitative estimate of the additional polarisation that is typically brought by the mixing in both cases, let us consider initially unpolarised light beams in a magnetic field region, in the absence of an initial propagating pseudoscalar field $\phi(0)$.
		As we have shown in Chap.~\ref{chap:mixing}, in such a case, the degree of linear polarisation evolves in the following way:\footnote{Note that, formally, one should take Faraday rotation into account at radio wavelengths.}
		\be
			p_{\mathrm{lin}}(z)=\frac{\frac{1}{2}\sin^2 2\theta\sin^2\left(\frac{1}{4}\frac{\Delta\mu^2}{\omega}z\right)}
					   {1 - \frac{1}{2}\sin^2 2\theta\sin^2\left(\frac{1}{4}\frac{\Delta\mu^2}{\omega}z\right)}.
			\tag{\ref{eq:plinrtheta}}
		\ee
		One can then drop the information associated with the propagation and simply check the maximum amount of linear polarisation that can be achieved in this region, namely,
		\be
			p_{\mathrm{lin}}\big|_{\mathrm{max}}(\omega)=\frac{\frac{1}{2}\sin^2 2\theta(\omega)}
							   {1 - \frac{1}{2}\sin^2 2\theta(\omega)}.
			\label{eq:plinmax}
		\ee
		Finally, we use the values of $\theta$ that we obtained for $\omega_1$ and for $\omega_2$, and replace them in Eq.~\eqref{eq:plinmax}.
		For an additional linear polarisation of $p_{\mathrm{lin}}\big|_{\mathrm{max}}(\omega_1)=2\%$ for what would be visible light, we only have at most a very tiny $p_{\mathrm{lin}}\big|_{\mathrm{max}}(\omega_2)=4\times10^{-10}\%$ in the other case, which is far smaller than what can be detected experimentally.
		The mixing of photons with spinless particles in external magnetic fields  thus cannot produce, for the same source, an observable effect in different energy regimes such as visible and radio. Note, of course, that an alignment sufficiently important in radio waves with this mechanism would imply too much polarisation in visible light, which would contradict the observations: the observed polarisation in visible light is indeed mainly of intrinsic origin, as discussed in Sec~\ref{sec:descriptionalignment}.

		While things can become more elaborate in more complex magnetic fields, the phenomenology we have discussed remains the same: an effect in radio wavelengths would be so limited that we should not expect to observe it, according to the scenario in which ALPs would have provided the mechanism responsible for coherent alignments in visible light. Additionally, other magnetic field configurations will not produce more polarisation than this toy model in general, as fluctuations tend to diminish the amount created via the mixing.\footnote{As we discussed in Chap~\ref{chap:mixing}, if one keeps $\phi(0)\neq0$, while the evolution of the polarisation due to the mixing is more complicated, the relevant parameters which drive the change of polarisation remain the two dimensionless quantities $\theta$ and $\frac{\Delta\mu^2}{\omega}z$.}

		In conclusion, pseudoscalar-photon mixing in external magnetic fields cannot explain the recent claim of very significant large-scale alignments of quasar polarisations in radio wavelengths either.
		It could be that the two very similar observations of large-scale coherent orientations of polarisation of radio waves and optical light from quasars require completely different physical explanations. Nevertheless, we stress that the spinless-particle scenario simply fails to reproduce polarisation data both in visible and in radio wavelengths. In the first case, the price to pay is the introduction of a circular polarisation problem which directly contradicts high-precision polarisation data. In the second case, it would lead to an extremely efficient mixing in visible light which would contradict high-precision linear polarisation data.

		On the other hand, while more quantitative predictions are still needed, we simply note that some effort has been done in new directions to try to explain the alignments in visible light~\cite{Urban:2011,*Antoniou:2010gw,*Poltis:2010yu,*Ciarcelluti:2012pc,*MosqueraCuesta:2011tz}, and that some of these could naturally explain alignments in radio waves, without generating any circular polarisation.

\end{appendices}

\backmatter

\chapterstyle{myOtherStyle}

\small

\bibliographystyle{mystyle}
\bibliography{alexbib}

\end{document}